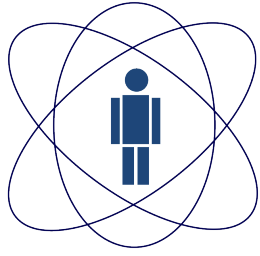

# Axions, Photons and Physics Beyond the Standard Model

*Autor:*

Jefferson Mendes Aguiar Paixão

*Orientador:*

Prof. Dr. José Abdalla Helayël-Neto

*Co-orientador:*

Prof. Dr. Mario Junior de Oliveira Neves

31 de julho de 2024

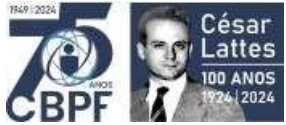
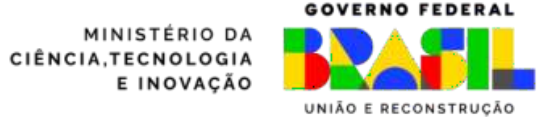

# "ÁXIONS, FÓTONS E FÍSICA ALÉM DO MODELO-PADRÃO"

## JEFFERSON MENDES AGUIAR PAIXÃO

Tese de Doutorado em Física apresentada no Centro Brasileiro de Pesquisas Físicas do Ministério da Ciência Tecnologia e Inovação. Fazendo parte da banca examinadora os seguintes professores:

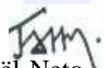

José Abdalla Helayël-Neto – Orientador/CBPF

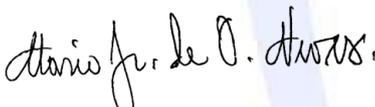

Mario Junior Neves – Coorientador/ UFRRJ

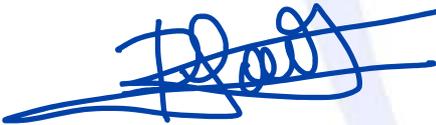

Patrício Alfredo Gaete Durán – UTFSM

Manoel Messias Ferreira Junior – UFMA

Antônio Duarte Pereira Júnior – UFF

Humberto Belich Junior – UFES

Rio de Janeiro, 22 de março de 2024.

*Dedico esta tese à memória de Kayo Sérgio Fontenele Paixão.*

# Agradecimentos

# Resumo

Nesta tese, reavaliamos alguns aspectos da eletrodinâmica axiônica, acoplando efeitos eletromagnéticos não lineares à física do axion. Apresentamos várias motivações para justificar o acoplamento do axion ao fóton em termos de uma extensão não linear geral do setor eletromagnético. Nossa ênfase no artigo se baseia na investigação dos tensores de permitividade e permeabilidade constitutivos, nos quais a contribuição do axion introduz uma dependência da frequência e do vetor de onda da radiação em propagação. Além disso, destacamos como a massa do axion e a constante de acoplamento axion-fóton contribuem para um comportamento dispersivo das ondas eletromagnéticas, em contraste com o que acontece no caso de extensões não lineares, quando índices refrativos efetivos aparecem que dependem apenas da direção da propagação em relação aos campos externos. O axion altera essa imagem produzindo índices refrativos com dependência no comprimento de onda. Aplicamos nossos resultados ao caso especial da Eletrodinâmica de Born-Infeld e mostramos que ela se torna birrefringente sempre que o axion está acoplado. Também incluímos em nossa análise o fenômeno de violação de Lorentz por meio de um termo Carroll-Field-Jackiw, no qual investigamos a influência desse setor nas relações de dispersão e, consequentemente, nos fenômenos ópticos.

# Abstract

In this thesis, we re-assess some aspects of axionic electrodynamics by coupling non-linear electromagnetic effects to axion physics. We present a number of motivations to justify the coupling of the axion to the photon in terms of a general non-linear extension of the electromagnetic sector. Our emphasis in the paper relies on the investigation of the constitutive permittivity and permeability tensors, for which the axion contribution introduces a dependence on the frequency and wave vector of the propagating radiation. Also, we point out how the axion mass and the axion-photon coupling constant contribute to a dispersive behavior of the electromagnetic waves, in contrast to what happens in the case of non-linear extensions, when effective refractive indices appear which depend only on the direction of the propagation with respect to the external fields. The axion changes this picture by yielding refractive indices with dependence on the wavelength. We apply our results to the special case of the Born-Infeld Electrodynamics and we show that it becomes birefringent whenever the axion is coupled. We also include in our analysis the Lorentz violation phenomenon through a Carroll-Field-Jackiw term, where we investigate the influence of this sector on the dispersion relations, and consequently, on the optical phenomena.

# List of Abbreviations

| | |
|---|---|
| **ALPS** | Axion-like Particles |
| **BI** | Born-Infeld |
| **CAST** | CERN Axion Solar Telescope |
| **CFJ** | Carroll-Field-Jackiw |
| **DM** | Dark Matter |
| **DR** | Dispersion Relation |
| **ED** | Electrodynamics |
| **EH** | Euler-Heisenberg |
| **JWST** | James Webb Space Telescope |
| **LSV** | Lorentz Symmetry Violation |
| **LQG** | Loop Quantum Gravity |
| **nEDM** | Neutron Electric Dipole Moment |
| **NED** | Nonlinear Electrodynamics |
| **PQ** | Peccei-Quinn |
| **PVLAS** | Polarisation of Vacuum with LASER |
| **QCD** | Quantum Chromodynamics |
| **QED** | Quantum Electrodynamics |
| **SME** | Standard Model Extension |
| **SM** | Standard Model |
| **VMB** | Vacuum Magnetic Birefringence |

# Contents

# Chapter 1

# Introduction

Axions are actively investigated in the literature ever since their proposal by Peccei and Quinn to solve the problem of strong CP-violation [1, 2]. More generally, inspired by axionic QCD, Axion-like Particles (ALPs) are treated as pseudo Nambu-Goldstone bosons that arise in various extensions of the Standard Model (SM), and are promising candidates for a dark matter portion in the universe [3, 4, 5]. Unlike of the axionic QCD, in which axions couple with the gluons, the scalar ALPs ($\phi$) couple with the photon through the interaction $g_{a\gamma} \phi (\mathbf{E} \cdot \mathbf{B})$, where $g_{a\gamma}$ is a coupling constant with length dimension. Several efforts have been made in an attempt to detect these particles, whether in astrophysical observations or in terrestrial experiments such as particle accelerators, or high-intensity lasers. It is important to emphasize the huge range of possibilities for the mass of the ALPs. In this sense, there are two perspectives in the search for ALPs, the first one takes into account the scattering processes that are capable of producing ALPs in the mass range eV $-$ TeV. The second one considers astrophysical observations, where it is capable to produce ALPs with an upper bound for the mass which may be in the range of $10^{-10}\,\text{eV} - 30\,\text{KeV}$ [6]. For instance, the Chandra's data analysis for the active galactic nucleus NGC 1275 at the center of the Perseus cluster provides the most stringent limit on the ALP-photon coupling constant for very light ALPs, *i.e.*, $g_{a\gamma} < (6.3 - 7.9) \times 10^{-13}\,\text{GeV}^{-1}$ for $m_a < 10^{-12}\,\text{eV}$ depending on the magnetic field, at 99.7% confidence level [7]. Recent searches for ALP-Photon resonant conversion on magnetar SGR J1745-2900 exclude couplings $g_{a\gamma} > 10^{-12}\,\text{GeV}^{-1}$ for $m_a \leq 10^{-6}\,\text{eV}$ [8]. Another well-established limit was obtained in CAST, which searches for axions coming from the solar core by converting the X-rays into axions via a magnetic field up to 9.5 T. They report the upper limit on the $g_{a\gamma} \simeq 0.66 \times 10^{-10}\,\text{GeV}^{-1}$ for $m_a < 0.02\,\text{eV}$ at 95% confidence level [9]. The figure 1.1 provides a more complete analysis of exclusion zones for ALP-photon coupling by ALP mass.

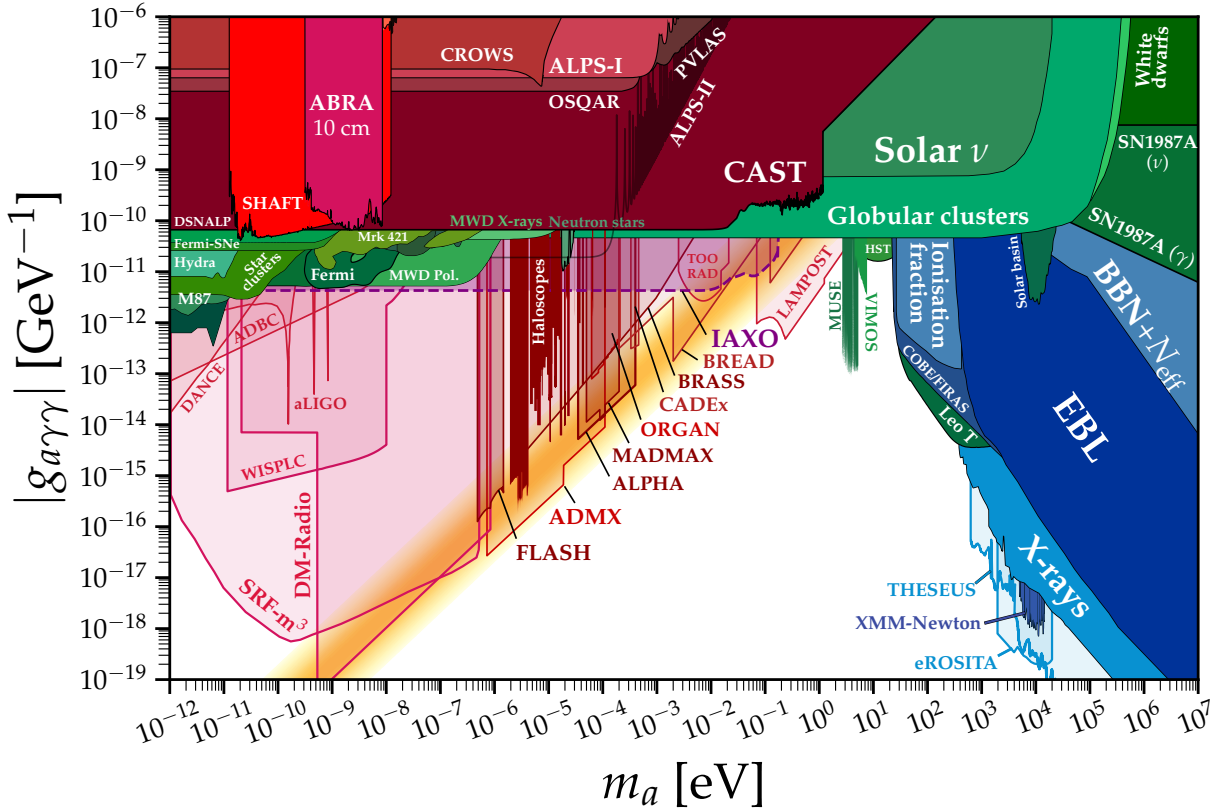

*Figura 1.1:* Exclusion zones for ALP-photon coupling constant versus ALP mass from several experiments and simulations. Source: [17]

For larger values of the ALPs masses, we must take into account the bounds obtained by experiments in particle accelerators. As example, axions can be produced in the reaction $\gamma\gamma \to \phi \to \gamma\gamma$ through the Primakoff-process at the LHC. In this case, there is a wide range for the ALPs masses, ranging from eV to TeV scale [10, 11, 12]. For lead-lead collisions, the exclusion limits for the ALP masses is $m_a \simeq (5-100)\,\text{GeV}$, for a coupling constant of $g_{a\gamma} \simeq 0.05\,\text{TeV}^{-1}$ at 95 % confidence level [13]. Since ALPs are searched for in experiments with intense magnetic fields, it is reasonable to expect that non-linear electrodynamics (NED) effects might be excited in this situation. They may arise whenever magnetic fields are close to the critical Schwinger magnetic field, $|\mathbf{B}|_S = m_e^2/q_e = 4.41 \times 10^9\,\text{T}$ [14]. In astrophysical searches for ALPs, there are cases of blazars and magnetars with magnetic fields of the order or higher than the critical Schwinger magnetic field [15, 16, 8]. In a recent work, it has been discovered that non-linearities can also arise at low magnetic fields in a QED level for the so-called Dirac materials. In this case, strong non-linear effects arise in dirac materials for a magnetic field strength of approximately 1 T [18]. Several non-linear theories are candidates to extension of the Maxwell electrodynamics (ED) in the literature [19, 20, 21, 22, 23, 24, 25]. One of the most known is the Born-Infeld (BI) NED that was originally proposed to remove the singularity of the electric field of a point-like charged particle at the origin [26]. It is worth mentioning that BI electrodynamics was investigated in a scenario involving axions, in which the analog of Snell's law was found considering an axionic domain wall [27]. Currently, the BI emerges in scenarios of superstring theory, quantum-gravitational models and magnetic monopoles [28, 29, 30, 31, 32, 33, 34]. The measurement of the light-by-light scattering at the ATLAS Collaboration of the LHC imposes a lower bound $\gtrsim$ 100 GeV on the BI parameter [35]. More stringent bounds on the BI parameter are also discussed in the electroweak model with the hypercharge sector associated with the BI model [36].

It is known that the vacuum can be considered as a non-linear optical medium and that this concept applies to the standard model of elementary particles [37]. In this sense, non-linear ED models may present dichroism and vacuum birefringence phenomena. In particular, the vacuum magnetic birefringence (VMB) is a macroscopic quantum effect predicted by QED [38, 39] in which the difference of the refractive indices between parallel and perpendicular polarized light is non-trivial in the presence of an external magnetic field. The Polarisation of Vacuum with LASER (PVLAS) experiment carried out 25 years of efforts in the search for the vacuum birefringence and dichroism, and although it did not reach the values predicted by the QED, it established the best limits known so far [41, 40, 42, 43]. Although the VBM has not yet been directly detected, the indirect evidence has been found in the neutron star RX J1856.5 − 3754, with magnetic fields on the order of $10^{13}$ Gauss (G) [44]. It is interesting to notice that the VBM phenomenon can be a tool for the detection of ALPs, where it is known that the conversion of photons into axions changes the polarization of the incident beam by means of a magnetic background field. Thereby, a measure of the birefringence can provide important limits on the axion mass and the coupling constant $g_{a\gamma}$ [46, 45]. In this direction, we point out that the birefringence effects related to the axion field in the presence of a laser beam were investigated in ref. [47]. We also highlight that the birefringence phenomena associated with a pure electric background field is due to the optical Kerr effect [48, 49].

At this stage, it should be mentioned that other extensions of the Maxwell electrodyna-mics (ED) coupled to the axion field have been investigated in the literature. For instance, in a seminal paper by Raffelt and Stodolsky [50], the authors included the non-linear effects of the Euler-Heisenberg electrodynamics and obtain the correspondent dispersion relations of the non-linear photon. We also highlight that the axion field theory was considered in connection with high-order derivative Podolsky ED [51], where the effective photonic theory and the inter-particle potential were carried out. Similarly, the axion field contributions were investigated in the context of non-commutative field theories [52, 53]. Furthermore, in the work of ref. [54], by using the Proca theory, the authors analyzed the influence of a massive photon and its effects on the axion-photon mixing. Later, the axion-Proca ED was obtained as an effective field theory in a Condensed Matter system [55]. Recently, new extensions involving a hidden photon (another massive dark matter candidate) coupled to the axion and photon fields have been proposed in the literature [56, 57]. The propagation effects in the presence of extra CPT-odd terms also motivate other extensions of the Maxwell ED [58]. The study of the constitutive relations on the wave propagation in bi-isotropic and anisotropic media has applications in material physics [59]. Axionlike couplings can be generated via quantum corrections in a Lorentz violation background [60].

In this thesis, we propose the study of a general NED coupled to an axionic scalar field, that we call $\phi$, coupled to the non-linear sector through the interaction $g\,\phi\,(\mathbf{E}_0 \cdot \mathbf{B}_0)$. We start the description of the model with a general non-linear kinetic sector. We use the following approach: we expand the 4-potential associated with electromagnetic (EM) fields $(\mathbf{E}_0, \mathbf{B}_0)$ around an EM background up to second order in the propagating EM fields. Thereby, we have a general linearized ED propagating in an EM background field coupled to an axionic scalar field. We explore the propagation effects for the plane wave solutions in which the dispersion relations, the group velocities, the electric permittivity, and the magnetic permeability are obtained in a uniform and constant EM background fields. The Born-Infeld (BI) non-linear theory is considered as an application of these results. Therefore, we discuss the properties of the wave propagation, like the dispersion relations, group velocities and the characteristics of the medium in the presence of an external magnetic field, and posteriorly, of an external electric field. We analyze the results in a regime of strong magnetic field for the BI theory. The birefringence phenomena is also investigated in the BI theory for the cases with magnetic and electric background fields, separately. In the case of the birefringence with a magnetic background, we calculate the axion coupling constant using the data of the PVLAS-FE experiment. For the birefringence with an electric background, we make a connection with the optical Kerr effect. Finally, following even further the perspective of an effective scenario in which different effects can be unified and interfere with each other, we investigate how the inclusion of Lorentz symmetry violation (LSV) induces new characteristics in the already consolidated non-linear ED coupled to an axionic model.

The topics covered in this thesis are organized as follows:

- In chapter 2 we introduce ALPs, we first follow the historical path of axions as a solution to the Strong CP Problem, proposed by Peccei and Quinn. After that, we introduce the ALPs from a perspective of the Lorentz symmetry violation (LSV), we show that in this case, the ALPs appear as a way to recover the Lorentz symmetry.

- In chapter 3, driven by the possibility of nonlinear contributions owing to axion production in strong magnetic fields, we review non-linear electrodynamic theories. We show an approach where the theory is described in terms of general coefficients to be defined, depending on the model. We obtain the equations of motion and probe some optical features through the dispersion relations. In addition, we show some examples of nonlinear Electrodynamics, namely Euler-Heisenberg [38], Born-Infeld [26] and the recent Mod-Max [67, 23, 68].

- In chapter 4, we describe the NED-axion model in an eletromagnectic background, and obtain the corresponding field equations for the axion and photon. Next, the section 4.1 focus on the properties of plane wave solutions and we organize our results in two subsections: the first one 4.1.1 considers the purely magnetic background case. The second one 4.1.2 discusses a purely electric background. In section 4.2, we apply all the results of the previous section to BI model, for a magnetic background in subsection 4.2.1; next, we go into a purely electric background in subsection 4.2.2. After that, in section 4.3, we investigate the birefringence of the axion-BI model for a wave in a magnetic subsec. 4.3.1 and in an electric background subsec. 4.3.2.

- In chapter 5 we include LSV in our analysis through a Carroll-Field-Jackiw (CFJ) term. We begin by offering motivations and justifications for contemplating an effective model that integrates diverse scenarios, specifically, ALP physics, NED and LSV. In section 5.1, the axionic non-linear theory is presented with the CFJ term in an electromagnetic background field. We follow an approach similar to the chapter 4, except that, in this case, we will only work with a magnetic background and there will be a subtlety in the definition of the refractive indices, due to the evidence of the effects of the LSV. Furthermore, we apply the results for the Euler-Heisenberg , the Born-Infeld, and the Mod-Max electrodynamics.

- Finally, chapter 6 casts our concluding comments and future perspectives.

Throughout this text, we adopt natural units $\hbar = c = 1$ with $4\pi\epsilon_0 = 1$, and the Minkowski metric $\eta^{\mu\nu} = \mathrm{diag}(+1, -1, -1, -1)$. The electric and magnetic fields have squared-energy mass dimension in which the conversion of Volt/m and Tesla (T) to the natural system is as follows: $1\,\mathrm{Volt/m} = 2.27 \times 10^{-24}\,\mathrm{GeV}^2$ and $1\,\mathrm{T} = 6.8 \times 10^{-16}\,\mathrm{GeV}^2$, respectively.

Chapter 2

# Axions and ALPs

Historically, axions emerged as an attempt to solve the problem of strong CP violation. In the standard model, the electroweak interaction is known to violate CP symmetry, product of charge conjugation C with parity P. Furthermore, the theoretical formulation of strong interactions, quantum chromodynamics (QCD), also shows that CP violation can occur may occur in a very small way. However, until today, the strong CP violation was never detected and since there is no reason for this symmetry to be preserved, this problem came to be known as the "strong CP problem" [72].

## 2.1 Strong CP Violation

The QCD represented by lagrangian:

$$\mathcal{L}_{QCD} = -\frac{1}{4g^2}(G^a_{\mu\nu})^2 - \sum_f \bar{q}_f(i\slashed{D} - m_f)q_f, \tag{2.1}$$

is incredibly successful with regard to theoretical calculations and boasts a extensive experimental confirmation. Here, $g$ represents the gauge coupling, $G$ is the the gluon field strength tensor, $\tilde{G}$ denotes its dual, and $q_f$ refers to quark flavors. However, in the 70s, there was an intriguing problem that put it in check. The Lagrangian of QCD has a global axial symmetry $U(1)_A$ that is broken spontaneously due to the formation of quark condensates. Consequently, there must have been a fourth Nambu-Goldstone boson besides the pions that was never detected. This puzzle became known as the "$U(1)_A$ problem" [73]. The solution to this question was obtained by t'Hooft who realized that the complex vacuum structure of QCD means that $U(1)_A$ is not a true symmetry at the quantum level [74] [1]. In fact, this anomaly makes the theory depend on a new parameter $\theta$, requiring the addition of a new term in the Lagrangian given by:

$$\mathcal{L}_\theta = \frac{\theta \, g^2}{32\pi^2} G^a_{\mu\nu} \widetilde{G}^{\mu\nu a}. \tag{2.2}$$

This term is known as $\theta-$term, it violates CP, and in many analyses, is neglected, since it can be written as a total divergence, and therefore ends up being irrelevant to the equations of motion. However, the term $\mathcal{L}_\theta$ has a physical consequence that must be taken into account, which is a small electric dipole moment for the neutron (nEDM). The big problem is that the experimental limits of the nEDM are of the order of $d_n < (0.0 \pm 1.1) \times 10^{-26} e\,\text{cm}$ [75], imposing an upper limit for the parameter $\theta$, of $\theta \ll 10^{-9}$. So the question arises, "if QCD violates CP symmetry, why does it happen in such a small way?"

Furthermore, when we include the massive terms of the quarks, the quark mass matrix is given by

$$\mathcal{L}_m = \bar{q}_{i\,R} M_{ij} q_{j\,L} + h.c, \tag{2.3}$$

imaginary phases emerge in the mass matrix $M$ that need to be eliminated. These phases can be eliminated from the mass matrix through chiral transformations applied to the quark fields. On the other hand, due to the chiral anomaly, this operation alters the CP-violatinge $\theta$-term as

$$\theta \rightarrow \bar{\theta} = \theta - \arg \det M, \tag{2.4}$$

where $\bar{\theta}$ is the coefficient of the term $G\tilde{G}$ of the complete theory. Phrasing differently, the puzzle concerning CP violation hinges on the perplexing fact that the angle $\bar{\theta}$ is exceedingly small.

Several approaches have been proposed to address this issue, but we will focus on the most renowned and widely disseminated one: Eliminating the parameter $\bar{\theta}$ dynamically by introducing an additional symmetry. The idea proposed by Peccei and Quinn was by introducing a global chiral symmetry, known as $U(1)_{PQ}$ [1]. This symmetry is spontaneously broken and the associated Nambu-Goldstone boson replaces the static CP-violating angle $\bar{\theta}$ with a dynamic CP-conserving field called axion. Under a $U(1)_{PQ}$ transformation,

---
[1] Symmetries that are realized classically but not quantumly are called anomalies. In this case, we say that $U(1)_A$ is a chiral anomaly

the axion field $a(x)$ transforms as

$$a(x) \to a(x) + \alpha f_a, \tag{2.5}$$

in which $f_a$ stands as the order parameter associated with the breaking of U(1)$_{PQ}$. The extended Lagrangian with the symmetry U(1)$_{PQ}$ will be

$$\mathcal{L}_{total} = \mathcal{L}_{SM} + \frac{\bar{\theta}\,g^2}{32\pi^2} G^a_{\mu\nu}\widetilde{G}^{\mu\nu a} + \xi \frac{a(x)}{f_a} \frac{g^2}{32\pi^2} G^a_{\mu\nu}\widetilde{G}^{\mu\nu a} + \frac{1}{2}(\partial_\mu a)^2 + \mathcal{L}_{int}(\partial_\mu a), \tag{2.6}$$

where the third term is the gluon-axion interaction potential due chiral anomaly of U(1)$_{PQ}$. The interaction terms incorporating axion derivatives do not play a significant role in our analysis of the strong CP problem. The minimum of the potential occurs when $\langle a \rangle = -\frac{f_a}{\xi}\bar{\theta}$. Thus, when the axion field takes its minimum value, the term $\bar{\theta}$ is canceled, eliminating CP violation in strong interactions. The axion mass can be estimated using the already known mass of pions through the following expression:

$$m_a^2 \approx \frac{m_\pi^2 f_\pi^2}{f_a^2}. \tag{2.7}$$

Additionally, several experiments have been conducted in the pursuit of detecting the axion. Figure 2.1 depicts the bounds established in the literature.

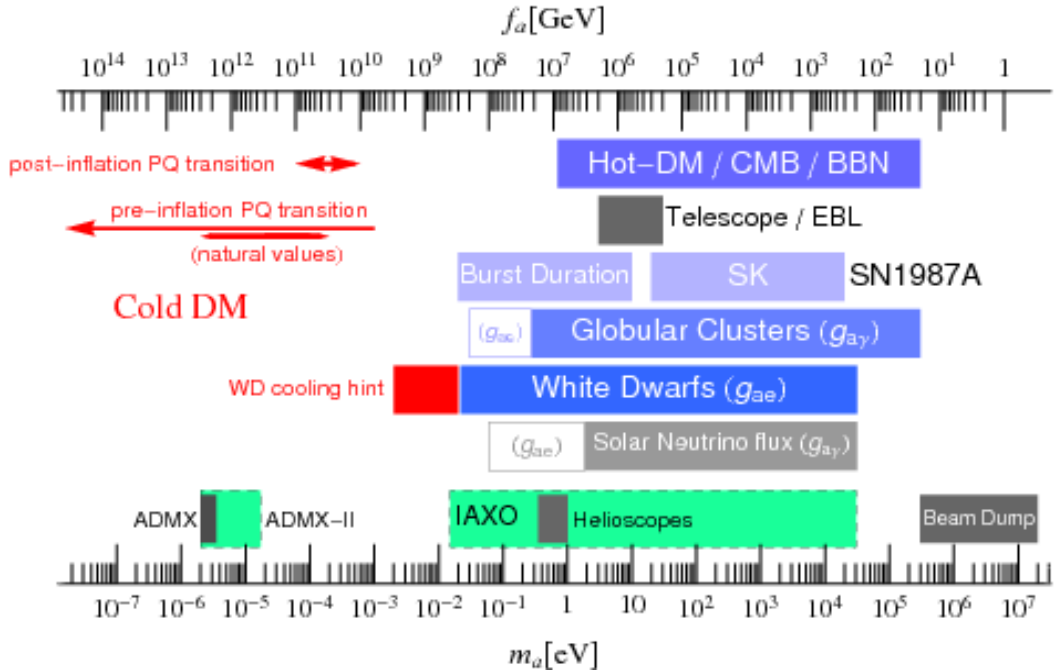

*Figura 2.1:* Exclusion zones for Axion $f_a$ constant versus $m$ from several experiments and simulations. Source: [79]

We find it crucial to allocate a section in this work to briefly explore the historical origins of the axion. For a more in-depth analysis, we suggest consulting the following reviews [76, 77, 78]. In this thesis, nevertheless, we aim at presenting a more heuristic approach, demonstrating the emergence of axionic electrodynamics. This perspective revolves around the spontaneous breaking of Lorentz symmetry.

## 2.2 ALPs from Lorentz Symmetry Violation

Although Lorentz invariance is a fundamental invariance principle in elementary particle physics and in Einstein's classical General Relativity, it is known that, for a quantum theory of gravity, such invariance may not hold any longer. For example, in string theories [82, 83, 85, 84, 86], it is estimated that there should be small violations of Lorentz symmetry next to the Planck energy scale, namely, $E_p = 10^{19}$GeV. This would occur in the early universe. At the end of the 1980s, there were efforts to build a model that incorporated the violation of Lorentz symmetry to the theories already consolidated in high-energy physics. During this period, Kosteletsky, Colladay and Samuel published a series of works in which they introduced an effective model known as the Standard-Model Extension (SME). This model describes a general action that encompasses the standard model, general relativity and terms that violate Lorentz symmetry. and CPT [87, 88].

In particular, an electrodynamic model that takes into account possible effects of a violation of Lorentz symmetry was developed by Carroll, Field and Jackiw. Known as Carroll-Field-Jackiw (CFJ) electrodynamics [89], is a generalization of a Chern-Simons term for $(3+1)$ dimensions. For the consistency of the model, a quadrivector is introduced that guarantees the gauge symmetry of the theory, but does not preserve the Lorentz and CPT symmetry. The CFJ term appears in the CPT-odd gauge sector of SME.

There is a vast literature on the CFJ electrodynamic model. In the work [90], limits were obtained for the CFJ Lorentz-breaking parameter in the time-like case through laboratory experiments such as quantum corrections to the spectrum of the hydrogen atom, electric dipole moment, as well as the interparticle potential between fermions. Studies on the possible effects of contributions of the CFJ model for the cosmic microwave background (CMB) were carried out in ref. [91]. Recently, in the supersymmetry scenario, the gauge boson-gaugino mixing was investigated by taking into account the effects of the LSV due to a CFJ term [93].

The CFJ term is written as follows

$$\mathcal{L}_{CFJ} = \frac{1}{4}\epsilon^{\mu\nu\kappa\lambda}\, v_\mu\, A_\nu\, F_{\kappa\lambda}\,. \tag{2.8}$$

However, an electrodynamics governed exclusively by this term is of no physical interest, as it does not provide dynamics to the photon. What is actually done is to investigate how the CFJ term affects conventional Maxwell electrodynamics. In this sense, the Lagrangian of interest is given by

$$\mathcal{L}_{CFJ} = -\frac{1}{4}F^2_{\mu\nu} + \frac{1}{4}\epsilon^{\mu\nu\kappa\lambda}\, v_\mu\, A_\nu\, F_{\kappa\lambda} - J_\mu\, A^\mu\,. \tag{2.9}$$

Firstly, we can investigate how the Lagrangian (2.9) behaves upon gauge transformations. Let us remember that gauge transformations are of the type:

$$A'_\mu = A_\mu + \partial_\mu \alpha \quad \Longrightarrow \quad \delta_g A_\mu = \partial_\mu \alpha\,. \tag{2.10}$$

Applying this variation to the equation (2.9), we have

$$\begin{aligned}
\delta_g \mathcal{L}_{CFJ} &= -\frac{1}{4}\delta_g F^2_{\mu\nu} + \frac{1}{4}\delta_g(\epsilon^{\mu\nu\kappa\lambda}\, v_\mu\, A_\nu\, F_{\kappa\lambda}) + \delta_g(J_\mu A^\mu) \\
&= -\frac{1}{2}F^{\mu\nu}\delta_g F_{\mu\nu} + \frac{1}{2}\epsilon^{\mu\nu\kappa\lambda}\, v_\mu\,(\delta_g A_\nu)\,\partial_\kappa A_\lambda + \frac{1}{2}\epsilon^{\mu\nu\kappa\lambda}\, v_\mu\, A_\nu\, \partial_\kappa(\delta_g A_\lambda) + \cancel{J^\mu \partial_\mu \alpha} \\
&= -\cancel{F^{\mu\nu}\partial_\mu\partial_\nu \alpha} + \frac{1}{2}\epsilon^{\mu\nu\kappa\lambda}\, v_\mu\,(\partial_\nu \alpha)\,\partial_\kappa A_\lambda + \frac{1}{2}\epsilon^{\mu\nu\kappa\lambda}\, v_\mu\, A_\nu\, \cancel{\partial_\kappa\partial_\lambda \alpha} \\
&= \partial_\nu\left(\frac{1}{2}\epsilon^{\mu\nu\kappa\lambda}\,\cancel{v_\mu\,\alpha\,\partial_\kappa A_\lambda}\right) - \frac{1}{2}\epsilon^{\mu\nu\kappa\lambda}\,(\partial_\nu v_\mu)\,\alpha\,\partial_\kappa A_\lambda - \frac{1}{2}\epsilon^{\mu\nu\kappa\lambda}\, v_\mu\,\alpha\,\cancel{\partial_\nu\partial_\kappa A_\lambda}, 
\end{aligned} \tag{2.11}$$

where in the third line, the terms cancel each other out due to the antisymmetry properties of the $F_{\mu\nu}$ and Levi-Civita tensors. While the first term of the fourth line cancels out due to the fact that it is a total divergence. Furthermore, the source term vanishes by integrating by parts and taking into account the conservation of the current. Therefore, the only way for the theory to be gauge invariant is if the condition

$$\partial_\mu v_\nu - \partial_\nu v_\mu = 0 \tag{2.12}$$

is satisfied. The equations of motion are

$$\partial_\mu F^{\mu\nu} + v_\mu \tilde{F}^{\mu\nu} = J^\nu, \tag{2.13}$$

where $\widetilde{F}^{\mu\nu} = \epsilon^{\mu\nu\alpha\beta}F_{\alpha\beta}/2$ is the dual tensor of $F_{B\mu\nu}$. In vector form the equations are

written below

$$\nabla \cdot \mathbf{E} - \mathbf{v} \cdot \mathbf{B} , = \rho \quad (2.14a)$$

$$\nabla \times \mathbf{B} + \mathbf{v} \times \mathbf{E} = v^0 \mathbf{B} + \frac{\partial \mathbf{E}}{\partial t} + \mathbf{J}, \quad (2.14b)$$

$$\nabla \cdot \mathbf{B} = 0 , \quad (2.14c)$$

$$\nabla \times \mathbf{E} + \frac{\partial \mathbf{B}}{\partial t} = \mathbf{0} , \quad (2.14d)$$

where the first pair of equations are obtained from (2.13) taking $\nu = 0$ and $\nu = i$, respectively. The modified Gauss law given by the equation (2.14a), shows that an electrostatic field can generate a small magnetic field, given that $v$ has a small magnitude. The second pair comes from Bianchi's identity, which remains unchanged.

The energy moment tensor of a model, in general, is obtained from the Lagrangian through the expression:

$$\Theta_{\mu\nu} = \sum_i \frac{\partial \mathcal{L}}{\partial(\partial_\mu \phi_i)} \partial^\mu \phi_i - \delta_{\mu\nu} \mathcal{L}, \quad (2.15)$$

where $\phi_i = (\phi_1, \phi_2, \dots)$ is the set of propagating fields that make up the model. However, we prefer to present an "unorthodox" method of obtaining the energy-momentum tensor from the equations of motion. This trick consists of algebraically manipulating the equations of motion, so that it is possible to identify the conservation law that the energy-momentum tensor obeys.

Let's begin by multiplying the equation (2.13) by $F_{\nu\rho}$. In doing so, using the relations

$$\tilde{F}^{\mu\nu} F_{\nu\rho} = -\frac{1}{4} \delta^\mu_\rho \tilde{F} \cdot F, \quad (2.16a)$$

$$\partial_\mu F_{\nu\rho} = -\partial_\nu F_{\rho\nu} - \partial_\rho F_{\mu\nu}, \quad (2.16b)$$

(note that eq. (2.16b) is basically Bianchi's identity) and after some manipulations, we obtained the following conservation law

$$\partial_\mu (\Theta_{CFJ})^\mu{}_\rho = \Omega_\rho, \quad (2.17)$$

where the energy-momentum tensor is written as

$$(\Theta_{CFJ})^\mu{}_\rho = F^{\mu\nu} F_{\nu\rho} + \frac{1}{4} \delta^\mu_\rho F^2 - \frac{1}{2} v_\rho \tilde{F}^{\mu\beta} A_\beta , \quad (2.18)$$

and the dissipative term is:

$$\Omega_\rho \equiv J^\nu F_{\nu\rho} - \frac{1}{2}(\partial_\mu v_\rho) F^{\mu\nu} A_\nu. \tag{2.19}$$

We can observe some points; the CFJ term will give new contributions to both the energy-momentum tensor and the dissipative part of the conservation law. Another important detail is that, in Maxwell's usual electrodynamics - making $v_\rho = 0$ in eq. (2.18) -, the EM tensor is symmetric, this implies that, in natural units, the momentum definitions and Poyinting vector coincide. In the case where we add Lorentz Violation, these definitions are different, since $(\Theta_{CFJ})^{0i} \neq (\Theta_{CFJ})^{i0}$. Let's see how conservation laws are written in vector form. Making $\rho = 0$ in eq. (2.17), we have the following expression

$$\partial_0 (\Theta_{CFJ})^0{}_0 + \partial_i (\Theta_{CFJ})^i{}_0 = J^\nu F_{\nu\rho} - \frac{1}{2}(\partial_\mu v_\rho) F^{\mu\nu} A_\nu, \tag{2.20}$$

or even

$$\partial_t u_{CFJ} + \nabla \cdot \mathbf{S}_{CFJ} = -\mathbf{J} \cdot \mathbf{E} + \frac{1}{2}(\partial_t v^0) \mathbf{B} \cdot \mathbf{A} - \frac{1}{2}(\nabla v^0) \cdot \mathbf{B} - \frac{1}{2}(\nabla v^0) \cdot (\mathbf{A} \times \mathbf{E}), \tag{2.21}$$

where the energy density is written as

$$u_{CFJ} \equiv (\Theta_{CFJ})^0{}_0 = \frac{1}{2}\left(\mathbf{E}^2 + \mathbf{B}^2\right) - \frac{1}{2} v^0 \mathbf{A} \cdot \mathbf{B}. \tag{2.22}$$

We can also identify the Poyinting vector, which is given by:

$$(\mathbf{S}_{CFJ})^i \equiv (\Theta_{CFJ})^i{}_0 = (\mathbf{E} \times \mathbf{B})^i - \frac{1}{2}\left(v^0 \Phi \mathbf{B}\right)^i - \frac{1}{2} v^0 (\mathbf{A} \times \mathbf{E})^i. \tag{2.23}$$

.

There is an effort to detect effects of violation of Lorentz symmetry in different types of experiments. Of particular interest for our purposes are astrophysical investigations of birefringence and scattering in cavity experiments [92]

Given this general view of the Lorentz violation, we will see below how ALPs can emerge in this scenario.

### 2.2.1 Axion-Like Particles (ALPs)

The way we propose to introduce an axion-like particle into the theory with LV discussed above, consists of using the condition (2.12), since a possible solution to this equation is that the 4-vector $v_\mu$ is the gradient of a scalar, that is, $v_\mu = \partial_\mu \phi$. However, in principle, such a scalar does not necessarily have the character of a field.

Furthermore, there is a microscopic motivation for this solution. Lorentz Violation effects are expected to manifest themselves close to the Planck scale, on the other hand, supersymmetric effects are expected from $10^{13}$ GeV. We can follow a line of thought in which both the Lorentz Violation and SUSY effects could have coexisted in a primordial universe. From this perspective, there are some investigations of a supersymmetric extension of Maxwell-Chern-Simons electrodynamics [93, 94]. The interesting fact is that in this scenario, the condition that the vector $v_\mu$ is the gradient of a scalar field arises naturally.

Following this approach, we can rewrite the Lagrangian (2.9) as

$$\mathcal{L} = -\frac{1}{4} F_{\mu\nu}^2 + \frac{1}{4} \epsilon^{\mu\nu\kappa\lambda} (\partial_\mu \phi) A_\nu F_{\kappa\lambda} - J_\mu A^\mu . \tag{2.24}$$

Integrating by parts we have

$$\begin{aligned}
\mathcal{L} &= -\frac{1}{4} F_{\mu\nu}^2 + \partial_\mu \left( \frac{1}{4} \epsilon^{\mu\nu\kappa\lambda} \phi A_\nu F_{\kappa\lambda} \right) - \frac{1}{4} \epsilon^{\mu\nu\kappa\lambda} \phi (\partial_\mu A_\nu) F_{\kappa\lambda} - \frac{1}{4} \epsilon^{\mu\nu\kappa\lambda} \phi A_\nu (\partial_\mu F_{\kappa\lambda}) - J_\mu A^\mu \\
&= -\frac{1}{4} F_{\mu\nu}^2 - \frac{1}{8} \epsilon^{\mu\nu\kappa\lambda} \phi F_{\mu\nu} F_{\kappa\lambda} - \frac{1}{2} \phi A_\nu (\partial_\mu \tilde{F}^{\mu\nu}) - J_\mu A^\mu \\
&= -\frac{1}{4} F_{\mu\nu}^2 - \frac{1}{4} \phi \tilde{F}^{\kappa\lambda} F_{\kappa\lambda} - J_\mu A^\mu ,
\end{aligned} \tag{2.25}$$

where, the second term of the first line is eliminated because it is a total divergence, while the fourth term of the second line vanishes due to the Bianchi identity, namely, $\partial_\mu \tilde{F}^{\mu\nu} = 0$.

We also highlight that there is other way of connection between the axionic theory and Lorentz and CPT-violation. In ref.[95] the author makes this connection by embedding CFJ electrodynamics in a premetric framework. It is used the fact that in CFJ electrodynamics the constraint $\partial_\mu v_\nu - \partial_\nu v_\mu = 0$ allows writing the Lorentz-breaking vector as the gradient of a scalar $v_\mu = \partial_\mu \phi$, so when doing this redefinition in the pre-metric Lagrangian, the axionic interaction term naturally arises. However, this approach does not provide dynamics to the axion. Moreover, the correspondent energy-momentum tensor does not depend on the axion field. Furthermore, the author discussed the relation between the birefringence phenomenon with Lorentz and CPT symmetry violation. It is possible to associate the non-observation of birefringence with the preservation of these symmetries. For more details on LSV, we indicate the review [96] and references therein.

If we are considering that $\phi$ is a field, it can be dynamical. Therefore, we introduce a Klein Gordon term, as well as a possible potential in the Lagrangian (2.25). Thus, we obtain the following expression

$$\mathcal{L} = -\frac{1}{4}F_{\mu\nu}^2 + \frac{1}{2}\partial_\mu\phi\partial^\mu\phi - V(\phi) - \frac{g}{4}\phi\, \tilde{F}^{\kappa\lambda}F_{\kappa\lambda} - J_\mu A^\mu\ . \qquad (2.26)$$

Note that in eq. (2.25) the quantity $\phi$ is dimensionless, but when we impose a dynamic for this scalar, the Klein-Gordon term - term of highest order derivative - fixes an inverse length dimension for $\phi$ . As a consequence, it is necessary to add a parameter $g$ with mass dimension $-1$ to the coupling term.

The equations of motion for the photon, based on the Lagrangian (2.26) are

$$\partial_\mu F^{\mu\nu} + g(\partial_\mu\phi)\tilde{F}^{\mu\nu} = J^\nu, \qquad (2.27)$$

while the ALP equation is given by

$$\Box\phi + V'(\phi) = g\,(\mathbf{E}\cdot\mathbf{B}). \qquad (2.28)$$

Taking $\nu = 0$ and $\nu = i$ in equation (2.27), we obtain the modified Gauss and Ampere laws, respectively. Therefore, in vector notation, Maxwell's equations are written as:

$$\nabla\cdot\mathbf{D} = 0\ ,\quad \nabla\times\mathbf{E} + \frac{\partial\mathbf{B}}{\partial t} = \mathbf{0}\ , \qquad (2.29a)$$

$$\nabla\cdot\mathbf{B} = 0\ ,\quad \nabla\times\mathbf{H} - \frac{\partial\mathbf{D}}{\partial t} = \mathbf{0}\ , \qquad (2.29b)$$

wherein

$$\mathbf{D} = \mathbf{E} + g\phi\mathbf{B}\ , \qquad (2.30a)$$

$$\mathbf{H} = \mathbf{B} - g\phi\mathbf{E}. \qquad (2.30b)$$

This approach in which ALPs emerge in a theory with violation of Lorentz symmetry, although not usual in the literature, demonstrates that ALPs can manifest in other physics scenarios beyond the standard model,transcending the traditional Peccei-Quinn mechanism.

# Chapter 3

# Non-linear electrodynamics

Nonlinear electrodynamics (NED) are generalizations of Maxwell's electrodynamics in scenarios of strong electromagnetic fields. These models have been extensively studied to this day, since the first model was presented by Max Born and Leopold Infeld in the 1930s to remove the divergence of the electron's self-energy by introducing an upper bound for the electric field at the origin [26].

In this chapter, we introduce a approach to describing a most general NED, which is formulated in terms of coefficients to be determined depending on the specific model under consideration. Additionally, we provide a brief overview of some NED models, namely Euler-Heisenberg, Born-Infeld, and ModMax.

We start up with the Lagrangian (density) of the model

$$\mathcal{L} = \mathcal{L}_{nl}(\mathcal{F}_0, \mathcal{G}_0) - J_\mu A^\mu ~, \tag{3.1}$$

where $\mathcal{L}_{nl}(\mathcal{F}_0, \mathcal{G}_0)$ denotes the most general Lagrangian of a non-linear electrodynamics that is function of the Lorentz- and gauge-invariant bilinears, a scalar and a pseudo-scalar, defined respectively by

$$\mathcal{F}_0 = -\frac{1}{4} F_{0\,\mu\nu} F_0^{\,\mu\nu} ~, \tag{3.2}$$

$$\mathcal{G}_0 = -\frac{1}{4} F_{0\,\mu\nu} \widetilde{F}_0^{\,\mu\nu}. \tag{3.3}$$

We consider the antisymmetric field strength tensor as $F_0^{\,\mu\nu} = \partial^\mu A_0^{\,\nu} - \partial^\nu A_0^{\,\mu}$, and the correspondent dual tensor is $\widetilde{F}_0^{\,\mu\nu} = \epsilon^{\mu\nu\alpha\beta} F_{0\alpha\beta}/2$, which satisfies the Bianchi identity $\partial_\mu \widetilde{F}_0^{\,\mu\nu} = 0$. Therefore, in terms of electromagnetic fields, the invariants are $\mathcal{F}_0 = \frac{1}{2}\left(\mathbf{E}_0^2 - \mathbf{B}_0^2\right)$ and $\mathcal{G}_0 = \mathbf{E}_0 \cdot \mathbf{B}_0$.

The action principle related to the Lagrangian (3.1) leads to the equations of motion with a classical external current $J^\nu$ :

$$\partial_\mu \left( \frac{\partial \mathcal{L}_{nl}}{\partial \mathcal{F}_0} F_0{}^{\mu\nu} + \frac{\partial \mathcal{L}_{nl}}{\partial \mathcal{G}_0} \widetilde{F}_0{}^{\mu\nu} \right) = J^\nu , \tag{3.4}$$

in which the current is conserved $\partial_\mu J^\mu = 0$.

The energy-momentum tensor is written as follows

$$\Theta^\mu{}_\nu = \frac{\partial \mathcal{L}_{nl}}{\partial \mathcal{F}_0} F_0{}^{\mu\kappa} F_{0\,\kappa\nu} - \delta^\mu_\nu \left( \frac{\partial \mathcal{L}_{nl}}{\partial \mathcal{G}_0} \mathcal{G}_0 - \mathcal{L}_{nl} \right) . \tag{3.5}$$

We expand the Abelian gauge field as $A_0{}^\mu = a^\mu + A_B{}^\mu$, with $a^\mu$ being the photon 4-potential, and $A_B{}^\mu$ denotes a background potential. In this sense, the tensor $F_0{}^{\mu\nu}$ is also written as the combination $F_0{}^{\mu\nu} = f^{\mu\nu} + F_B{}^{\mu\nu}$, in which $f^{\mu\nu} = \partial^\mu a^\nu - \partial^\nu a^\mu = \left( -e^i,\, -\epsilon^{ijk} b^k \right)$ is the EM field strength tensor that propagates in the space-time, and $F_B{}^{\mu\nu} = \left( -E^i,\, -\epsilon^{ijk} B^k \right)$ corresponds to the EM background field. The notation of the 4-vector and tensors with index (B) indicates that it is associated with the background. Also note that in vector notation, we adopt lowercase letters for the propagating fields (e.g. **e** and **b**) and uppercase letters for the background fields (e.g. **E** and **B**). So, in this notation, for simplicity we omit the index (B). The reader should pay attention to this detail in order to avoid confusion in the notation, as in chapter 2, we used uppercase letters to represent the propagating fields. At this stage, we consider the general case in which the background depends on the space-time coordinates. Under this prescription, we also expand the Lagrangian (3.1) around the background up to fourth-order in the propagating field $a^\mu$ to yield the expression

$$\begin{aligned}\mathcal{L}^{(2)} &= -\frac{1}{4} c_1 f_{\mu\nu}^2 - \frac{1}{4} c_2 f_{\mu\nu} \widetilde{f}^{\mu\nu} - \frac{1}{2} G_B^{\mu\nu} f_{\mu\nu} + \frac{1}{8} Q_B^{\mu\nu\kappa\lambda} f_{\mu\nu} f_{\kappa\lambda} + \frac{1}{8} R_B^{\mu\nu\kappa\lambda\rho\sigma} f_{\mu\nu} f_{\kappa\lambda} f_{\rho\sigma} \\ &\quad + \frac{1}{16} S_B^{\mu\nu\kappa\lambda\rho\sigma\omega\tau} f_{\mu\nu} f_{\kappa\lambda} f_{\rho\sigma} f_{\omega\tau} - J_\mu (a^\mu + A_B{}^\mu) + \mathcal{L}_{nl}(\mathcal{F}_B, \mathcal{G}_B) , \end{aligned} \tag{3.6}$$

where the background tensors are defined by

$$G_B^{\mu\nu} = c_1 F_B^{m u\nu} + c_2 \widetilde{F}_B^{\mu\nu} , \tag{3.7}$$

$$Q_B^{\mu\nu\kappa\lambda} = d_1 F_B^{\mu\nu} F_B^{\kappa\lambda} + d_2 \widetilde{F}_B^{\mu\nu} \widetilde{F}_B^{\kappa\lambda} + d_3 F_B^{\mu\nu} \widetilde{F}_B^{\kappa\lambda} + d_3 \widetilde{F}_B^{\mu\nu} F_B^{\kappa\lambda} , \tag{3.8}$$

$$\begin{aligned}R_B^{\mu\nu\kappa\lambda\rho\sigma} =\ & d_1 F_B^{\mu\nu} \eta^{\kappa\lambda} \eta^{\rho\sigma} + \frac{1}{2} d_2 \widetilde{F}_B^{\mu\nu} \epsilon^{\kappa\lambda\rho\sigma} + \frac{1}{2} d_3 F_B^{\mu\nu} \epsilon^{\kappa\lambda\rho\sigma} + d_3 \widetilde{F}_B^{\mu\nu} \eta^{\kappa\lambda} \eta^{\rho\sigma} \\ & - \frac{1}{6} m_1 F_B^{\mu\nu} F_B^{\kappa\lambda} F_B^{\rho\sigma} - \frac{1}{6} m_2 \widetilde{F}_B^{\mu\nu} \widetilde{F}_B^{\kappa\lambda} \widetilde{F}_B^{\rho\sigma} - \frac{1}{2} m_3 F_B^{\mu\nu} F_B^{\kappa\lambda} \widetilde{F}_B^{\rho\sigma} \\ & - \frac{1}{2} m_3 F_B^{\mu\nu} \widetilde{F}_B^{\kappa\lambda} \widetilde{F}_B^{\rho\sigma} , \end{aligned} \tag{3.9}$$

$$\begin{aligned}S_B^{\mu\nu\kappa\lambda\rho\sigma\alpha\beta} =\ & \frac{1}{2} d_1 F_B^{\mu\nu} \eta^{\kappa\lambda} \eta^{\rho\sigma} + \frac{1}{8} d_2 \epsilon^{\mu\nu\kappa\lambda} \epsilon + \frac{1}{2} d_3 \eta^{\mu\kappa} \eta^{\nu\lambda} \epsilon^{\rho\sigma\alpha\beta} - \frac{1}{2} m_1 F_B^{\mu\nu} F_B^{\kappa\lambda} \eta^{\rho\sigma} \eta^{\alpha\beta} \\ & - \frac{1}{4} m_2 \widetilde{F}_B^{\mu\nu} \widetilde{F}_B^{\kappa\lambda} \epsilon^{\rho\sigma\alpha\beta} - \frac{1}{4} m_3 F_B^{\mu\nu} F_B^{\kappa\lambda} \epsilon^{\rho\sigma\alpha\beta} - m_3 F_B^{\mu\nu} \widetilde{F}_B^{\kappa\lambda} \eta^{\rho\alpha} \eta^{\sigma\beta} - \frac{1}{2} m_4 F_B^{\mu\nu} \widetilde{F}_B^{\kappa\lambda} \epsilon^{\rho\sigma\alpha\beta} \\ & - \frac{1}{2} m_4 \eta^{\mu\kappa} \eta^{\nu\lambda} \widetilde{F}_B^{\rho\sigma} \widetilde{F}_B^{\alpha\beta} + \frac{1}{24} n_1 F_B^{\mu\nu} F_B^{\kappa\lambda} F_B^{\rho\sigma} F_B^{\alpha\beta} + \frac{1}{24} n_1 \widetilde{F}_B^{\mu\nu} \widetilde{F}_B^{\kappa\lambda} \widetilde{F}_B^{\rho\sigma} \widetilde{F}_B^{\alpha\beta} \\ & + \frac{1}{6} n_3 F_B^{\mu\nu} F_B^{\kappa\lambda} F_B^{\rho\sigma} \widetilde{F}_B^{\alpha\beta} + \frac{1}{4} n_3 F_B^{\mu\nu} F_B^{\kappa\lambda} F_B^{\rho\sigma} \widetilde{F}_B^{\alpha\beta} + \frac{1}{6} n_5 F_B^{\mu\nu} F_B^{\kappa\lambda} \widetilde{F}_B^{\rho\sigma} \widetilde{F}_B^{\alpha\beta} , \end{aligned} \tag{3.10}$$

and $\mathcal{L}_{nl}(\mathcal{F}_B, \mathcal{G}_B)$ is the non-linear Lagrangian as function of the Lorentz invariants $\mathcal{F}_B = -\frac{1}{4} F_{B\mu\nu}^2 = \mathbf{E}^2 - \mathbf{B}^2$ and $\mathcal{G}_B = -\frac{1}{4} F_{B\mu\nu} \widetilde{F}_B^{\mu\nu} = \mathbf{E} \cdot \mathbf{B}$, both in terms of the EM background field, and $\widetilde{F}_B^{\mu\nu} = \epsilon^{\mu\nu\alpha\beta} F_{B\alpha\beta}/2 = \left( -B^i, \epsilon^{ijk} E^k \right)$ is the dual tensor of $F_{B\mu\nu}$. Furthermore, the coefficients $c_1$, $c_2$, $d_1$, $d_2$ and $d_3$ are evaluated at $\mathbf{E}$ and $\mathbf{B}$, as follows:

$$\begin{aligned}c_1 &= \left.\frac{\partial \mathcal{L}_{nl}}{\partial \mathcal{F}_0}\right|_{\mathbf{E},\mathbf{B}} , \ c_2 = \left.\frac{\partial \mathcal{L}_{nl}}{\partial \mathcal{G}_0}\right|_{\mathbf{E},\mathbf{B}} , \ d_1 = \left.\frac{\partial^2 \mathcal{L}_{nl}}{\partial \mathcal{F}_0^2}\right|_{\mathbf{E},\mathbf{B}} , \ d_2 = \left.\frac{\partial^2 \mathcal{L}_{nl}}{\partial \mathcal{G}_0^2}\right|_{\mathbf{E},\mathbf{B}} , \ d_3 = \left.\frac{\partial^2 \mathcal{L}_{nl}}{\partial \mathcal{F}_0 \partial \mathcal{G}_0}\right|_{\mathbf{E},\mathbf{B}} , \\ m_1 &= \left.\frac{\partial^3 \mathcal{L}_{nl}}{\partial \mathcal{F}_0^3}\right|_{\mathbf{E},\mathbf{B}} , \ m_2 = \left.\frac{\partial^3 \mathcal{L}_{nl}}{\partial \mathcal{G}_0^3}\right|_{\mathbf{E},\mathbf{B}} , \ m_3 = \left.\frac{\partial^3 \mathcal{L}_{nl}}{\partial \mathcal{F}_0^2 \partial \mathcal{G}_0}\right|_{\mathbf{E},\mathbf{B}} , \ m_4 = \left.\frac{\partial^3 \mathcal{L}_{nl}}{\partial \mathcal{F} \partial \mathcal{G}_0^2}\right|_{\mathbf{E},\mathbf{B}} , \\ n_1 &= \left.\frac{\partial^4 \mathcal{L}_{nl}}{\partial \mathcal{F}_0^4}\right|_{\mathbf{E},\mathbf{B}} , \ n_2 = \left.\frac{\partial^4 \mathcal{L}_{nl}}{\partial \mathcal{G}_0^4}\right|_{\mathbf{E},\mathbf{B}} , \ n_3 = \left.\frac{\partial^4 \mathcal{L}_{nl}}{\partial \mathcal{F}_0^3 \partial \mathcal{G}_0}\right|_{\mathbf{E},\mathbf{B}} , \ n_4 = \left.\frac{\partial^4 \mathcal{L}_{nl}}{\partial \mathcal{F}_0^2 \partial \mathcal{G}_0^2}\right|_{\mathbf{E},\mathbf{B}} , \ n_5 = \left.\frac{\partial^4 \mathcal{L}_{nl}}{\partial \mathcal{F} \partial \mathcal{G}_0^3}\right|_{\mathbf{E},\mathbf{B}} . \end{aligned} \tag{3.11}$$

In general, for the purposes of this thesis, which involve investigations into propagation effects, we will only be interested in coefficients up to second order. Because, in these scenarios, the field equations are linearized, *i.e*, coming from a second-order Lagrangian. We chose to keep these results in order to record a more general case, in order to be useful for future readers. An example of the usefulness of these results could be the investigation of photon photon scattering in the presence of an intense external magnetic field. That said, from here on, we will take $m_i = n_j = 0$, for $(i = 1, 2, 3, 4; j = 1, 2, 3, 4, 5)$.

Using the action principle with respect to $a^\mu$, the Lagrangian (4.6) yields the EM field equations with source $J^\mu$

$$\partial^\mu \left[ c_1 f_{\mu\nu} + c_2 \widetilde{f}_{\mu\nu} - \frac{1}{2} Q_{B\mu\nu\kappa\lambda} f^{\kappa\lambda} \right] = -\partial^\mu G_{B\mu\nu} + J_\nu , \tag{3.12}$$

and the Bianchi identity remains the same one for the photon field, namely, $\partial_\mu \widetilde{f}^{\mu\nu} = 0$. Notice that, when we fix $c_1 = 1$ and $d_1 = d_2 = d_3 = 0$, the non-linear effects disappear, and we have the usual Maxwell ED. Moreover, the Maxwell equations also are recovered in eq. (3.12) by taking the aforementioned considerations and turn-off the background fields, $F_{B\mu\nu} = 0$.

## 3.1 The dispersion relations

In this section, we obtain the dispersion relations (DRs) associated with the photon field in the presence of a uniform and constant electromagnetic background. Thereby, from now on, all the coefficients defined in eq. (3.11) are not dependent on the space-time coordinates. We start with the equations written in terms of the fields $\mathbf{e}$ and $\mathbf{b}$. For the study of the wave propagation, we just consider the linear terms in $\mathbf{e}$ and $\mathbf{b}$, as well as the equations with no current and source, $\mathbf{J} = \mathbf{0}$ and $\rho = 0$. Under these conditions, the modified electrodynamics from eq. (3.12) and Bianchi identity is read below:

$$\nabla \cdot \mathbf{D} = 0 \ , \quad \nabla \times \mathbf{e} + \frac{\partial \mathbf{b}}{\partial t} = \mathbf{0} \ , \tag{3.13a}$$

$$\nabla \cdot \mathbf{b} = 0 \ , \quad \nabla \times \mathbf{H} - \frac{\partial \mathbf{D}}{\partial t} = \mathbf{0} \ , \tag{3.13b}$$

where the vectors $\mathbf{D}$ and $\mathbf{H}$ are given by

$$\mathbf{D} = c_1 \mathbf{e} + d_1 \mathbf{E} (\mathbf{E} \cdot \mathbf{e}) + d_2 \mathbf{B} (\mathbf{B} \cdot \mathbf{e}) - d_1 \mathbf{E} (\mathbf{B} \cdot \mathbf{b}) + d_2 \mathbf{B} (\mathbf{E} \cdot \mathbf{b}) \ , \tag{3.14a}$$

$$\mathbf{H} = c_1 \mathbf{b} - d_1 \mathbf{B} (\mathbf{B} \cdot \mathbf{b}) - d_2 \mathbf{E} (\mathbf{E} \cdot \mathbf{b}) + d_1 \mathbf{B} (\mathbf{E} \cdot \mathbf{e}) - d_2 \mathbf{E} (\mathbf{B} \cdot \mathbf{e}) \ . \tag{3.14b}$$

Observe that, in eqs. (3.14a) and (3.14b), we have eliminated the terms with dependence on the coefficient $d_3$, since all models treated in this thesis do not depend on this coefficient. We write the Fourier transform for the fields $\mathbf{e}$ and $\mathbf{b}$ and $\widetilde{\phi}$ such that the field equations (3.13a) and (3.13b) in momentum space are given by :

$$\mathbf{k} \cdot \mathbf{D}_0 = 0 \ , \quad \mathbf{k} \times \mathbf{e}_0 - \omega \, \mathbf{b}_0 = \mathbf{0} \ , \tag{3.15a}$$

$$\mathbf{k} \cdot \mathbf{b}_0 = 0 \ , \quad \mathbf{k} \times \mathbf{H}_0 + \omega \, \mathbf{D}_0 = \mathbf{0} \ . \tag{3.15b}$$

In terms of the electric and magnetic amplitudes $e_{0i}$ and $b_{0i}$, we obtain the following amplitudes for $\mathbf{D}$ and $\mathbf{H}$:

$$D_{0i} = \varepsilon_{ij} \, e_{0j} + \sigma_{ij} \, b_{0j} \ , \tag{3.16a}$$

$$H_{0i} = -\sigma_{ji} \, e_{0j} + (\mu^{-1})_{ij} \, b_{0j} \ , \tag{3.16b}$$

in which the electric permittivity symmetric tensor $\varepsilon_{ij}$ and $\sigma_{ij}$ are defined by

$$\varepsilon_{ij} = c_1\,\delta_{ij} + d_1\,E_i\,E_j + d_2\,B_i\,B_j\,, \tag{3.17a}$$

$$\sigma_{ij} = -d_1\,E_i\,B_j + d_2\,B_i\,E_j\,. \tag{3.17b}$$

In addition, $\mu^{-1}$ stands for the inverse of the magnetic permeability symmetric tensor, with the components

$$(\mu^{-1})_{ij} = c_1\,\delta_{ij} - d_1\,B_i\,B_j - d_2\,E_i\,E_j\,. \tag{3.18}$$

The inverse of eq. (3.18) yields the following expression for the magnetic permeability

$$\mu_{ij} = \frac{1}{c_1}\frac{(1 - d_B\,\mathbf{B}^2 - d_E\,\mathbf{E}^2)\,\delta_{ij} + d_B\,B_i\,B_j + d_E\,E_i\,E_j + d_B\,d_E\,(\mathbf{E}\times\mathbf{B})_i\,(\mathbf{E}\times\mathbf{B})_j}{1 - d_B\,\mathbf{B}^2 - d_E\,\mathbf{E}^2 + d_B\,d_E\,(\mathbf{E}\times\mathbf{B})^2}\,, \tag{3.19}$$

where we adopted the shorthand notations

$$d_B := \frac{d_1}{c_1} \quad\text{and}\quad d_E := \frac{d_2}{c_1}\,, \tag{3.20}$$

for simplicity. In both the cases in which $\mathbf{E} = \mathbf{0}$ or $\mathbf{B} = \mathbf{0}$, we have $\sigma_{ij} = 0$.

Using the equations in momentum space (3.15a)–(3.15b) and the constitutive relations (3.16a) and (3.16b), we obtain the wave equation for the components of the electric amplitude :

$$M^{ij}\,\mathbf{e}_0^{\,j} = 0\,, \tag{3.21}$$

where the matrix elements $M^{ij}$ are given by

$$\begin{aligned}M^{ij} =\ & a\,\delta^{ij} + b\,k^i\,k^j + c_B\,B^i\,B^j + c_E\,E^i\,E^j + d_B\,(\mathbf{B}\cdot\mathbf{k})\left(B^i\,k^j + B^j\,k^i\right)\\ & + d_E\,(\mathbf{E}\cdot\mathbf{k})\left(E^i\,k^j + E^j\,k^i\right) - d_B\,\omega\left[E^i\,(\mathbf{B}\times\mathbf{k})^j + E^j\,(\mathbf{B}\times\mathbf{k})^i\right]\\ & + d_E\,\omega\left[B^i\,(\mathbf{E}\times\mathbf{k})^j + B^j\,(\mathbf{E}\times\mathbf{k})^i\right]\,,\end{aligned} \tag{3.22}$$

whose the coefficients $a$, $b$, $c_B$ and $c_E$ are defined by

$$a = \omega^2 - \mathbf{k}^2 + d_B\,(\mathbf{k}\times\mathbf{B})^2 + d_E\,(\mathbf{k}\times\mathbf{E})^2\,, \tag{3.23a}$$

$$b = 1 - d_B\,\mathbf{B}^2 - d_E\,\mathbf{E}^2\,, \tag{3.23b}$$

$$c_B = d_E\,\omega^2 - d_B\,\mathbf{k}^2\,,\quad c_E = d_B\,\omega^2 - d_E\,\mathbf{k}^2\,. \tag{3.23c}$$

The non-trivial solution of eq. (3.21) implies that the determinant of the matrix $M^{ij}$ vanishes. It is challenging to computationally analyze the situation when both $\mathbf{E} \neq \mathbf{0}$ and $\mathbf{B} \neq \mathbf{0}$. Frequency solutions are feasible for the cases where $\mathbf{E} = \mathbf{0}$ or $\mathbf{B} = \mathbf{0}$. In what follows, we investigate these two particular cases separately.

### 3.1.1 The magnetic background case

Let us consider $\mathbf{E} = \mathbf{0}$ in the matrix elements (3.22):

$$M^{ij}\big|_{\mathbf{E}=0} = a_B\, \delta^{ij} + b_B\, k^i k^j + c_B\, B^i B^j + d_B\, (\mathbf{B} \cdot \mathbf{k})\left(B^i k^j + B^j k^i\right),  \tag{3.24}$$

where $a_B = \omega^2 - \mathbf{k}^2 + d_B\, (\mathbf{k} \times \mathbf{B})^2$ and $b_B = 1 - d_B\, \mathbf{B}^2$. The determinant of (3.24) is

$$\det(M)\big|_{\mathbf{E}=0} = a_B\,\Big\{\, a_B^2 + 2\, a_B\, d_B\, (\mathbf{B}\cdot\mathbf{k})^2 + a_B\left(c_B \mathbf{B}^2 + b_B \mathbf{k}^2\right) + \\ + b_B\, c_B\, (\mathbf{k}\times\mathbf{B})^2 - d_B^2\, (\mathbf{B}\cdot\mathbf{k})^2 (\mathbf{k}\times\mathbf{B})^2 \,\Big\}, \tag{3.25}$$

and imposing that $\det(M)\big|_{\mathbf{E}=0} = 0$, we obtain the first solution $a_B = 0$, that yields

$$\omega_{1B}(\mathbf{k}) = |\mathbf{k}|\,\sqrt{1 - \frac{d_1}{c_1}(\mathbf{B}\times\hat{\mathbf{k}})^2}\,. \tag{3.26}$$

The other solutions come from the polynomial equation

$$\omega^2\left(P_B\, \omega^4 + Q_B\, \omega^2 + R_B\right) = 0\,, \tag{3.27}$$

where the coefficients are defined by

$$P_B = 1 + \frac{d_2}{c_1}\mathbf{B}^2\,, \tag{3.28a}$$

$$Q_B = -2\,\mathbf{k}^2 - \frac{d_2}{c_1}\mathbf{B}^2\mathbf{k}^2 - \frac{d_2}{c_1}(\mathbf{k}\cdot\mathbf{B})^2\,, \tag{3.28b}$$

$$R_B = \mathbf{k}^4 + \frac{d_2}{c_1}\mathbf{k}^2(\mathbf{k}\cdot\mathbf{B})^2\,. \tag{3.28c}$$

The second solution is the trivial $\omega = 0$. Solving the above polynomial equation, one can show that the non-trivial solutions correspond to

$$\omega_{2B} = |\mathbf{k}|\,\sqrt{1 - \frac{d_2\,(\mathbf{B}\times\hat{\mathbf{k}})^2}{c_1 + d_2\mathbf{B}^2}} \tag{3.29a}$$

$$\omega_{3B} = |\mathbf{k}|\,. \tag{3.29b}$$

The usual photon frequencies are recovery in the limit $d_1 \to 0$, $d_2 \to 0$, or, equivalently, if we turn off the external magnetic field, $|\mathbf{B}| \to 0$. Note that, if the model does not depend on the coefficient $d_2$, the second solution also reduces to the usual Maxwell relations. This results are also obtained in ref. [64].

The refractive index associated with the DRs are defined by

$$n_{iB} = \frac{|\mathbf{k}|}{\omega_{iB}}\quad (i = 1, 2, 3)\,. \tag{3.30}$$

We point out that, for the DR in eq. (3.26), the refractive index only depends on the direction of the magnetic field $\mathbf{B}$ with the wave propagating direction $\mathbf{k}$.

Since we have three solutions for the frequencies, each solution has a different group velocity. For the frequency in eq. (3.26), we obtain

$$\mathbf{v}_{gB}\big|_{\omega=\omega_{1B}} = \frac{c_1\,\hat{\mathbf{k}} + d_1\,\mathbf{B}\times(\mathbf{B}\times\hat{\mathbf{k}})}{c_1\,\sqrt{1 - \frac{d_1}{c_1}(\mathbf{B}\times\hat{\mathbf{k}})^2}}. \tag{3.31}$$

The polynomial equation (3.27) has the correspondent group velocity:

$$\mathbf{v}_{gB} = \hat{\mathbf{k}}\frac{d\omega}{dk} = \frac{-\hat{\mathbf{k}}}{2\omega\,(2\omega^2 P_B + Q_B)}\left(\frac{dR_B}{dk} + \omega^2\frac{dQ_B}{dk}\right), \tag{3.32}$$

where $\omega$ is now evaluated at the DRs $\omega = \omega_{2B}$ and $\omega = \omega_{3B}$, in which $k \equiv |\mathbf{k}|$. Using the definitions of $P_B$, $Q_B$ and $R_B$, the expression (3.32) is read below

$$\mathbf{v}_{gB} = \frac{\mathbf{k}}{\omega}\left[\frac{2c_1(\omega^2-\mathbf{k}^2) - d_2(\mathbf{B}\cdot\mathbf{k})^2 + d_2\omega^2\mathbf{B}^2 + d_2(\omega^2-\mathbf{k}^2)(\mathbf{B}\cdot\hat{\mathbf{k}})^2}{2c_1(\omega^2-\mathbf{k}^2) + d_2\mathbf{B}^2(2\omega^2-\mathbf{k}^2) - d_2(\mathbf{B}\cdot\mathbf{k})^2}\right]. \tag{3.33}$$

Substituting the frequencies $\omega_{2B}$ and $\omega_{3B}$ in eq. (3.33), the group velocities are given by :

$$\mathbf{v}_{gB}\big|_{\omega=\omega_{2B}} = \frac{c_1\,\hat{\mathbf{k}} + d_2\,\mathbf{B}\,(\mathbf{B}\cdot\hat{\mathbf{k}})}{(c_1+d_2\,\mathbf{B}^2)\sqrt{1-\frac{d_2\,(\mathbf{B}\times\hat{\mathbf{k}})^2}{c_1+d_2\mathbf{B}^2}}}. \tag{3.34a}$$

$$\mathbf{v}_{gB}\big|_{\omega=\omega_{3B}} = \hat{\mathbf{k}}. \tag{3.34b}$$

The results (3.31) and (3.34a) show the dependence of the group velocity on the angle between the magnetic background $\mathbf{B}$ and the propagation direction $\hat{\mathbf{k}}$. It is important to highlight that the Maxwell limit recovers the known results for the group velocities (3.31) and (3.33), *i.e.*, $\mathbf{v}_g = c\,\hat{\mathbf{k}}$ (with $c=1$), when $d_1 = d_2 = 0$ and $c_1 = 1$.

### 3.1.2 The electric background case

The electric background case is obtained with $\mathbf{B} = 0$ in eq. (3.22) :

$$M^{ij}\big|_{\mathbf{B}=0} = a_E\,\delta^{ij} + b_E\,k^i k^j + c_E\,E^i E^j + d_E\,(\mathbf{E}\cdot\mathbf{k})\left(E^i k^j + E^j k^i\right), \tag{3.35}$$

where $a_E = \omega^2 - \mathbf{k}^2 + d_E\,(\mathbf{k}\times\mathbf{E})^2$ and $b_E = 1 - d_E\,\mathbf{E}^2$. The correspondent determinant is similar to the result (3.25) :

$$\det(M)\big|_{\mathbf{B}=0} = a_E\,\Big\{a_E^2 + 2\,a_E\,d_E\,(\mathbf{E}\cdot\mathbf{k})^2 + a_E\left(c_E\mathbf{E}^2 + b_E\mathbf{k}^2\right)$$
$$+ b_E\,c_E\,(\mathbf{k}\times\mathbf{E})^2 - d_E^2\,(\mathbf{E}\cdot\mathbf{k})^2\,(\mathbf{k}\times\mathbf{E})^2\Big\}. \tag{3.36}$$

The null determinant in eq. (3.36) implies the first condition $a_E = 0$, or equivalently $\omega^2 - \mathbf{k}^2 + d_E (\mathbf{k} \times \mathbf{E})^2 = 0$, that yields the solutions $\omega_{1E}^{\pm} = \pm \omega_{1E}(\mathbf{k})$ and $\omega_{2E}^{\pm} = \pm \omega_{2E}(\mathbf{k})$, where the DRs are given by

$$\omega_{1E.}(\mathbf{k}); = \sqrt{\mathbf{k}^2 - \frac{d_2}{c_1}(\mathbf{E} \times \mathbf{k})^2}, \tag{3.37a}$$

$$\omega_{2E.}(\mathbf{k}) = |\mathbf{k}|. \tag{3.37b}$$

The second condition for eq. (3.36) to be null leads to the polynomial equation

$$\omega^2 \left[ \left(1 + \frac{d_1}{c_1}\mathbf{E}^2\right) \omega^2 - \mathbf{k}^2 - \frac{d_1}{c_1}(\mathbf{E} \cdot \mathbf{k})^2 \right] = 0. \tag{3.38}$$

The first root in eq. (3.38) is $\omega = 0$, and the non-trivial solutions are $\omega_{3E}^{\pm} = \pm \omega_{3E}(\mathbf{k})$, with

$$\omega_{3E}(\mathbf{k}) = |\mathbf{k}| \sqrt{1 - \frac{d_1 (\mathbf{E} \times \hat{\mathbf{k}})^2}{c_1 + d_1 \mathbf{E}^2}}. \tag{3.39}$$

The refractive index in an electric background field is

$$n_{iE} = \frac{|\mathbf{k}|}{\omega_{iE}} \quad (i = 1, 2, 3). \tag{3.40}$$

The DR (3.39) yields a refractive index that depends on the direction of $\mathbf{E}$ with the $\mathbf{k}$-wave propagation. From the condition $a_E = 0$, the correspondent group velocity is given by

$$\mathbf{v}_{gE} = \frac{\mathbf{k}}{\omega} \left[ 1 - \frac{d_2}{c_1}(\mathbf{E} \times \hat{\mathbf{k}})^2 \right], \tag{3.41}$$

where $\omega$ must be evaluated at the dispersion relations $\omega_{1E}$ and $\omega_{2E}$. Substituting the frequencies (3.37a) and (3.37b) in eq. (3.41), we obtain the results

$$\mathbf{v}_{gE}\big|_{\omega=\omega_{1E}} = \frac{c_1 \hat{\mathbf{k}} + d_1 \mathbf{E} \times (\mathbf{E} \times \hat{\mathbf{k}})}{c_1 \sqrt{1 - \frac{d_1}{c_1}(\mathbf{E} \times \hat{\mathbf{k}})^2}}, \tag{3.42a}$$

$$\mathbf{v}_{gE}\big|_{\omega=\omega_{2E}} = \hat{\mathbf{k}}. \tag{3.42b}$$

The third possible solution for the group velocity comes from the eq. (3.38). In this case, we obtain the group velocity

$$\mathbf{v}_{gE} = \frac{\mathbf{k}}{\omega} \left[ 1 - \frac{d_1 (\mathbf{E} \times \hat{\mathbf{k}})^2}{c_1 + d_1 \mathbf{E}^2} \right]. \tag{3.43}$$

Using the dispersion relation (3.39) in eq. (3.43), the correspondent group velocity reads

$$\mathbf{v}_{gE}\big|_{\omega=\omega_{3E}} = \frac{c_1 \hat{\mathbf{k}} + d_1 \mathbf{E} (\mathbf{E} \cdot \hat{\mathbf{k}})}{(c_1 + d_1 \mathbf{E}^2) \sqrt{1 - \frac{d_1 (\mathbf{E} \times \hat{\mathbf{k}})^2}{c_1 + d_1 \mathbf{E}^2}}}. \tag{3.44}$$

In the limit, $d_1 = d_2 = 0$ and $c_1 = 1$, the group velocities (3.42a) and (3.44) reduce to the usual Maxwell case $\mathbf{v}_{gE} = c\,\hat{\mathbf{k}}$ (with $c = 1$). Analogously to the magnetic background case, the results obtained in this section also depends on the angle between the electric background and the wave propagation direction. In all the results, the dispersion relations and the group velocities depend on the coefficients $c_1$, $d_1$ and $d_2$, which are fixed by the non-linear ED as functions of the magnetic or electric background fields.

## 3.2 Optical Properties

### 3.2.1 Birrefrigence

Let us recall Maxwell's equations in momenta space:

$$\mathbf{k} \cdot \mathbf{D}_0 = 0 \quad, \quad \mathbf{k} \times \mathbf{e}_0 - \omega\,\mathbf{b}_0 = \mathbf{0}\,, \tag{3.45a}$$

$$\mathbf{k} \cdot \mathbf{b}_0 = 0 \quad, \quad \mathbf{k} \times \mathbf{H}_0 + \omega\,\mathbf{D}_0 = \mathbf{0}\,, \tag{3.45b}$$

In the case that the background field is purely magnetic, we have the following expressions:

$$D_{0i}(\mathbf{k}, \omega) = \varepsilon_{ij}\, e_{0j}\,, \tag{3.46a}$$

$$H_{0i}(\mathbf{k}, \omega) = (\mu^{-1})_{ij}\, b_{0j}\,, \tag{3.46b}$$

in which

$$\varepsilon_{ij} = c_1\,\delta_{ij} + d_2\, B_i\, B_j\,, \tag{3.47a}$$

$$(\mu^{-1})_{ij} = c_1\,\delta_{ij} - d_1\, B_i\, B_j \tag{3.47b}$$

Substituting the equations (3.46a), (3.46b) and the Faraday-Lenz law (3.45a) into the Ampere-Mawell equation (3.45b), we can obtain an equation of motion in terms only of the amplitude of the electric field and of the permittivity and permeability tensors. Thus, we will have:

$$\mathbf{k} \times \left[\,(\mu)^{-1} \cdot (\mathbf{k} \times \mathbf{e}_0)\,\right] + \omega^2\,(\epsilon \cdot \mathbf{e}_0) = 0. \tag{3.48}$$

(i) Let's look at the situation in which the amplitude of the electric field is parallel to the background Magnetic field, that is, $\mathbf{e}_0 \parallel \mathbf{B}$. We can make the following choice:

$\mathbf{k} = k\,\hat{\mathbf{x}}$ , $\mathbf{e}_0 = e_{03}\,\hat{\mathbf{z}}$ and $\mathbf{B} = B\,\hat{\mathbf{z}}$ . Carrying out the explicit calculations, we obtain that

$$\begin{pmatrix} 0 \\ 0 \\ -k^2 e_3 \mu_{22}^{-1} \end{pmatrix} + \begin{pmatrix} 0 \\ 0 \\ \omega^2 e_3 \epsilon_{33} \end{pmatrix} = \begin{pmatrix} 0 \\ 0 \\ 0 \end{pmatrix} . \tag{3.49}$$

In this case, the wave equation (3.48) yields the relation $\mu_{22}(k,\omega)\,\varepsilon_{33}(k,\omega)\,\omega^2 = k^2$, where the parallel refractive index is defined by

$$n_\parallel^{(B)}(k,\omega) = \sqrt{\mu_{22}(k,\omega)\,\varepsilon_{33}(k,\omega)} \ . \tag{3.50}$$

(ii) Analogously, let's consider the situation in which the amplitude of the electric field is perpendicular to the background field, $\mathbf{e}_0 \perp \mathbf{B}$. We keep the wave vector fixed, $\mathbf{k} = k\,\hat{\mathbf{x}}$, but now $\mathbf{e}_0 = e_2\,\hat{\mathbf{y}}$ and $\mathbf{B} = B\,\hat{\mathbf{z}}$. In this way, we will have

$$\begin{pmatrix} 0 \\ -k^2 e_2 \mu_{33}^{-1} \\ 0 \end{pmatrix} + \begin{pmatrix} 0 \\ \omega^2 e_2 \epsilon_{22} \\ 0 \end{pmatrix} = \begin{pmatrix} 0 \\ 0 \\ 0 \end{pmatrix} . \tag{3.51}$$

In this case, we have the relation $\mu_{33}(k,\omega)\,\varepsilon_{22}(k,\omega)\,\omega^2 = k^2$, in which the perpendicular refractive index is

$$n_\perp^{(B)}(k,\omega) = \sqrt{\mu_{33}(k,\omega)\,\varepsilon_{22}(k,\omega)} \ . \tag{3.52}$$

The quantity called birefringence is an optical property of an anisotropic medium expressed by the dependence of the refractive index on the polarization and direction of propagation of an electromagnetic wave. The phenomenon of birefringence manifests itself by the difference between the refractive indices of eqs. (3.52) and (3.50) , as defined below,

$$\Delta n \equiv n_\parallel - n_\perp \ . \tag{3.53}$$

### 3.2.2   *Euler-Heseinberg Electrodynamics*

The first model describing non-linear electrodynamics was investigated by H. Euler and B. Kockel [65] in 1934 . The main objective was to adapt Maxwell's equations in a vacuum so that they were compatible with the quantum phenomenon of creation of virtual electron-positron pairs predicted by Dirac [66] in 1931. The effective Lagrangian of this model is written as

$$\mathcal{L}_{EH}(\mathcal{F},\mathcal{G}) = \mathcal{F} + \frac{2\alpha^2}{45m_e^4}\left(4\mathcal{F}^2 + 7\mathcal{G}^2\right), \tag{3.54}$$

where $\alpha = e^2 = (137)^{-1} = 0.00729$ is the fine structure constant. Later, the Lagrangian (3.54) was generalized by W.Heisenberg and H. Euler [38] in 1936 by a non-perturbative expression given by

$$\mathcal{L}_{EH}^{\text{full}}(\mathcal{F},\mathcal{G}) = \mathcal{F} - \frac{1}{8\pi^2}\int_0^\infty \frac{ds}{s^3}e^{-m^2 s}\left[(es)^2\mathcal{G}\frac{\mathbb{R}e\cosh es\sqrt{-\mathcal{F}+i\mathcal{G}}}{\mathbb{I}m\cosh es\sqrt{-\mathcal{F}+i\mathcal{G}}} + \frac{2}{3}(es)^2\mathcal{F} - 1\right], \tag{3.55}$$

where $m = 0.5\text{MeV}$ is the mass of the electron. In this sense, the Lagrangian (3.54) becomes a specific case for the above expression expanded to the fourth order in $F^{\mu\nu}$ in a scenario where the photons have low energy, i.e, $\hbar\omega << mc^2$.

Using the general field equation (3.4) for a nonlinear ED, the equations of motion for the Lagrangian (3.54) can be written as follows

$$\partial_\mu\left[(1-2\beta\mathcal{F})F^{\mu\nu} - \gamma\mathcal{G}\tilde{F}^{\mu\nu}\right] = 0. \tag{3.56}$$

In vector notation, we can rewrite eqs. (3.56) as

$$\nabla \cdot \mathbf{D} = 0, \tag{3.57a}$$

$$\frac{\partial \mathbf{D}}{\partial t} - \nabla \times \mathbf{H} = 0, \tag{3.57b}$$

$$\nabla \cdot \mathbf{B} = 0, \tag{3.57c}$$

$$\frac{\partial \mathbf{B}}{\partial t} + \nabla \times \mathbf{E} = 0, \tag{3.57d}$$

where the vectors $\mathbf{D}$ and $\mathbf{H}$ are defined, respectively, by

$$\mathbf{D} = (1-2\beta\mathcal{F})\mathbf{E} + \gamma\mathcal{G}\mathbf{B}, \tag{3.58}$$

$$\mathbf{H} = (1-2\beta\mathcal{F})\mathbf{B} - \gamma\mathcal{G}\mathbf{E}. \tag{3.59}$$

$$\tag{3.60}$$

It is worth remembering that the equations (3.57c) and (3.57d) given by the Bianchi identities do not change.

### 3.2.3  Born-Infeld Electrodynamics

The BI electrodynamics is described by the Lagrangian

$$\mathcal{L}_{BI}(\mathcal{F}_0, \mathcal{G}_0) = \beta^2 \left[ 1 - \sqrt{1 - 2\frac{\mathcal{F}_0}{\beta^2} - \frac{\mathcal{G}_0^2}{\beta^4}} \right] , \qquad (3.61)$$

where $\beta$ is a scale parameter with dimension of squared energy (in natural units). It can be interpreted as a critical field in the theory, and has the same dimension of the electromagnetic field. The BI Lagrangian is CP-invariant since it depends on the $\mathcal{G}_0^2$. The limit of $\beta \to \infty$ recovers the Maxwell Lagrangian. According to the original work where the model was proposed, a maximum electric field produces a finite self-energy for the electron that fixes the BI-parameter at $\beta = 1.187 \times 10^{20}$ V/m, that in MeV scale is $\sqrt{\beta} = 16$ MeV [26]. However, more recent investigations have yielded stronger constraints on the $\beta$ parameter. For example, the measurement by ATLAS of the light-by-light scattering in Pb-Pb collisions constraints a stringent lower bound on the $\beta$-parameter [35], namely,

$$\sqrt{\beta} \gtrsim 100 \,\text{GeV} , \qquad (3.62)$$

in the case of a pure quantum electrodynamics with the Born-Infeld theory associated with the Abelian $U(1)$ symmetry.

Originally, the BI model was proposed to remove the singularity of the electric field of a point-like charged particle at the origin [26]. Also, the BI model comes naturally from string theories. For example, in the works [28, 29], the reader may find the microscopic origin of the Born-Infeld action as described in the framework of string theory. The equations of motion from the Lagrangian (3.61) are written below

$$\partial_\mu \left[ \frac{1}{\sqrt{\Omega}} \left( F^{\mu\nu} + \frac{1}{\beta^2} \mathcal{G} \tilde{F}^{\mu\nu} \right) \right] = 0 , \qquad (3.63)$$

where

$$\Omega = 1 - 2\frac{\mathcal{F}_0}{\beta^2} - \frac{\mathcal{G}_0^2}{\beta^4} . \qquad (3.64)$$

In vector form, we can join the equation (3.63) with the Bianchi identities (which do not change), and obtain the Maxwell equations which are given by the following expressions:

$$\nabla \cdot \mathbf{D} = \rho, \tag{3.65a}$$

$$\nabla \times \mathbf{E} + \partial_t \mathbf{B} = 0, \tag{3.65b}$$

$$\nabla \cdot \mathbf{H} = 0, \tag{3.65c}$$

$$\nabla \times \mathbf{H} - \partial_t \mathbf{D} = \mathbf{J}. \tag{3.65d}$$

where the vectors $\mathbf{D}$ and $\mathbf{H}$ are defined, respectively, as follows

$$\mathbf{D} = \frac{1}{\sqrt{\Omega}} \left( \mathbf{E} + \frac{\mathcal{G}}{\beta^2} \mathbf{B} \right), \tag{3.66a}$$

$$\mathbf{H} = \frac{1}{\sqrt{\Omega}} \left( \mathbf{B} - \frac{\mathcal{G}}{\beta^2} \mathbf{E} \right). \tag{3.66b}$$

Using eq. (3.5), we determine the Born-Infeld energy moment tensor, which can be written as

$$(\Theta_{BI})^\mu_\nu = \frac{1}{\sqrt{\Omega}} F^{\mu\kappa} F_{\kappa\lambda} + \delta^\mu_\nu \left( \frac{1}{\beta^2 \sqrt{\Omega}} \mathcal{G} - \mathcal{L} \right). \tag{3.67}$$

Let us investigate what the problem of a point charge is like in Born-Infeld electrodynamics. Let us start by considering a stationary point charge $q$ located at the origin. In this case, there are no magnetic fields, that is, $\mathbf{B} = \mathbf{H} = 0$. Then the field equations (3.65a) reduce to

$$\nabla \cdot \mathbf{D} = 0, \tag{3.68a}$$

$$\nabla \times \mathbf{E} = 0. \tag{3.68b}$$

Since the electric field of a point charge is spherically symmetric and only depends on the radius, we can use spherical coordinates, so that eq. (3.68a) will simply be

$$\frac{1}{r^2} \frac{d}{dr}(r^2 D_r) = 0, \tag{3.69}$$

whose solution is

$$D_r = \frac{K}{r^2}, \tag{3.70}$$

where $K$ is a constant to be determined. We can use Gauss's law

$$\oint_s \mathbf{D} \cdot d\mathbf{a} = q. \tag{3.71}$$

to obtain it. Applying (3.70), we are led to $4\pi K = q$. Therefore,

$$D_r = \frac{q}{4\pi r^2}. \tag{3.72}$$

On the other hand, $D_r$ is the radial component of the vector $\mathbf{D}$ given by

$$D_r = \frac{E_r}{\sqrt{1 - \frac{E_r^2}{\beta^2}}} = -\frac{\Phi'(r)}{\sqrt{1 - \frac{(\Phi'(r))^2}{\beta^2}}}, \tag{3.73}$$

where $\mathbf{E} = -\partial_r \Phi = \Phi'(r)$ and $\Phi$ is the scalar potential. Combining the eqs. (3.72) and (3.73), we will have

$$\frac{q}{4\pi r^2} = -\frac{\Phi'(r)}{\sqrt{1 - \frac{(\Phi'(r))^2}{\beta^2}}}. \tag{3.74}$$

The variables in this equation can be easily separated, so that

$$d\Phi = -\frac{q}{4\pi r_0} \frac{dr}{\sqrt{1 + (r/r_0)^4}}, \tag{3.75}$$

where $r_0 \equiv |q|/4\pi\beta$. Performing integration and taking the reference point at infinity we are led to

$$\Phi = \frac{q}{4\pi r_0} f\left(\frac{r}{r_0}\right), \tag{3.76}$$

with

$$f(x) = \int_x^\infty \frac{du}{\sqrt{1 + u^4}}, \quad \text{with,} \quad u \equiv \frac{r}{r_0}. \tag{3.77}$$

This is the potential of a point charge in the Born-Infeld theory. Note that it is always finite, unlike the Coulomb potential (which is recovered when $x \gg 1$) which diverges at the origin. The integral (3.77) is solved by replacing $u = \tan \beta/2$. Thus

$$\begin{aligned} f(x) &= \frac{1}{2} \int_{\beta'}^\pi \frac{d\beta}{\sqrt{1 - \frac{1}{2}\sin^2 \beta}} \\ &= \frac{1}{2} \int_0^\pi \frac{d\beta}{\sqrt{1 - \frac{1}{2}\sin^2 \beta}} + \frac{1}{2} \int_{\beta'}^0 \frac{d\beta}{\sqrt{1 - \frac{1}{2}\sin^2 \beta}}, \end{aligned} \tag{3.78}$$

where $\beta' = 2\arctan x$. The integrals above are the well-known elliptic integrals of the first kind, which are defined as

$$F(\phi, k) = \int_0^\phi \frac{d\theta}{\sqrt{1 - k^2 \sin^2 \theta}}. \tag{3.79}$$

Therefore, we can write eq. (3.78) as

$$\begin{aligned} f(x) &= \frac{1}{2}F\left(\pi, \frac{1}{\sqrt{2}}\right) - \frac{1}{2}F\left(\beta', \frac{1}{\sqrt{2}}\right) \\ &= f(0) - \frac{1}{2}F\left(\beta', \frac{1}{\sqrt{2}}\right). \end{aligned} \quad (3.80)$$

The first integral occurs in the situation where $x = 0$. Calculating, we get that

$$f(0) = \frac{1}{2}\int_0^\pi \frac{\mathrm{d}\beta}{\sqrt{1 - \frac{1}{2}\sin^2\beta}} = 1.85. \quad (3.81)$$

So, at the origin, the potential is given by

$$\Phi(0) = 1.85 \frac{q}{4\pi r_0}. \quad (3.82)$$

This is the maximum value of $f(x)$ and consequently, the maximum value of the potential. Finally, the electric field for a point particle in Born-Infeld electrodynamics is obtained by

$$\mathbf{E} = -\frac{\mathrm{d}\Phi}{\mathrm{d}r}\hat{\mathbf{r}} = \frac{1}{\sqrt{1 + (\frac{r}{r_0})^4}}\hat{\mathbf{r}}. \quad (3.83)$$

Note that there are also no singularities in the electric field. The figure 3.1 shows a plot comparing Born-Infeld potential with the Coulomb potential.

### 3.2.4 ModMax Electrodynamics

A recent model that has been widely investigated in the literature is the ModMax electrodynamics [67, 23, 68]. The ModMax Lagrangian is given by

$$\begin{aligned} \mathcal{L}_{MM}(\mathcal{F}, \mathcal{G}) &= \cosh\gamma\,\mathcal{F} + \sinh\gamma\sqrt{\mathcal{F}^2 + \mathcal{G}^2} + J_\mu A^\mu, \\ &= \frac{\cosh\gamma}{2}(\mathbf{E}^2 - \mathbf{B}^2) + \frac{\sinh\gamma}{2}\sqrt{(\mathbf{E}^2 - \mathbf{B}^2)^2 + 4(\mathbf{E}\cdot\mathbf{B})^2} + \rho\phi + \mathbf{J}\cdot\mathbf{A}, \end{aligned} \quad (3.84)$$

where $\gamma$ is a adimensional parameter.

This non-linear ED has been well motivated in the literature due to the conformal invariance. Thus, it is the only non-linear ED that preserve the duality and the conformal symmetries in the same Lagrangian. Causality and unitarity conditions restrict $\gamma$ to be positive. The equations of motion from (3.84) are axpressed as

$$\partial_\mu\left(\cosh\gamma\, F^{\mu\nu} + \sinh\gamma\,\frac{\mathcal{F}F^{\mu\nu} + \mathcal{G}\tilde{F}^{\mu\nu}}{\sqrt{\mathcal{F}^2 + \mathcal{G}^2}}\right) = J^\nu. \quad (3.85)$$

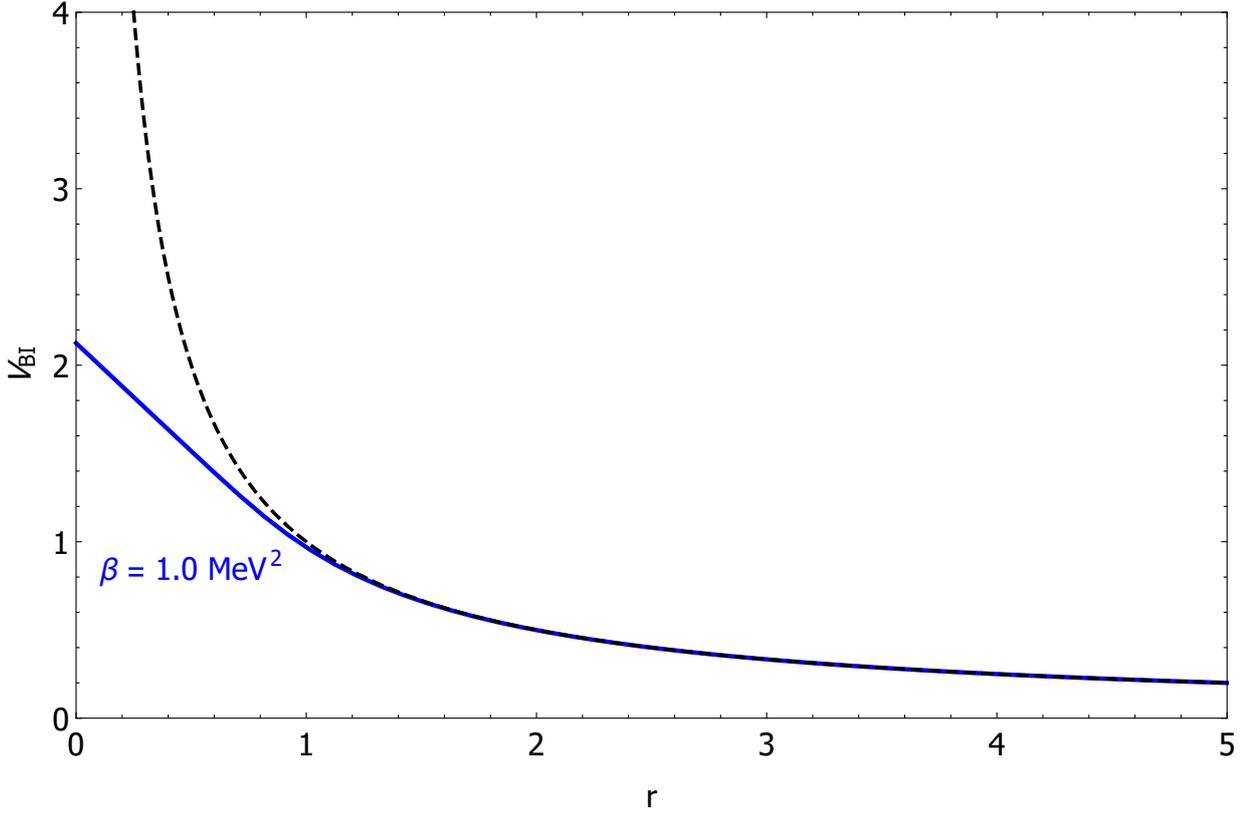

Figura 3.1: Comparison of Born-Infeld and Coulomb Potentials

Note that the continuity equation $\partial_\nu J^\nu = 0$ is satisfied. Furthermore, the Bianchi identities remain unchanged, that is, $\partial_\mu \tilde{F}^{\mu\nu} = 0$. The equations of motion can be expressed in terms of the electric and magnetic fields, namely

$$\nabla \cdot \mathbf{D} = \rho, \tag{3.86a}$$
$$\nabla \times \mathbf{E} + \partial_t \mathbf{B} = 0, \tag{3.86b}$$
$$\nabla \cdot \mathbf{H} = 0, \tag{3.86c}$$
$$\nabla \times \mathbf{H} - \partial_t \mathbf{D} = \mathbf{J}. \tag{3.86d}$$

where the vectors $\mathbf{D}$ and $\mathbf{H}$ are given by

$$\mathbf{D} = \cosh\gamma \, \mathbf{E} + \sinh\gamma \, \frac{\mathcal{F}\mathbf{E} + \mathcal{G}\mathbf{B}}{\sqrt{\mathcal{F}^2 + \mathcal{G}^2}}, \tag{3.87a}$$
$$\mathbf{H} = \cosh\gamma \, \mathbf{B} + \sinh\gamma \, \frac{\mathcal{F}\mathbf{B} - \mathcal{G}\mathbf{E}}{\sqrt{\mathcal{F}^2 + \mathcal{G}^2}}. \tag{3.87b}$$

Multiplying the Bianchi identity by $F_{\mu\nu}$, and using the Eq.(3.85) we arrive at the following conservation law

$$\partial_\mu \Theta^{\mu\nu}_{MM} = J_\alpha F^{\alpha\nu}, \tag{3.88}$$

where the energy-momentum tensor of the ModMax ED is given by

$$\Theta^{\mu\nu}_{MM} = \left(F^{\mu\rho}F_\rho{}^\nu - \frac{1}{4}\eta^{\mu\nu}\mathcal{F}\right)\left(\cosh\gamma + \frac{\sinh\gamma\mathcal{F}}{\sqrt{\mathcal{F}^2+\mathcal{G}^2}}\right). \qquad (3.89)$$

we can observe that the energy-momentum tensor is symmetric. This is due to the fact that ModMax electrodynamics is a conformal invariance model. Furthermore, if we ignore the external source, $J^\mu = 0$, we guarantee the continuity equation $\partial_\mu \Theta^{\mu\nu}_{MM} = 0$. The components of the energy-momentum tensor can be described as

$$\Theta^{00}_{MM} = \frac{1}{2}(\mathbf{E}^2+\mathbf{B}^2)\left[\cosh\gamma + \frac{\sinh\gamma(\mathbf{E}^2-\mathbf{B}^2)}{\sqrt{(\mathbf{E}^2-\mathbf{B}^2)^2+4(\mathbf{E}\cdot\mathbf{B})^2}}\right], \qquad (3.90a)$$

$$\Theta^{0i}_{MM} = (\mathbf{E}\times\mathbf{B})^i\left[\cosh\gamma + \frac{\sinh\gamma(\mathbf{E}^2-\mathbf{B}^2)}{\sqrt{(\mathbf{E}^2-\mathbf{B}^2)^2+4(\mathbf{E}\cdot\mathbf{B})^2}}\right], \qquad (3.90b)$$

$$\Theta^{ij}_{MM} = \left[E^i E^j + B^i B^j - \delta^{ij}\frac{1}{2}(\mathbf{E}^2+\mathbf{B}^2)\right]\left(\cosh\gamma + \frac{\sinh\gamma\mathcal{F}}{\sqrt{\mathcal{F}^2+\mathcal{G}^2}}\right). \qquad (3.90c)$$

Note that the expressions (3.90a) and (3.90b) are not well defined in the case of zero electromagnetic fields. This is due to the fact that we have the invariants $\mathcal{F}$ and $\mathcal{G}$ in the denominator. We can get around this problem if we work by defining the Hamiltonian from the following Legendre transform

$$\mathcal{H}(\mathbf{D},\mathbf{B})_{MM} = \mathbf{E}\cdot\mathbf{D} - \mathcal{L}_{MM}(\mathbf{E},\mathbf{B}), \qquad (3.91)$$

where $\mathbf{D}$ is the displacement vector defined by $\mathbf{D} \equiv \frac{\partial \mathcal{L}}{\partial \mathbf{E}}$. Note that we can make the following mechanical analogy: $\mathbf{D}$ can be identified as the canonical momentum $\mathbf{p}$ and $\mathbf{E}$ as the time derivative of a generalized coordinate $\mathbf{q}$. Using the Lagrangian (3.84) we get the expression

$$\mathcal{H}(\mathbf{D},\mathbf{B})_{MM} = \frac{1}{2}\left[\cosh\gamma(\mathbf{D}^2+\mathbf{B}^2) - \sinh\gamma\sqrt{(\mathbf{D}^2-\mathbf{B}^2)^2+4(\mathbf{D}\cdot\mathbf{B})^2}\right], \qquad (3.92)$$

which is positive definite if the parameter $\gamma$ obeys the constraint

$$\tanh\gamma < \sqrt{\frac{(\mathbf{D}^2+\mathbf{B}^2}{(\mathbf{D}^2-\mathbf{B}^2)^2+4(\mathbf{D}\cdot\mathbf{B})^2}}. \qquad (3.93)$$

In a theory where the energy-momentum tensor is symmetric, $\Theta_{0i} = \Theta_{i0}$, the definitions of momentum and Poynting vector coincide. In this sense, using the equation (3.91) we can write the Poynting vector, in terms of the vectors $\mathbf{D}$ and $\mathbf{B}$. Thus, we will have

$$\mathbf{S}_p = \mathbf{D}\times\mathbf{B}. \qquad (3.94)$$

An interesting investigation that combines the Born-Infeld and ModMax models is also addressed in the work [67], where the authors describe a generalized BI electrodynamics with dual symmetry, which in the weak field limit, reduces to Mod-Max.

The non-linear models presented in this session are just some of the many existing in the literature. Other models can be found in the references: [19, 20, 21, 22, 23, 24, 25, 69, 70, 71, 32]. Except for the Born-Infeld and Euler-Heisenberg models that come from fundamental theories such as string theory and QED, respectively, the vast majority of these models are considered effective theories. The overview is that in many cases Maxwellian electrodynamics is not always sufficient to describe certain phenomena. Furthermore, these models - which at an inattentive glance may seem like a mere exercise in mathematical physics - play an important role in several areas, such as gravitation, cosmology, topological materials and higher dimensional models. The work [23] and the references contained in it provide a passport to enter the vast universe of non-linear electrodynamic theories.

# Chapter 4

# Non-linear electrodynamics coupled to an axionic scalar field

We start up with the Lagrangian (density) of the model

$$\mathcal{L} = \mathcal{L}_{nl}(\mathcal{F}_0, \mathcal{G}_0) + \frac{1}{2}\left(\partial_\mu \phi\right)^2 - \frac{1}{2}m^2\,\phi^2 + g\,\phi\,\mathcal{G}_0 - J_\mu\,A^\mu \ , \tag{4.1}$$

where $\phi$ corresponds to the axion scalar field with mass $m$, and $g$ is the non-minimal coupling constant (with length dimension) of the axion with the electromagnetic field, *i.e.*, the usual coupling with the $\mathcal{G}_0$-invariant. There are many investigations and experiments to constrain the possible regions in the space of the parameters $g$ and $m$, which still remain with a wide range of values, depending on the phenomenological scale in analysis. For more details, we point out the recent reviews [97, 98].

The action principle leads to the following equations of motion with a classical external current $J^\nu$ :

$$\partial_\mu \left( \frac{\partial \mathcal{L}_{nl}}{\partial \mathcal{F}_0} F_0^{\,\mu\nu} + \frac{\partial \mathcal{L}_{nl}}{\partial \mathcal{G}_0} \widetilde{F}_0^{\,\mu\nu} \right) = -g\,(\partial_\mu \phi)\,\widetilde{F}_0^{\,\mu\nu} + J^\nu \ , \tag{4.2a}$$

$$\left(\Box + m^2\right)\phi = g\,\mathcal{G}_0 \ , \tag{4.2b}$$

in which the current is conserved $\partial_\mu J^\mu = 0$.

We expand the Lagrangian (4.1) around the background up to second order in the propagating field $a^\mu$ to yield the expression

$$\begin{aligned}
\mathcal{L}^{(2)} &= -\frac{1}{4}c_1\,f_{\mu\nu}^2 - \frac{1}{4}c_2\,f_{\mu\nu}\widetilde{f}^{\mu\nu} - \frac{1}{2}G_{B\mu\nu}\,f^{\mu\nu} + \frac{1}{8}Q_{B\mu\nu\kappa\lambda}\,f^{\mu\nu}f^{\kappa\lambda} + \frac{1}{2}\left(\partial_\mu\phi\right)^2 - \frac{1}{2}m^2\,\phi^2 \\
&\quad - \frac{1}{2}g\,\phi\,\widetilde{F}_{B\mu\nu}\,f^{\mu\nu} + g\,\phi\,\mathcal{G}_B - J_\mu\,a^\mu - J_\mu\,A_B^{\,\mu} + \mathcal{L}_{nl}\left(\mathcal{F}_B, \mathcal{G}_B\right) \ ,
\end{aligned} \tag{4.3}$$

where the background tensors are defined by

$$\begin{aligned} G_{B\mu\nu} &= c_1 \, F_{B\mu\nu} + c_2 \, \widetilde{F}_{B\mu\nu} \, , \\ Q_{B\mu\nu\kappa\lambda} &= d_1 \, F_{B\mu\nu} \, F_{B\kappa\lambda} + d_2 \, \widetilde{F}_{B\mu\nu} \, \widetilde{F}_{B\kappa\lambda} + d_3 \, F_{B\mu\nu} \widetilde{F}_{B\kappa\lambda} + d_3 \, \widetilde{F}_{B\mu\nu} F_{B\kappa\lambda} \, , \end{aligned} \quad (4.4)$$

and $\mathcal{L}_{nl}\left(\mathcal{F}_B, \mathcal{G}_B\right)$ is the non-linear Lagrangian as function of the Lorentz invariants $\mathcal{F}_B = -\frac{1}{4} F_{B\mu\nu}^2 = \mathbf{E}^2 - \mathbf{B}^2$ and $\mathcal{G}_B = -\frac{1}{4} F_{B\mu\nu} \widetilde{F}_B{}^{\mu\nu} = \mathbf{E} \cdot \mathbf{B}$, both in terms of the EM background field, and $\widetilde{F}_B{}^{\mu\nu} = \epsilon^{\mu\nu\alpha\beta} F_{B\alpha\beta}/2 = \left(-B^i, \, \epsilon^{ijk} E^k\right)$ is the dual tensor of $F_{B\mu\nu}$. Furthermore, the coefficients $c_1$, $c_2$, $d_1$, $d_2$ and $d_3$ are evaluated at $\mathbf{E}$ and $\mathbf{B}$, as follows :

$$c_1 = \left.\frac{\partial \mathcal{L}_{nl}}{\partial \mathcal{F}_0}\right|_{\mathbf{E},\mathbf{B}}, \; c_2 = \left.\frac{\partial \mathcal{L}_{nl}}{\partial \mathcal{G}_0}\right|_{\mathbf{E},\mathbf{B}}, \; d_1 = \left.\frac{\partial^2 \mathcal{L}_{nl}}{\partial \mathcal{F}_0^2}\right|_{\mathbf{E},\mathbf{B}}, \; d_2 = \left.\frac{\partial^2 \mathcal{L}_{nl}}{\partial \mathcal{G}_0^2}\right|_{\mathbf{E},\mathbf{B}}, \; d_3 = \left.\frac{\partial^2 \mathcal{L}_{nl}}{\partial \mathcal{F}_0 \partial \mathcal{G}_0}\right|_{\mathbf{E},\mathbf{B}} \quad (4.5)$$

that depend on the EM field magnitude and it may also be functions of the space-time coordinates. Following the previous definitions of the tensors, we have that $G_{B\mu\nu} = -G_{B\nu\mu}$, and $Q_{B\mu\nu\kappa\lambda}$ is symmetric under exchange $\mu\nu \leftrightarrow \kappa\lambda$, and antisymmetric when $\mu \leftrightarrow \nu$ or $\kappa \leftrightarrow \lambda$. Note that the current $J^\mu$ also couples to the external potential $A_B{}^\mu$, but this term and $\mathcal{L}_{nl}\left(\mathcal{F}_B, \mathcal{G}_B\right)$ are irrelevant for the field equations in which we are interested. If we consider the scalar potential as $V(\phi) = m^2 \phi^2/2 - g\,\phi\,\mathcal{G}_B$, it has a minimal at $\phi_0 = g\,\mathcal{G}_B/m^2$. Writing $\phi = \widetilde{\phi} + \phi_0$, the term $g\,\phi\,\mathcal{G}_B$ can be eliminated in the Lagrangian (4.3) :

$$\begin{aligned} \mathcal{L}^{(2)} &= -\frac{1}{4} c_1 \, f_{\mu\nu}^2 - \frac{1}{4} c_2 \, f_{\mu\nu} \widetilde{f}^{\mu\nu} - \frac{1}{2} H_{B\mu\nu} \, f^{\mu\nu} + \frac{1}{8} Q_{B\mu\nu\kappa\lambda} \, f^{\mu\nu} \, f^{\kappa\lambda} + \frac{1}{2} (\partial_\mu \widetilde{\phi})^2 - \frac{1}{2} m^2 \, \widetilde{\phi}^2 \\ &\quad - \frac{1}{2} g \, \widetilde{\phi} \, \widetilde{F}_{B\mu\nu} \, f^{\mu\nu} - J_\mu \, a^\mu - J_\mu \, A_B{}^\mu + \mathcal{L}_{nl}\left(\mathcal{F}_B, \mathcal{G}_B\right) + \frac{g^2 \, \mathcal{G}_B^2}{2m^2} \, , \end{aligned} \quad (4.6)$$

where $H_{B\mu\nu} = G_{B\mu\nu} + g^2 \, \mathcal{G}_B \, \widetilde{F}_{B\mu\nu}/m^2$. In this context, the scalar field $\widetilde{\phi}$ is reinterpreted as the axion field with mass $m$. It should be noted that $\phi_0$ is non-trivial only in the presence of both electric and magnetic background fields.

Using the action principle in relation to $a^\mu$, the Lagrangian (4.6) yields the EM field equations with source $J^\mu$

$$\partial^\mu \left[c_1 \, f_{\mu\nu} + c_2 \, \widetilde{f}_{\mu\nu} - \frac{1}{2} Q_{B\mu\nu\kappa\lambda} \, f^{\kappa\lambda}\right] = -g \, (\partial^\mu \widetilde{\phi}) \, \widetilde{F}_{B\mu\nu} - \partial^\mu H_{B\mu\nu} + J_\nu \, , \quad (4.7)$$

and the Bianchi identity remains the same one for the photon field, namely, $\partial_\mu \widetilde{f}^{\mu\nu} = 0$. The action principle related to $\widetilde{\phi}$ in eq. (4.6), now yields the axion field equation evaluated at the EM background :

$$\left(\Box + m^2\right) \widetilde{\phi} = -\frac{1}{2} g \, \widetilde{F}_{B\mu\nu} \, f^{\mu\nu} \, . \quad (4.8)$$

Notice that, when we fix $c_1 = 1$ and $d_1 = d_2 = d_3 = 0$, the non-linear effects disappear, and we have the usual Maxwell ED coupled to the axion field and EM background. In the limit $g \to 0$, the axion uncouples the photon field, and we have a simple combination of a massive free scalar field with the Maxwell ED. Moreover, the Maxwell equations also are recovered in eq. (4.7) by taking the aforementioned considerations and turn-off the background fields, $F_{B\mu\nu} = 0$.

## 4.1 The dispersion relations in the presence of magnetic and electric background fields

In this section, we obtain the dispersion relations (DRs) associated with the axion and photon fields in the presence of a uniform and constant electromagnetic background. Thereby, from now on, all the coefficients defined in eq. (4.5) are not dependent on the space-time coordinates. We start with the equations written in terms of the fields $\mathbf{e}$ and $\mathbf{b}$. For the study of the wave propagation, we just consider the linear terms in $\mathbf{e}$, $\mathbf{b}$ and $\widetilde{\phi}$, as well as the equations with no current and source, $\mathbf{J} = \mathbf{0}$ and $\rho = 0$. Under these conditions, the modified electrodynamics from eq. (4.7) and Bianchi identity is read below :

$$\nabla \cdot \mathbf{D} = 0 \quad , \quad \nabla \times \mathbf{e} + \frac{\partial \mathbf{b}}{\partial t} = \mathbf{0} , \tag{4.9a}$$

$$\nabla \cdot \mathbf{b} = 0 \quad , \quad \nabla \times \mathbf{H} - \frac{\partial \mathbf{D}}{\partial t} = \mathbf{0} , \tag{4.9b}$$

where the vectors $\mathbf{D}$ and $\mathbf{H}$ are given by

$$\mathbf{D} = c_1 \mathbf{e} + d_1 \mathbf{E} (\mathbf{E} \cdot \mathbf{e}) + d_2 \mathbf{B} (\mathbf{B} \cdot \mathbf{e}) - d_1 \mathbf{E} (\mathbf{B} \cdot \mathbf{b}) + d_2 \mathbf{B} (\mathbf{E} \cdot \mathbf{b}) + g\,\widetilde{\phi}\,\mathbf{B}, \tag{4.10a}$$

$$\mathbf{H} = c_1 \mathbf{b} - d_1 \mathbf{B} (\mathbf{B} \cdot \mathbf{b}) - d_2 \mathbf{E} (\mathbf{E} \cdot \mathbf{b}) + d_1 \mathbf{B} (\mathbf{E} \cdot \mathbf{e}) - d_2 \mathbf{E} (\mathbf{B} \cdot \mathbf{e}) - g\,\widetilde{\phi}\,\mathbf{E}. \tag{4.10b}$$

Observe that, in eqs. (4.10a) and (4.10b), we have eliminated the terms with dependence on the coefficient $d_3$, since $d_3 = 0$ in the non-linear ED model in which we will consider ahead. The axion field equation (4.8) leads to

$$\left(\Box + m^2\right) \widetilde{\phi} = g\,(\mathbf{e} \cdot \mathbf{B}) + g\,(\mathbf{b} \cdot \mathbf{E}) . \tag{4.11}$$

We write the Fourier integrals for the fields $\mathbf{e}$, $\mathbf{b}$ and $\widetilde{\phi}$ such that the field equations

(4.9a), (4.9b) and (4.11) in momentum space are given by :

$$\mathbf{k} \cdot \mathbf{D}_0 = 0 \ , \quad \mathbf{k} \times \mathbf{e}_0 - \omega \, \mathbf{b}_0 = \mathbf{0} \ , \tag{4.12a}$$

$$\mathbf{k} \cdot \mathbf{b}_0 = 0 \ , \quad \mathbf{k} \times \mathbf{H}_0 + \omega \, \mathbf{D}_0 = \mathbf{0} \ , \tag{4.12b}$$

$$\left(\mathbf{k}^2 - \omega^2 + m^2\right) \widetilde{\phi}_0 = g\left(\mathbf{B} \cdot \mathbf{e}_0\right) + g\left(\mathbf{E} \cdot \mathbf{b}_0\right) \ , \tag{4.12c}$$

where the amplitudes $D_{0i}$ and $H_{0i}$ are functions of the $\mathbf{k}$-wave vector and the frequency $\omega$. In terms of the electric and magnetic amplitudes $e_{0i}$ and $b_{0i}$, we obtain

$$D_{0i}(\mathbf{k},\omega) = \varepsilon_{ij}(\mathbf{k},\omega)\, e_{0j} + \sigma_{ij}(\mathbf{k},\omega)\, b_{0j} \ , \tag{4.13a}$$

$$H_{0i}(\mathbf{k},\omega) = -\sigma_{ji}(\mathbf{k},\omega)\, e_{0j} + (\mu^{-1})_{ij}(\mathbf{k},\omega)\, b_{0j} \ , \tag{4.13b}$$

in which the electric permittivity symmetric tensor $\varepsilon_{ij}(\mathbf{k},\omega)$ and $\sigma_{ij}(\mathbf{k},\omega)$ are defined by

$$\varepsilon_{ij}(\mathbf{k},\omega) = c_1\, \delta_{ij} + d_1\, E_i\, E_j + d_2\, B_i\, B_j + \frac{g^2\, B_i\, B_j}{\mathbf{k}^2 - \omega^2 + m^2} \ , \tag{4.14a}$$

$$\sigma_{ij}(\mathbf{k},\omega) = -d_1\, E_i\, B_j + d_2\, B_i\, E_j + \frac{g^2\, B_i\, E_j}{\mathbf{k}^2 - \omega^2 + m^2} \ . \tag{4.14b}$$

In addition, $\mu^{-1}$ stands for the inverse of the magnetic permeability symmetric tensor, with the components

$$(\mu^{-1})_{ij} = c_1\, \delta_{ij} - d_1\, B_i\, B_j - d_2\, E_i\, E_j - \frac{g^2\, E_i\, E_j}{\mathbf{k}^2 - \omega^2 + m^2} \ . \tag{4.15}$$

The inverse of eq. (4.15) yields the following expression for the magnetic permeability

$$\mu_{ij}(\mathbf{k},\omega) = \frac{1}{c_1} \frac{(1 - d_B\, \mathbf{B}^2 - d_E\, \mathbf{E}^2)\, \delta_{ij} + d_B\, B_i\, B_j + d_E\, E_i\, E_j + d_B\, d_E\, (\mathbf{E} \times \mathbf{B})_i\, (\mathbf{E} \times \mathbf{B})_j}{1 - d_B\, \mathbf{B}^2 - d_E\, \mathbf{E}^2 + d_B\, d_E\, (\mathbf{E} \times \mathbf{B})^2} \ , \tag{4.16}$$

where we adopted the shorthand notations

$$d_B := \frac{d_1}{c_1} \quad \text{and} \quad d_E := \frac{d_2}{c_1} + \frac{g^2/c_1}{\mathbf{k}^2 + m^2 - \omega^2} \ , \tag{4.17}$$

for simplicity of the equations. In both the cases in which $\mathbf{E} = \mathbf{0}$ or $\mathbf{B} = \mathbf{0}$, we have $\sigma_{ij}(\mathbf{k},\omega) = 0$. Moreover, it is important to note that the dependence of electric permittivity and magnetic permeability on $\mathbf{k}$ and $\omega$ is exclusively due to the presence of the axion coupling, see eqs. (4.14a) and (4.16), as well as the above definition for the coefficient $d_E$.

The eigenvalues of the electric permittivity matrix are given by

$$\lambda_{1\varepsilon} = c_1 \ , \tag{4.18a}$$

$$\lambda_{2\varepsilon} = c_1 \left(1 + \frac{d_B\, \mathbf{E}^2}{2} + \frac{d_E\, \mathbf{B}^2}{2}\right) - c_1 \sqrt{\left(\frac{d_B\, \mathbf{E}^2}{2} - \frac{d_E\, \mathbf{B}^2}{2}\right)^2 + d_B\, d_E\, (\mathbf{E} \cdot \mathbf{B})^2} \tag{4.18b}$$

$$\lambda_{3\varepsilon} = c_1 \left(1 + \frac{d_B\, \mathbf{E}^2}{2} + \frac{d_E\, \mathbf{B}^2}{2}\right) + c_1 \sqrt{\left(\frac{d_B\, \mathbf{E}^2}{2} - \frac{d_E\, \mathbf{B}^2}{2}\right)^2 + d_B\, d_E\, (\mathbf{E} \cdot \mathbf{B})^2} \tag{4.18c}$$

The correspondent eigenvectors are known as the optical axes of the system. If these eigenvalues are positive, it satisfy the conditions

$$c_1 > 0 \quad \text{and} \quad 1 + d_E\,\mathbf{B}^2 + d_B\,\mathbf{E}^2 + d_B\,d_E\,(\mathbf{E}\times\mathbf{B})^2 > 0 \;, \tag{4.19}$$

and the electric permittivity matrix will be defined positive.

The eigenvalues of the magnetic permeability matrix are

$$\lambda_{1\mu} = \frac{1}{c_1} \;, \tag{4.20a}$$

$$\lambda_{2\mu} = \frac{1}{2c_1}\frac{2 - d_B\,\mathbf{B}^2 - d_E\,\mathbf{E}^2 - \sqrt{(d_B\,\mathbf{B}^2 - d_E\,\mathbf{E}^2)^2 + 4\,d_B\,d_E\,(\mathbf{E}\cdot\mathbf{B})^2}}{1 - d_B\,\mathbf{B}^2 - d_E\,\mathbf{E}^2 + d_B\,d_E\,(\mathbf{E}\times\mathbf{B})^2} \;, \tag{4.20b}$$

$$\lambda_{3\mu} = \frac{1}{2c_1}\frac{2 - d_B\,\mathbf{B}^2 - d_E\,\mathbf{E}^2 + \sqrt{(d_B\,\mathbf{B}^2 - d_E\,\mathbf{E}^2)^2 + 4\,d_B\,d_E\,(\mathbf{E}\cdot\mathbf{B})^2}}{1 - d_B\,\mathbf{B}^2 - d_E\,\mathbf{E}^2 + d_B\,d_E\,(\mathbf{E}\times\mathbf{B})^2} \;, \tag{4.20c}$$

where the permeability is positive if we impose the conditions

$$c_1 > 0 \quad \text{and} \quad 1 - d_B\,\mathbf{B}^2 - d_E\,\mathbf{E}^2 + d_B\,d_E\,(\mathbf{E}\times\mathbf{B})^2 > 0 \;. \tag{4.21}$$

However, situations with negative eigenvalues are also acceptable and this would indicate, according to the references [61, 62, 63], that the vacuum manifests the behaviour of a category of metamaterial.

If we just consider the magnetic background field ($\mathbf{E} = \mathbf{0}$), the electric permittivity has two degenerated eigenvalues, $\lambda_{1\varepsilon} = \lambda_{2\varepsilon}$. Analogously, if the background is purely electric ($\mathbf{B} = \mathbf{0}$), the magnetic permeability has the two degenerated eigenvalues, $\lambda_{1\mu} = \lambda_{2\mu}$. In the limit $g \to 0$, the conditions (4.19) and (4.21) are reduced to $c_1^2 + c_1\,d_1\,\mathbf{B}^2 + c_1\,d_2\,\mathbf{E}^2 + d_1\,d_2\,(\mathbf{E}\times\mathbf{B})^2 > 0$ and $c_1^2 - c_1\,d_1\,\mathbf{B}^2 - c_1\,d_2\,\mathbf{E}^2 + d_1\,d_2\,(\mathbf{E}\times\mathbf{B})^2 > 0$, respectively. For the particular case of Maxwell ED coupled to the axion field in which $c_1 = 1$ and $d_1 = d_2 = 0$ in eqs. (4.19) and (4.21), we arrive at the following constraints on the frequency: $-\sqrt{\mathbf{k}^2 + m^2 + g^2\,\mathbf{E}^2} < \omega < \sqrt{\mathbf{k}^2 + m^2 + g^2\,\mathbf{E}^2}$ (for the positive permittivity) and $-\sqrt{\mathbf{k}^2 + m^2 - g^2\,\mathbf{E}^2} < \omega < \sqrt{\mathbf{k}^2 + m^2 - g^2\,\mathbf{E}^2}$ (for the positive permeability).

Using the equations in momentum space (4.12a)−(4.12c) and the constitutive relations (4.13a) and (4.13b), we obtain the wave equation for the components of the electric amplitude:

$$M^{ij}\,\mathbf{e}_0^{\,j} = 0 \;, \tag{4.22}$$

where the matrix elements $M^{ij}$ are given by

$$\begin{aligned} M^{ij} &= a\,\delta^{ij} + b\,k^i\,k^j + c_B\,B^i\,B^j + c_E\,E^i\,E^j + d_B\,(\mathbf{B}\cdot\mathbf{k})\left(B^i\,k^j + B^j\,k^i\right) \\ &\quad + d_E\,(\mathbf{E}\cdot\mathbf{k})\left(E^i\,k^j + E^j\,k^i\right) - d_B\,\omega\left[E^i\,(\mathbf{B}\times\mathbf{k})^j + E^j\,(\mathbf{B}\times\mathbf{k})^i\right] \\ &\quad + d_E\,\omega\left[B^i\,(\mathbf{E}\times\mathbf{k})^j + B^j\,(\mathbf{E}\times\mathbf{k})^i\right]\,, \end{aligned} \quad (4.23)$$

whose the coefficients $a$, $b$, $c_B$ and $c_E$ are defined by

$$a = \omega^2 - \mathbf{k}^2 + d_B\,(\mathbf{k}\times\mathbf{B})^2 + d_E\,(\mathbf{k}\times\mathbf{E})^2\,, \quad (4.24\text{a})$$

$$b = 1 - d_B\,\mathbf{B}^2 - d_E\,\mathbf{E}^2\,, \quad (4.24\text{b})$$

$$c_B = d_E\,\omega^2 - d_B\,\mathbf{k}^2\,,\quad c_E = d_B\,\omega^2 - d_E\,\mathbf{k}^2\,. \quad (4.24\text{c})$$

The non-trivial solution of eq. (4.22) implies that the determinant of the matrix $M^{ij}$ is null. It is difficult to analyze the situation with both $\mathbf{E} \neq \mathbf{0}$ and $\mathbf{B} \neq \mathbf{0}$. The frequency solutions are feasible for the cases with $\mathbf{E} = \mathbf{0}$ or $\mathbf{B} = \mathbf{0}$. In what follows, we investigate these two particular cases separately.

### 4.1.1 The magnetic background case

Let us consider $\mathbf{E} = \mathbf{0}$ in the matrix elements (4.23):

$$M^{ij}\big|_{\mathbf{E}=0} = a_B\,\delta^{ij} + b_B\,k^i\,k^j + c_B\,B^i\,B^j + d_B\,(\mathbf{B}\cdot\mathbf{k})\left(B^i\,k^j + B^j\,k^i\right)\,, \quad (4.25)$$

where $a_B = \omega^2 - \mathbf{k}^2 + d_B\,(\mathbf{k}\times\mathbf{B})^2$ and $b_B = 1 - d_B\,\mathbf{B}^2$. The determinant of (4.25) is

$$\begin{aligned} \det(M)\big|_{\mathbf{E}=0} &= a_B\,\big\{\,a_B^2 + 2\,a_B\,d_B\,(\mathbf{B}\cdot\mathbf{k})^2 + a_B\left(c_B\mathbf{B}^2 + b_B\mathbf{k}^2\right) + \\ &\quad + b_B\,c_B\,(\mathbf{k}\times\mathbf{B})^2 - d_B^2\,(\mathbf{B}\cdot\mathbf{k})^2\,(\mathbf{k}\times\mathbf{B})^2\,\big\}\,, \end{aligned} \quad (4.26)$$

and imposing that $\det(M)\big|_{\mathbf{E}=0} = 0$, we obtain the first solution $a_B = 0$, that yields

$$\omega_{1B}(\mathbf{k}) = |\mathbf{k}|\,\sqrt{1 - \frac{d_1}{c_1}\,(\mathbf{B}\times\hat{\mathbf{k}})^2}\,. \quad (4.27)$$

The other solutions come from the polynomial equation

$$\omega^2\left(P_B\,\omega^4 + Q_B\,\omega^2 + R_B\right) = 0\,, \quad (4.28)$$

where the coefficients are defined by

$$P_B = 1 + \frac{d_2}{c_1}\mathbf{B}^2\,, \quad (4.29\text{a})$$

$$Q_B = -2\,\mathbf{k}^2 - m^2 - \frac{d_2}{c_1}\mathbf{B}^2\left(\mathbf{k}^2 + m^2\right) - \frac{d_2}{c_1}(\mathbf{k}\cdot\mathbf{B})^2 - \frac{g^2}{c_1}\mathbf{B}^2\,, \quad (4.29\text{b})$$

$$R_B = \mathbf{k}^2\left(\mathbf{k}^2 + m^2\right) + \frac{d_2}{c_1}\left(\mathbf{k}^2 + m^2\right)(\mathbf{k}\cdot\mathbf{B})^2 + \frac{g^2}{c_1}(\mathbf{k}\cdot\mathbf{B})^2\,. \quad (4.29\text{c})$$

The second solution is the trivial $\omega = 0$. Solving the above polynomial equation, one can show that the non-trivial solutions correspond to

$$\omega_{2B}^2 = \mathbf{k}^2 + \frac{m^2}{2} + \frac{g^2\mathbf{B}^2 - d_2(\mathbf{B} \times \mathbf{k})^2}{2(c_1 + d_2\mathbf{B}^2)}$$

$$-\sqrt{\left[\mathbf{k}^2 + \frac{m^2}{2} + \frac{g^2\mathbf{B}^2 - d_2(\mathbf{B} \times \mathbf{k})^2}{2(c_1 + d_2\mathbf{B}^2)}\right]^2 - \mathbf{k}^2 \frac{g^2(\mathbf{B}\cdot\hat{\mathbf{k}})^2 + (\mathbf{k}^2 + m^2)(c_1 + d_2(\mathbf{B}\cdot\hat{\mathbf{k}})^2)}{c_1 + d_2\mathbf{B}^2}} \quad (4.30\text{a})$$

$$\omega_{3B}^2 = \mathbf{k}^2 + \frac{m^2}{2} + \frac{g^2\mathbf{B}^2 - d_2(\mathbf{B} \times \mathbf{k})^2}{2(c_1 + d_2\mathbf{B}^2)}$$

$$+\sqrt{\left[\mathbf{k}^2 + \frac{m^2}{2} + \frac{g^2\mathbf{B}^2 - d_2(\mathbf{B} \times \mathbf{k})^2}{2(c_1 + d_2\mathbf{B}^2)}\right]^2 - \mathbf{k}^2 \frac{g^2(\mathbf{B}\cdot\hat{\mathbf{k}})^2 + (\mathbf{k}^2 + m^2)(c_1 + d_2(\mathbf{B}\cdot\hat{\mathbf{k}})^2)}{c_1 + d_2\mathbf{B}^2}} \quad (4.30\text{b})$$

The equations (4.30a) and (4.30b) indicate that the dispersive character of the refractive index is exclusively due to the presence of the axion. The non-linearity alone does not yield dispersion, as the equation above show.

At this stage, it is interesting to analyse the approximation of the very small axion coupling constant. Using $g^2|\mathbf{B}| \ll 1$, the previous frequencies are reduced to

$$\omega_{2B}(\mathbf{k}) \simeq \sqrt{\mathbf{k}^2 + m^2} + \mathcal{O}(g^2) \quad \text{and} \quad \omega_{3B}(\mathbf{k}) \simeq |\mathbf{k}|\sqrt{1 - \frac{d_2(\mathbf{B}\times\hat{\mathbf{k}})^2}{c_1 + d_2\mathbf{B}^2}} + \mathcal{O}(g^2). \quad (4.31)$$

In this approximation, $\omega_{2B}$ leads to the usual DR for a massive particle, while the results for $\omega_{1B}$ and $\omega_{3B}$, eqs. (4.27) and (4.31), confirm the DRs obtained in ref. [64] for a general non-linear ED in a uniform magnetic background. The refractive index associated with the DRs are defined by

$$n_{iB} = \frac{|\mathbf{k}|}{\omega_{iB}} \quad (i = 1, 2, 3). \quad (4.32)$$

We point out that, for the DR in eq. (4.27), the refractive index only depends on the direction of the magnetic field $\mathbf{B}$ with the wave propagating direction $\mathbf{k}$. On the other hand, for the DRs in eqs. (4.30a) and (4.30b), the refractive index depends on the wavelength ($\lambda = 2\pi/|\mathbf{k}|$) due to the presence of the axion mass ($m$) and the coupling constant ($g$). In the limit $m \to 0$ and $g \to 0$, all the refractive indices do not depend on the wavelength in the non-linear EDs.

Since we have three solutions for the frequencies, each solution has a different group velocity. For the frequency in eq. (4.27), we obtain

$$\mathbf{v}_{gB}|_{\omega=\omega_{1B}} = \hat{\mathbf{k}}\sqrt{1 - \frac{d_1}{c_1}(\hat{\mathbf{k}}\times\mathbf{B})^2}. \quad (4.33)$$

The polynomial equation (4.28) has the correspondent group velocity:

$$\mathbf{v}_{gB} = \hat{\mathbf{k}}\frac{d\omega}{dk} = \frac{-\hat{\mathbf{k}}}{2\omega\left(2\omega^2 P_B + Q_B\right)}\left(\frac{dR_B}{dk} + \omega^2\frac{dQ_B}{dk}\right), \quad (4.34)$$

where $\omega$ is now evaluated at the DRs $\omega = \omega_{2B}$ and $\omega = \omega_{3B}$, in which $k \equiv |\mathbf{k}|$. Using the definitions of $P_B$, $Q_B$ and $R_B$, the expression (4.34) is read below

$$\mathbf{v}_{gB} = \frac{\mathbf{k}}{\omega}\left[\frac{2c_1(\omega^2 - \mathbf{k}^2) - c_1 m^2 - d_2(\mathbf{B}\cdot\mathbf{k})^2 + d_2\omega^2\mathbf{B}^2 + d_2(\omega^2 - \mathbf{k}^2 - m^2)(\mathbf{B}\cdot\hat{\mathbf{k}})^2 - g^2(\mathbf{B}\cdot\hat{\mathbf{k}})^2}{2c_1(\omega^2 - \mathbf{k}^2) - c_1 m^2 + d_2\mathbf{B}^2(2\omega^2 - \mathbf{k}^2 - m^2) - d_2(\mathbf{B}\cdot\hat{\mathbf{k}})^2 - g^2\mathbf{B}^2}\right]. \quad (4.35)$$

Substituting the frequencies $\omega_{2B}$ and $\omega_{3B}$ in eq. (4.35), the group velocities in the approximation $g^2\,|\mathbf{B}| \ll 1$ are given by :

$$\mathbf{v}_{gB}\big|_{\omega=\omega_{2B}} \simeq \hat{\mathbf{k}}\sqrt{1 - \frac{d_2\,(\mathbf{B}\times\hat{\mathbf{k}})^2}{c_1 + d_2\,\mathbf{B}^2}} + \mathcal{O}(g^2), \quad (4.36a)$$

$$\mathbf{v}_{gB}\big|_{\omega=\omega_{3B}} \simeq \frac{\mathbf{k}}{\sqrt{\mathbf{k}^2 + m^2}} + \mathcal{O}(g^2). \quad (4.36b)$$

The results (4.33) and (4.36a) show the dependence of the group velocity on the angle between the magnetic background $\mathbf{B}$ and the propagation direction $\hat{\mathbf{k}}$. It is important to highlight that the Maxwell limit recovers the known results for the group velocities (4.33) and (4.35), *i.e.*, $\mathbf{v}_g = c\,\hat{\mathbf{k}}$ (with $c = 1$), when $d_1 = d_2 = 0$ and $c_1 = 1$.

### 4.1.2  The electric background case

The electric background case is obtained with $\mathbf{B} = 0$ in eq. (4.23) :

$$M^{ij}\big|_{\mathbf{B}=0} = a_E\,\delta^{ij} + b_E\,k^i\,k^j + c_E\,E^i\,E^j + d_E\,(\mathbf{E}\cdot\mathbf{k})\left(E^i\,k^j + E^j\,k^i\right), \quad (4.37)$$

where $a_E = \omega^2 - \mathbf{k}^2 + d_E\,(\mathbf{k}\times\mathbf{E})^2$ and $b_E = 1 - d_E\,\mathbf{E}^2$. The correspondent determinant is similar to the result (4.26) :

$$\det(M)\big|_{\mathbf{B}=0} = a_E\left\{a_E^2 + 2\,a_E\,d_E\,(\mathbf{E}\cdot\mathbf{k})^2 + a_E\left(c_E\mathbf{E}^2 + b_E\mathbf{k}^2\right)\right.$$
$$\left. + b_E\,c_E\,(\mathbf{k}\times\mathbf{E})^2 - d_E^2\,(\mathbf{E}\cdot\mathbf{k})^2\,(\mathbf{k}\times\mathbf{E})^2\right\}. \quad (4.38)$$

The null determinant in eq. (4.38) implies the first condition $a_E = 0$, or equivalently $\omega^2 - \mathbf{k}^2 + d_E\,(\mathbf{k}\times\mathbf{E})^2 = 0$, that yields the solutions $\omega_{1E}^{\pm} = \pm\omega_{1E}(\mathbf{k})$ and $\omega_{2E}^{\pm} = \pm\omega_{2E}(\mathbf{k})$, where the DRs are given by

$$\omega_{1E}(\mathbf{k}) = \sqrt{\mathbf{k}^2 + \frac{m^2}{2} - \frac{d_2}{2c_1}(\mathbf{E}\times\mathbf{k})^2 - \sqrt{\left[\frac{m^2}{2} + \frac{d_2}{2c_1}(\mathbf{E}\times\mathbf{k})^2\right]^2 + \frac{g^2}{c_1}(\mathbf{E}\times\mathbf{k})^2}}, \quad (4.39\text{a})$$

$$\omega_{2E}(\mathbf{k}) = \sqrt{\mathbf{k}^2 + \frac{m^2}{2} - \frac{d_2}{2c_1}(\mathbf{E}\times\mathbf{k})^2 + \sqrt{\left[\frac{m^2}{2} + \frac{d_2}{2c_1}(\mathbf{E}\times\mathbf{k})^2\right]^2 + \frac{g^2}{c_1}(\mathbf{E}\times\mathbf{k})^2}}. \quad (4.39\text{b})$$

The second condition for eq. (4.38) to be null leads to the polynomial equation

$$\omega^2\left[\left(1 + \frac{d_1}{c_1}\mathbf{E}^2\right)\omega^2 - \mathbf{k}^2 - \frac{d_1}{c_1}(\mathbf{E}\cdot\mathbf{k})^2\right] = 0. \quad (4.40)$$

The first root in eq. (4.40) is $\omega = 0$, and the non-trivial solutions are $\omega_{3E}^{\pm} = \pm\omega_{3E}(\mathbf{k})$, with

$$\omega_{3E}(\mathbf{k}) = |\mathbf{k}|\sqrt{1 - \frac{d_1(\mathbf{E}\times\hat{\mathbf{k}})^2}{c_1 + d_1\mathbf{E}^2}}. \quad (4.41)$$

Therefore, we obtain three possible DRs for the axionic non-linear ED in an electric background. However, only the frequencies $\omega_{1E}(\mathbf{k})$ and $\omega_{2E}(\mathbf{k})$ contain contributions of the axion field.

The refractive index in an electric background field is

$$n_{iE} = \frac{|\mathbf{k}|}{\omega_{iE}} \quad (i = 1, 2, 3). \quad (4.42)$$

The DR (4.41) yields a refractive index that depends on the direction of $\mathbf{E}$ with the $\mathbf{k}$-wave propagation. In the case of the DRs (4.39a) and (4.39b), the refractive indices depend on the wavelength if we consider $m \neq 0$.

From the condition $a_E = 0$, the correspondent group velocity is given by

$$\mathbf{v}_{gE} = \frac{\mathbf{k}}{\omega}\left[1 + \frac{g^2}{c_1}\frac{(\mathbf{E}\times\mathbf{k})^2}{(\mathbf{k}^2 - \omega^2 + m^2)^2}\right]^{-1} \times$$
$$\times \left[1 - \frac{d_2}{c_1}(\mathbf{E}\times\hat{\mathbf{k}})^2 - \frac{g^2}{c_1}\frac{(\mathbf{E}\times\hat{\mathbf{k}})^2}{\mathbf{k}^2 - \omega^2 + m^2} + \frac{g^2}{c_1}\frac{(\mathbf{E}\times\mathbf{k})^2}{(\mathbf{k}^2 - \omega^2 + m^2)^2}\right], \quad (4.43)$$

where $\omega$ must be evaluated at the dispersion relations $\omega_{1E}$ and $\omega_{2E}$. Substituting the frequencies (4.39a) and (4.39b) in eq. (4.43), we obtain the results

$$\mathbf{v}_{gE}|_{\omega=\omega_{1E}} \simeq \hat{\mathbf{k}}\sqrt{1 - \frac{d_2}{c_1}(\mathbf{E}\times\hat{\mathbf{k}})^2} + \mathcal{O}(g^2), \quad (4.44\text{a})$$

$$\mathbf{v}_{gE}|_{\omega=\omega_{2E}} \simeq \frac{\mathbf{k}}{\sqrt{\mathbf{k}^2 + m^2}}\left[1 - \frac{d_2}{c_1}(\mathbf{E}\times\hat{\mathbf{k}})^2\right] + \mathcal{O}(g^2). \quad (4.44\text{b})$$

The third possible solution for the group velocity comes from the eq. (4.40). In this case, we obtain the group velocity

$$\mathbf{v}_{gE} = \frac{\mathbf{k}}{\omega} \left[ 1 - \frac{d_1 (\mathbf{E} \times \hat{\mathbf{k}})^2}{c_1 + d_1 \mathbf{E}^2} \right] . \qquad (4.45)$$

Using the dispersion relation (4.41) in eq. (4.45), the correspondent group velocity reads

$$\mathbf{v}_{gE}|_{\omega=\omega_{3E}} = \hat{\mathbf{k}} \sqrt{1 - \frac{d_1 (\mathbf{E} \times \hat{\mathbf{k}})^2}{c_1 + d_1 \mathbf{E}^2}} . \qquad (4.46)$$

In the Maxwell limit, $d_1 = d_2 = 0$ and $c_1 = 1$, the group velocities (4.44a) and (4.46) reduce to the usual case when $g \to 0$ : $\mathbf{v}_{gE} = c\,\hat{\mathbf{k}}$ (with $c = 1$). Still in this limit, the group velocity (4.44b) is reduced to the result of a wave-particle of mass $m$. Analogously to the magnetic background case, the results obtained in this section also depends on the angle between the electric background and the wave propagation direction. In all the results, the dispersion relations and the group velocities depend on the coefficients $c_1$, $d_1$ and $d_2$, which are fixed by the non-linear ED as functions of the magnetic or electric background fields.

## 4.2 Application to the axion-Born-Infeld model

In this section, we apply the Born-Infeld (BI) theory as an example of non-linear electrodynamics in the model (4.1). The main aspects of this model are developed in subsection 3.2.3. Therefore, we can discuss the results of the previous section applied to a well-known non-linear ED in the literature.

Keeping this in mind, we shall consider $\sqrt{\beta} = 100$ GeV as the scale of the BI theory in our future analysis. In what follows, let us examine the dispersion relations and group velocities, as well as the properties of the permittivity and permeability tensors of the axion-BI model for the cases of magnetic background field (subsection 4.2.1), and posteriorly, in the presence of an electric background field (subsection 4.2.2).

### 4.2.1 The axion-BI model in a magnetic background

From eq. (4.5), we obtain the correspondent coefficients $c_1$, $d_1$ and $d_2$ for the BI theory,

$$c_1^{BI}\big|_{\mathbf{E}=0,\mathbf{B}} = \frac{\beta}{\sqrt{\beta^2 + \mathbf{B}^2}} \ , \ d_1^{BI}\big|_{\mathbf{E}=0,\mathbf{B}} = \frac{\beta}{(\beta^2 + \mathbf{B}^2)^{3/2}} \ , \ d_2^{BI}\big|_{\mathbf{E}=0,\mathbf{B}} = \frac{1}{\beta\sqrt{\beta^2 + \mathbf{B}^2}} \ , \quad (4.47)$$

with $c_2 = d_3 = 0$. All these coefficients are positive and the axion-BI model reduces to the usual Maxwell theory coupled to the axion in the limit $\beta \to \infty$, *i. e.*, $\lim_{\beta \to \infty} c_1^{BI}\big|_{\mathbf{E}=0,\mathbf{B}} = 1$ and $\lim_{\beta \to \infty} d_1^{BI}\big|_{\mathbf{E}=0,\mathbf{B}} = \lim_{\beta \to \infty} d_2^{BI}\big|_{\mathbf{E}=0,\mathbf{B}} = 0$ . Substituting these coefficients in the results (4.27), (4.30a) and (4.30b), we obtain the dispersion relations in a uniform and constant magnetic field $\mathbf{B}$ :

$$\omega_{1B}^{(BI)}(\mathbf{k}) = |\mathbf{k}|\sqrt{1 - \frac{(\mathbf{B}\times\hat{\mathbf{k}})^2}{\beta^2 + \mathbf{B}^2}} . \tag{4.48a}$$

$$\left[\omega_{2B}^{(BI)}(\mathbf{k})\right]^2 = \mathbf{k}^2 + \frac{m^2}{2} - \frac{(\mathbf{B}\times\mathbf{k})^2}{2(\beta^2 + \mathbf{B}^2)} + \frac{g^2\mathbf{B}^2}{2\sqrt{1 + \mathbf{B}^2/\beta^2}}$$
$$- \left\{ \left[\mathbf{k}^2 + \frac{m^2}{2} - \frac{(\mathbf{B}\times\mathbf{k})^2}{2(\beta^2 + \mathbf{B}^2)} + \frac{g^2\mathbf{B}^2}{2\sqrt{1 + \mathbf{B}^2/\beta^2}}\right]^2 \right.$$
$$\left. -\mathbf{k}^2(\mathbf{k}^2 + m^2)\left[1 - \frac{(\mathbf{B}\times\hat{\mathbf{k}})^2}{\beta^2 + \mathbf{B}^2}\right] - \frac{g^2(\mathbf{B}\cdot\mathbf{k})^2}{\sqrt{1 + \mathbf{B}^2/\beta^2}} \right\}^{1/2}, \tag{4.48b}$$

$$\left[\omega_{3B}^{(BI)}(\mathbf{k})\right]^2 = \mathbf{k}^2 + \frac{m^2}{2} - \frac{(\mathbf{B}\times\mathbf{k})^2}{2(\beta^2 + \mathbf{B}^2)} + \frac{g^2\mathbf{B}^2}{2\sqrt{1 + \mathbf{B}^2/\beta^2}}$$
$$+ \left\{ \left[\mathbf{k}^2 + \frac{m^2}{2} - \frac{(\mathbf{B}\times\mathbf{k})^2}{2(\beta^2 + \mathbf{B}^2)} + \frac{g^2\mathbf{B}^2}{2\sqrt{1 + \mathbf{B}^2/\beta^2}}\right]^2 \right.$$
$$\left. -\mathbf{k}^2(\mathbf{k}^2 + m^2)\left[1 - \frac{(\mathbf{B}\times\hat{\mathbf{k}})^2}{\beta^2 + \mathbf{B}^2}\right] - \frac{g^2(\mathbf{B}\cdot\mathbf{k})^2}{\sqrt{1 + \mathbf{B}^2/\beta^2}} \right\}^{1/2} . \tag{4.48c}$$

At this stage, it is interesting to consider particular limits. For instance, under an intense magnetic background field, *i. e.*, $|\mathbf{B}| \gg \beta$, the dispersion relations are reduced to $\omega_{1B}^{(BI)}(\mathbf{k}) = \omega_{2B}^{(BI)}(\mathbf{k}) \simeq |\mathbf{k}\cdot\hat{\mathbf{B}}|$ and $\omega_{3B}^{(BI)}(\mathbf{k}) \simeq \sqrt{\mathbf{k}^2 + m^2}$, in which we have also considered $g^2\beta \ll 1$ in the frequencies $\omega_{2B}^{(BI)}$ and $\omega_{3B}^{(BI)}$. In the strong magnetic field regime, the DRs $\omega_{1B}^{(BI)}$ and $\omega_{2B}^{(BI)}$ are not dependent on the magnetic field magnitude, and depend only on the angle between the wave vector $\mathbf{k}$ and the direction of the magnetic background field. On the other hand, when $\beta \to \infty$, we recover the photon DR in the frequency $\omega_{1B}^{(BI)}$, while $\omega_{2B}^{(BI)}$ and $\omega_{3B}^{(BI)}$ reduce to the expressions :

$$\lim_{\beta \to \infty} \omega_{2B}^{(BI)}(\mathbf{k}) = \sqrt{\mathbf{k}^2 + \frac{m^2 + g^2\mathbf{B}^2}{2} - \sqrt{\left(\frac{m^2 + g^2\mathbf{B}^2}{2}\right)^2 + g^2(\mathbf{B}\times\mathbf{k})^2}} , \tag{4.49a}$$

$$\lim_{\beta \to \infty} \omega_{3B}^{(BI)}(\mathbf{k}) = \sqrt{\mathbf{k}^2 + \frac{m^2 + g^2\mathbf{B}^2}{2} + \sqrt{\left(\frac{m^2 + g^2\mathbf{B}^2}{2}\right)^2 + g^2(\mathbf{B}\times\mathbf{k})^2}} . \tag{4.49b}$$

The dependence of the DRs with the angle between $\mathbf{B}$ and $\mathbf{k}$ remains in eqs. (4.49a) and

(4.49b). In the regime $g^2|\mathbf{B}| \ll 1$, the previous frequencies lead to

$$\omega_{2B}(\mathbf{k}) \simeq |\mathbf{k}| \left[ 1 - \frac{g^2}{2m^2} \left( \mathbf{B} \times \hat{\mathbf{k}} \right)^2 \right] , \qquad (4.50a)$$

$$\omega_{3B}(\mathbf{k}) \simeq \sqrt{\mathbf{k}^2 + m^2} + \frac{g^2 \mathbf{B}^2}{2\sqrt{\mathbf{k}^2 + m^2}} + \frac{g^2}{2m^2} \frac{(\mathbf{B} \times \mathbf{k})^2}{\sqrt{\mathbf{k}^2 + m^2}} . \qquad (4.50b)$$

Using the results (4.33) and (4.36a), the group velocity for the axion-BI model is

$$\mathbf{v}_{gBI}\Big|_{\omega_{1B}^{(BI)}} = \mathbf{v}_{gBI}\Big|_{\omega_{2B}^{(BI)}} = \hat{\mathbf{k}} \sqrt{1 - \frac{(\mathbf{B} \times \hat{\mathbf{k}})^2}{\beta^2 + \mathbf{B}^2}} . \qquad (4.51)$$

As expected, the $\beta \to \infty$ limit recovers the result of usual electrodynamics group velocity $\mathbf{v}_g = \hat{\mathbf{k}}$, in natural units. Under a strong magnetic field, the group velocities also depend on the angle of $\hat{\mathbf{k}}$ with the $\hat{\mathbf{B}}$-direction :

$$\mathbf{v}_{gBI}\Big|_{\omega_{1B}^{(BI)}} = \mathbf{v}_{gBI}\Big|_{\omega_{2B}^{(BI)}} \simeq \hat{\mathbf{k}} \, |\hat{\mathbf{k}} \cdot \hat{\mathbf{B}}| = \hat{\mathbf{k}} \, |\cos\theta| . \qquad (4.52)$$

Substituting the coefficients (4.47) in the eigenvalues of the electric permittivity matrix, we obtain

$$\lambda_{1\varepsilon}^{(BI)}\Big|_{\mathbf{E}=\mathbf{0}} = \lambda_{2\varepsilon}^{(BI)}\Big|_{\mathbf{E}=\mathbf{0}} = \frac{\beta}{\sqrt{\beta^2 + \mathbf{B}^2}} , \qquad (4.53a)$$

$$\lambda_{3\varepsilon}^{(BI)}\Big|_{\mathbf{E}=\mathbf{0}} = \sqrt{1 + \frac{\mathbf{B}^2}{\beta^2}} + \frac{g^2 \mathbf{B}^2}{\mathbf{k}^2 + m^2 - \omega^2} . \qquad (4.53b)$$

Similarly, the eigenvalues of the magnetic permeability are

$$\lambda_{1\mu}^{(BI)}\Big|_{\mathbf{E}=\mathbf{0}} = \lambda_{2\mu}^{(BI)}\Big|_{\mathbf{E}=\mathbf{0}} = \sqrt{1 + \frac{\mathbf{B}^2}{\beta^2}} , \quad \lambda_{3\mu}^{(BI)}\Big|_{\mathbf{E}=\mathbf{0}} = \left(1 + \frac{\mathbf{B}^2}{\beta^2}\right)^{3/2} . \qquad (4.54)$$

With these expressions, we conclude that the eigenvalues of eqs. (4.53a) and (4.54) are positive, while that in eq. (4.53b) is positive if the $\omega$-frequency satisfies the inequality :

$$-\sqrt{\mathbf{k}^2 + m^2 + \frac{g^2 \mathbf{B}^2}{\sqrt{1 + \mathbf{B}^2/\beta^2}}} < \omega < \sqrt{\mathbf{k}^2 + m^2 + \frac{g^2 \mathbf{B}^2}{\sqrt{1 + \mathbf{B}^2/\beta^2}}} . \qquad (4.55)$$

In the limit $\beta \to \infty$, the results (4.53a) and (4.54) go to one, but in eq. (4.53b), we obtain a contribution of the axion coupling with the Maxwell ED :

$$\lim_{\beta \to \infty} \lambda_{3\varepsilon}^{(BI)}\Big|_{\mathbf{E}=\mathbf{0}} = 1 + \frac{g^2 \mathbf{B}^2}{\mathbf{k}^2 + m^2 - \omega^2} . \qquad (4.56)$$

*4.2.2 The axion-BI model in an electric background*

In this case, the coefficients $c_1$, $d_1$ and $d_2$ at $\mathbf{B} = 0$ are given by

$$c_1^{BI}\big|_{\mathbf{E},\mathbf{B}=0} = \frac{\beta}{\sqrt{\beta^2 - \mathbf{E}^2}} \ , \ d_1^{BI}\big|_{\mathbf{E},\mathbf{B}=0} = \frac{\beta}{(\beta^2 - \mathbf{E}^2)^{3/2}} \ , \ d_2^{BI}\big|_{\mathbf{E},\mathbf{B}=0} = \frac{1}{\beta\sqrt{\beta^2 - \mathbf{E}^2}} \ , \quad (4.57)$$

and $c_2 = d_3 = 0$, in which the magnitude of the electric background must satisfy the condition $\beta > |\mathbf{E}|$ for these coefficients to be real. When $|\mathbf{E}| > \beta$, the coefficients are complex and can bring interesting consequences for the DRs and the wave propagation. Phenomenologically, the constraint of $|\mathbf{E}| > \beta$ is not usual since that, in general, electric fields with strong magnitude of $|\mathbf{E}| > 100$ GeV are not suitable, as it can destabilize the vacuum, generating electron-positron pairs. Using these coefficients in the dispersion relations (4.39a), (4.39b) and (4.41), we arrive at the following frequencies for the axion-BI model in a uniform and constant electric background :

$$\omega_{1E}^{(BI)}(\mathbf{k}) = \sqrt{\mathbf{k}^2 + \frac{m^2}{2} - \frac{(\mathbf{E} \times \mathbf{k})^2}{2\beta^2} - \sqrt{\left[\frac{m^2}{2} + \frac{(\mathbf{E} \times \mathbf{k})^2}{2\beta^2}\right]^2 + g^2(\mathbf{E} \times \mathbf{k})^2 \sqrt{1 - \frac{\mathbf{E}^2}{\beta^2}}}}, \quad (4.58a)$$

$$\omega_{2E}^{(BI)}(\mathbf{k}) = \sqrt{\mathbf{k}^2 + \frac{m^2}{2} - \frac{(\mathbf{E} \times \mathbf{k})^2}{2\beta^2} + \sqrt{\left[\frac{m^2}{2} + \frac{(\mathbf{E} \times \mathbf{k})^2}{2\beta^2}\right]^2 + g^2(\mathbf{E} \times \mathbf{k})^2 \sqrt{1 - \frac{\mathbf{E}^2}{\beta^2}}}}, \quad (4.58b)$$

$$\omega_{3E}^{(BI)}(\mathbf{k}) = |\mathbf{k}|\sqrt{1 - \frac{(\mathbf{E} \times \hat{\mathbf{k}})^2}{\beta^2}} \ , \quad (4.58c)$$

where $\beta > |\mathbf{E} \times \hat{\mathbf{k}}|$ in the frequency (4.58c). Notice that the Maxwell limit yields the photon DR in eq. (4.58c). Moreover, the solutions (4.58a) and (4.58b) are, respectively, reduced to

$$\lim_{\beta \to \infty} \omega_{1E}^{(BI)}(\mathbf{k}) = \sqrt{\mathbf{k}^2 + \frac{m^2}{2} - \sqrt{\frac{m^4}{4} + g^2 (\mathbf{E} \times \mathbf{k})^2}} \ , \quad (4.59a)$$

$$\lim_{\beta \to \infty} \omega_{2E}^{(BI)}(\mathbf{k}) = \sqrt{\mathbf{k}^2 + \frac{m^2}{2} + \sqrt{\frac{m^4}{4} + g^2 (\mathbf{E} \times \mathbf{k})^2}} \ . \quad (4.59b)$$

For a weak electric field, the previous DRs assume the form

$$\omega_{1E}(\mathbf{k}) \simeq |\mathbf{k}| \left[1 - \frac{g^2}{2m^2}\left(\mathbf{E} \times \hat{\mathbf{k}}\right)^2\right] \ , \quad (4.60a)$$

$$\omega_{2E}(\mathbf{k}) \simeq \sqrt{\mathbf{k}^2 + m^2} + \frac{g^2}{2m^2}\frac{(\mathbf{E} \times \mathbf{k})^2}{\sqrt{\mathbf{k}^2 + m^2}} \ . \quad (4.60b)$$

The group velocities associated with these dispersion relations can be read below :

$$\mathbf{v}_g\big|_{\omega_{1E}^{(BI)}} = \mathbf{v}_g\big|_{\omega_{3E}^{(BI)}} = \hat{\mathbf{k}}\sqrt{1 - \frac{(\mathbf{E}\times\hat{\mathbf{k}})^2}{\beta^2}}\,, \tag{4.61a}$$

$$\mathbf{v}_g\big|_{\omega_{2E}^{(BI)}} = \frac{\mathbf{k}}{\sqrt{\mathbf{k}^2+m^2}}\left[1 - \frac{(\mathbf{E}\times\hat{\mathbf{k}})^2}{\beta^2}\right]\,, \tag{4.61b}$$

where the condition $\beta > |\mathbf{E}\times\hat{\mathbf{k}}|$ constraints the velocity (4.61b) in the same direction of the wave propagation.

Analogously to the magnetic background case, we insert the expressions (4.57) in the eigenvalues of the electric permittivity and magnetic permeability matrices. Thereby, we obtain for the electric permittivity :

$$\lambda_{1\varepsilon}^{(BI)}\Big|_{\mathbf{B=0}} = \lambda_{2\varepsilon}^{(BI)}\Big|_{\mathbf{B=0}} = \frac{\beta}{\sqrt{\beta^2-\mathbf{E}^2}} \quad,\quad \lambda_{3\varepsilon}^{(BI)}\Big|_{\mathbf{B=0}} = \frac{\beta^3}{(\beta^2-\mathbf{E}^2)^{3/2}}\,. \tag{4.62}$$

In addition, for the magnetic permeability, we arrive at the following eigenvalues

$$\lambda_{1\mu}^{(BI)}\Big|_{\mathbf{B=0}} = \lambda_{2\mu}^{(BI)}\Big|_{\mathbf{B=0}} = \sqrt{1-\frac{\mathbf{E}^2}{\beta^2}}\,, \tag{4.63a}$$

$$\lambda_{3\mu}^{(BI)}\Big|_{\mathbf{B=0}} = \left[\sqrt{1-\frac{\mathbf{E}^2}{\beta^2}} - \frac{g^2\mathbf{E}^2}{\mathbf{k}^2+m^2-\omega^2}\right]^{-1}\,. \tag{4.63b}$$

Observe that the eigenvalues (4.62) and (4.63a) are real with the condition $\beta > |\mathbf{E}|$. The eigenvalue (4.63b) is positive if the $\omega$-frequency satisfies the condition :

$$-\sqrt{\mathbf{k}^2+m^2-\frac{g^2\mathbf{E}^2}{\sqrt{1-\mathbf{E}^2/\beta^2}}} < \omega < \sqrt{\mathbf{k}^2+m^2-\frac{g^2\mathbf{E}^2}{\sqrt{1-\mathbf{E}^2/\beta^2}}}\,. \tag{4.64}$$

## 4.3  The birefringence in the non-linear ED axion model

The vacuum birefringence is one of the phenomena present in some non-linear EDs. We back to the wave equation (4.22), to investigate the birefringence in a uniform and constant magnetic background, and also for the electric background case in the linearized axion-BI model.

### 4.3.1  Birefringence in the axion-BI model with a magnetic background field

The birefringence analysis requires that we impose some conditions on the wave propagation. We assume the propagation direction $\mathbf{k} = k\,\hat{\mathbf{x}}$ and the magnetic background $\mathbf{B} = B\,\hat{\mathbf{z}}$. In the first situation, we consider the electric wave amplitude parallel to $\mathbf{B}$, with $\mathbf{e}_0 = e_{03}\,\hat{\mathbf{z}}$. In this case, the wave equation (4.22) (with $\mathbf{E} = \mathbf{0}$) yields the relation $\mu_{22}(k,\omega)\,\varepsilon_{33}(k,\omega)\,\omega^2 = k^2$, where the parallel refractive index is defined by

$$n_{\parallel}^{(B)}(k,\omega) = \sqrt{\mu_{22}(k,\omega)\,\varepsilon_{33}(k,\omega)} = \sqrt{1 + \frac{d_2}{c_1}B^2 + \frac{g^2}{c_1}\frac{B^2}{k^2 - \omega^2 + m^2}} \ . \tag{4.65}$$

The second situation is when the electric wave amplitude is perpendicular to the magnetic background field, with $\mathbf{e}_0 = e_{02}\,\hat{\mathbf{y}}$. In this case, the wave equation leads to the relation $\mu_{33}(k,\omega)\,\varepsilon_{22}(k,\omega)\,\omega^2 = k^2$, in which the perpendicular refractive index is

$$n_{\perp}^{(B)}(k,\omega) = \sqrt{\mu_{33}(k,\omega)\,\varepsilon_{22}(k,\omega)} = \left[1 - \frac{d_1}{c_1}B^2\right]^{-1/2} \ . \tag{4.66}$$

Using the coefficients of the axion-BI model in eq. (4.47), the difference between these two refractive indices, $\Delta n_{BI}^{(B)}(k,\omega) = n_{\parallel}^{(B)}(k,\omega) - n_{\perp}^{(B)}(k,\omega)$, is given by

$$\Delta n_{BI}^{(B)}(k,\omega) = \sqrt{1 + \frac{B^2}{\beta^2} + \frac{g^2\,B^2}{k^2 - \omega^2 + m^2}\sqrt{1 + \frac{B^2}{\beta^2}}} - \sqrt{1 + \frac{B^2}{\beta^2}} \ . \tag{4.67}$$

From eq. (4.65), we expect that the variation of the refractive index depends on the $\omega$-frequency. Thereby, we can have the birefringence phenomena associated with the three DRs from (4.48a), (4.48b) and (4.48c), respectively. The limit $g \to 0$ recovers the well-known result in which the pure BI theory does not exhibit birefringence, $i.\ e.$, $\lim_{g \to 0} \Delta n_{BI}^{(B)}(k,\omega) = 0$. Substituting the dispersion relations (4.48a), (4.48b) and (4.48c) in eq. (4.67), we obtain the differences of the refractive indices :

$$\Delta n_{BI}^{(B)}(k)\Big|_{\omega_{1B}^{(BI)}} = \Delta n_{BI}^{(B)}(k)\Big|_{\omega_{2B}^{(BI)}} \simeq \frac{g^2 B}{2}\frac{B\,(B^2 + \beta^2)}{m^2\,(B^2 + \beta^2) + B^2\,k^2} \ , \tag{4.68a}$$

$$\Delta n_{BI}^{(B)}(k)\Big|_{\omega_{3B}^{(BI)}} \simeq \frac{k}{\sqrt{k^2 + m^2}} - \sqrt{1 + \frac{B^2}{\beta^2}} \ , \tag{4.68b}$$

where we have used the approximation $g^2\,B \ll 1$. Notice that in eq. (4.68a), a very small (residual) birefringence remains in the model with $g^2$-dependence. The result (4.68b) recovers the usual Maxwell ED when $\beta \to \infty$, and if we consider a very small mass for the axion-particle $m \simeq 0$.

In the case of $\beta \gg B$, which is a very reasonable approximation, since in the worst case scenario, the smallest value of the beta parameter is estimated to be in the order of $\beta \approx 10^6 \, \text{MeV}^2$ [26]. If we take Schwinger's critical electric field as a comparison, we would have something around $|B_S| \approx 3 \times 10^{-3} \, \text{MeV}^2$. Thus, the equation (4.68a) becomes:

$$\frac{|\Delta n_{BI}^{(B)}|}{B^2} \simeq \frac{g^2}{2m^2} \,. \tag{4.69}$$

This equation provides a connection between the VMB phenomenon and the ratio of the ALPs parameter space, namely, ALP coupling constant by mass. Note that there is no dependence on the Born-Infeld parameter, only on the magnitude of the magnetic field.

The PVLAS-FE experiment presented the following result for the vacuum magnetic birefringence in ALPs [40]:

$$\frac{\Delta n^{\text{PVLAS-FE}}}{B^2} = (+19 \pm 27) \times 10^{-24} \, \text{T}^{-2} \,. \tag{4.70}$$

In this situation, we have that $B = 2.5$ T and the wavelength of $\lambda = 1064$ nm (or equivalently, $k = 0.185$ eV). Let us consider the axion mass at $m = 1$ meV and BI parameter of $\sqrt{\beta} = 100$ GeV. Using the results in eqs. (4.68a) and (4.70), we estimate the axion coupling constant as

$$g \simeq 9.065 \times 10^{-9} \, \text{GeV}^{-1} \,, \tag{4.71}$$

which is consistent with the upper bound $g < 6.4 \times 10^{-8} \, \text{GeV}^{-1}$ (95% C.L.) reported in this experiment.

### 4.3.2 Birefringence in the axion-BI model with an electric background field

The case with an electric background is similar to the previous section. We consider the same wave propagation direction, with the electric background $\mathbf{E} = E\hat{\mathbf{z}}$. In the first situation in which the electric wave amplitude $\mathbf{e}_0$ is parallel to $\mathbf{E}$, the correspondent refractive index is

$$n_\parallel^{(E)} = \sqrt{1 + \frac{d_1}{c_1} E^2} \,. \tag{4.72}$$

On the other hand, when $\mathbf{e}_0$ is perpendicular to $\mathbf{E}$, the refractive index leads to

$$n_\perp^{(E)}(k,\omega) = \left[1 - \frac{d_2}{c_1} E^2 - \frac{g^2}{c_1} \frac{E^2}{k^2 - \omega^2 + m^2}\right]^{-1/2} \,. \tag{4.73}$$

Thereby, the difference between the parallel and perpendicular refractive indices in the axion-BI model is

$$\Delta n_{BI}^{(E)}(k,\omega) = \left(1 - \frac{E^2}{\beta^2}\right)^{-1/2} - \left[1 - \frac{E^2}{\beta^2} - \frac{g^2 E^2}{k^2 - \omega^2 + m^2}\sqrt{1 - \frac{E^2}{\beta^2}}\right]^{-1/2} \,. \tag{4.74}$$

The limit of the pure BI electrodynamics ($g \to 0$) recovers the result of vanishing birefringence in eq. (4.74). Furthermore, note that the difference (4.74) depends on the $\omega$-function, that is function of the electric field and $\beta$-parameter. Substituting the dispersion relations (4.58a), (4.58b) and (4.58c) in eq. (4.74), we obtain

$$\Delta n_{BI}^{(E)}(k)\Big|_{\omega_{1E}^{(BI)}} = \Delta n_{BI}^{(E)}(k)\Big|_{\omega_{3E}^{(BI)}} \simeq -\frac{g^2\,E}{2}\frac{E\,\beta^2}{(1-E^2/\beta^2)\,(m^2\beta^2+E^2k^2)}\,, \quad (4.75a)$$

$$\Delta n_{BI}^{(E)}(k)\Big|_{\omega_{2E}^{(BI)}} \simeq \left(1-\frac{E^2}{\beta^2}\right)^{-1/2} - \frac{k}{\sqrt{k^2+m^2}}\,, \quad (4.75b)$$

in which the approximation for a weak electric field, $g^2 E \ll 1$, is applied. The result (4.75a) shows a very small effect of the birefringence with the $g^2$ dependence. In eq. (4.75b), the birefringence is null when $\beta \to \infty$ and the axion mass is approximately zero. In the approximation $\beta \gg Ek/m$, the result (4.75a) yields the electric birefringence

$$\frac{|\Delta n_{BI}^{(E)}|}{E^2} \simeq \frac{g^2}{2m^2}\,. \quad (4.76)$$

Therefore, this result can be used to constrain the parameter space $(m,g)$ through the optical Kerr effect.

# Chapter 5

# Non-linear axion-photon mixing coupled to CFJ Lorentz symmetry violation

Before going on and starting to work out the developments of this chapter, we would like to call into question our motivation to bring together three different physical scenarios beyond the Standard Model in a single Lagrangian, namely: axions, non-linear electrodynamic extensions and Lorentz-symmetry violating physics (LSV is here realized by means of the Carroll-Field-Jackiw term). The usual procedure is to consider each of these physical situations separately, once we expect that their respective individual effects correspond to tiny corrections to current physics. Connecting these three diverse physics in a single action might appear as a waste of efforts or, simply, an exercise to mix up different effects. Nevertheless, what we truly wish by coupling axions to non-linear electrodynamics and LSV physics is to show how the parameters associated to the axion and LSV sectors couple to external electric and magnetic fields whenever non-linearity is considered. Actually, the main effort we endeavor is to inspect how the magnetic background field may broaden the effects of the tiny axionic and LSV parameters on physical properties such as birefringence, refractive indices, dichroism and group velocity. This is investigated with the help of the dispersion relations we shall derive in different situations characterized by particular configurations of external fields. And to enforce our claim to consider the simultaneous presence of these three sorts of effects, we gather some works in Refs. [99, 100, 101, 102]. In these papers, non-linear quantum electrodynamics, axion electrodynamics and LSV are studied in Condensed Matter scenarios such as Dirac and Weyl semimetals and topological magnetic materials. We are then motivated to assume that topological materials appear as a natural laboratory that justify the inspection of how the effects of non-linearity, axi-ons and LSV interfere with each other. Cosmology provides another viable scenario that may justifies efforts in the quest for the interference between the three effects we are here discussing. In Refs. [104, 105, 103], we cast reference works that support our proposal. Finally, knowing that non-linearity, axions and LSV are issues currently investigated in connection with astrophysical structures [106, 107, 108], we can also elect Astrophysics as another field of interest to study the concomitant presence of these three issues and how they affect each other in the study of the propagation of electromagnetic waves in the QED vacuum.

In this chapter, we investigate the propagation effects of a general axionic non-linear ED in presence of a CFJ term. As mentioned, the CFJ introduces the 4-vector that breaks the Lorentz symmetry, and the isotropy of the space-time. We introduce a uniform magnetic field expanding the propagating field of the model up to second order around this background field. The properties of the medium are discussed in presence of the magnetic background. We obtain the dispersion relations of the linearized theory in terms of the magnetic background, the CFJ 4-vector, and the axion coupling constant. The case of a space-like quadrivector is analysed, such that the plane wave frequencies are functions of the wave vector (**k**), the magnetic background (**B**), and the CFJ background vector (**v**). Thereby, we consider two cases : (a) when **k**, **B** and **v** are perpendiculars, and (b) when **k** is parallel to **B**, but both vectors remain perpendicular to **v**. The solutions of these cases define the perpendicular and parallel frequencies, respectively. Using these dispersion relations, we calculate the birefringence through the perpendicular and parallel refractive indices. We apply our results to the non-linear electrodynamics of Euler-Heisenberg [38], Born-Infeld [26], and Modified Maxwell (ModMax) [67, 23, 68].

## 5.1 The non-linear axion-photon electrodynamics including the Carroll-Field-Jackiw term

We initiate with the description of the model whose Lagrangian density reads as follows :

$$\mathcal{L} = \mathcal{L}_{nl}(\mathcal{F}_0, \mathcal{G}_0) + \frac{1}{2}\,(\partial_\mu \phi)^2 - \frac{1}{2}\,m^2\,\phi^2 + g\,\phi\,\mathcal{G}_0 + \frac{1}{4}\,\epsilon^{\mu\nu\kappa\lambda}\,v_\mu\,A_{0\nu}\,F_{0\kappa\lambda} - J_\mu\,A_0^{\ \mu}\,, \quad (5.1)$$

where the CFJ term introduces the background 4-vector $v^\mu = (v^0, \mathbf{v})$ whose components do not depend on the space-time coordinates. It has mass dimension in natural units and is responsible for the Lorentz symmetry violation in the gauge sector of the model. In addition, $\phi$ is the axion scalar field with mass $m$, and $g$ is the non-minimal coupling constant (with length dimension) of the axion with the electromagnetic field, *i.e.*, the usual coupling with the $\mathcal{G}_0$-invariant in the axion-photon model.

Similar to the approach in previous chapters, we expand the Abelian gauge field, $A_0{}^\mu = a^\mu + A_B{}^\mu$, around the background up to second order in the propagating field $a^\mu$ to yield the expression

$$\begin{aligned}\mathcal{L}^{(2)} &= -\frac{1}{4} c_1 f_{\mu\nu}^2 - \frac{1}{4} c_2 f_{\mu\nu} \widetilde{f}^{\mu\nu} + \frac{1}{8} Q_{B\mu\nu\kappa\lambda} f^{\mu\nu} f^{\kappa\lambda} \\ &+ \frac{1}{2} \left(\partial_\mu \widetilde{\phi}\right)^2 - \frac{1}{2} m^2 \widetilde{\phi}^2 - \frac{1}{2} g \widetilde{\phi} \widetilde{F}_{B\mu\nu} f^{\mu\nu} + \frac{1}{4} \epsilon^{\mu\nu\kappa\lambda} v_\mu a_\nu f_{\kappa\lambda} - \bar{J}_\nu a^\nu\,,\end{aligned} \quad (5.2)$$

where $\bar{J}_\nu = J_\nu - \partial^\mu (H_{B\mu\nu}) - v^\mu \widetilde{F}_{B\mu\nu}$ represents an effective external current that couples to the photon field; it includes an eventual matter current and the contributions that stem from the background electromagnetic fields. The tensors associated with this electromagnetic background are defined in what follows:

$$\begin{aligned}H_{B\mu\nu} &= c_1 F_{B\mu\nu} + c_2 \widetilde{F}_{B\mu\nu} + \frac{g^2}{m^2} \mathcal{G}_B \widetilde{F}_{B\mu\nu}\,, & (5.3\mathrm{a}) \\ Q_{B\mu\nu\kappa\lambda} &= d_1 F_{B\mu\nu} F_{B\kappa\lambda} + d_2 \widetilde{F}_{B\mu\nu} \widetilde{F}_{B\kappa\lambda} + d_3 F_{B\mu\nu} \widetilde{F}_{B\kappa\lambda} + d_3 \widetilde{F}_{B\mu\nu} F_{B\kappa\lambda}\,. & (5.3\mathrm{b})\end{aligned}$$

The axion field is shifted as $\phi \to \widetilde{\phi} + \phi_0$ in order to eliminate the $g\,\phi\,\mathcal{G}_B$ term that would appear in the Lagrangian (5.2). The coefficients $c_1$, $c_2$, $d_1$, $d_2$ and $d_3$ are evaluated at $\mathbf{E}$ and $\mathbf{B}$, as follows :

$$c_1 = \left.\frac{\partial \mathcal{L}_{nl}}{\partial \mathcal{F}_0}\right|_{\mathbf{E},\mathbf{B}},\ c_2 = \left.\frac{\partial \mathcal{L}_{nl}}{\partial \mathcal{G}_0}\right|_{\mathbf{E},\mathbf{B}},\ d_1 = \left.\frac{\partial^2 \mathcal{L}_{nl}}{\partial \mathcal{F}_0^2}\right|_{\mathbf{E},\mathbf{B}},\ d_2 = \left.\frac{\partial^2 \mathcal{L}_{nl}}{\partial \mathcal{G}_0^2}\right|_{\mathbf{E},\mathbf{B}},\ d_3 = \left.\frac{\partial^2 \mathcal{L}_{nl}}{\partial \mathcal{F}_0 \partial \mathcal{G}_0}\right|_{\mathbf{E},\mathbf{B}} \quad (5.4)$$

that depend on the EM field magnitude and may also be functions of the space-time coordinates. Following the previous definitions, the background tensors satisfy the properties $H_{B\mu\nu} = -H_{B\nu\mu}$, whereas $Q_{B\mu\nu\kappa\lambda}$ is symmetric under exchange $\mu\nu \leftrightarrow \kappa\lambda$, and antisymmetric under $\mu \leftrightarrow \nu$ and $\kappa \leftrightarrow \lambda$. Note that the current $J^\mu$ couples to the external potential $A_B{}^\mu$, but this term and $\mathcal{L}_{nl}(\mathcal{F}_B, \mathcal{G}_B)$ are irrelevant for the field equations in which we are interested.

Using the minimal action principle by varying $a^\mu$, the Lagrangian (5.2) yields the EM field equations with source $\bar{J}^\mu$

$$\partial^\mu \left[ c_1 f_{\mu\nu} + c_2 \widetilde{f}_{\mu\nu} - \frac{1}{2} Q_{B\mu\nu\kappa\lambda} f^{\kappa\lambda}\right] + v^\mu \widetilde{f}_{\mu\nu} = -g\,(\partial^\mu \widetilde{\phi})\,\widetilde{F}_{B\mu\nu} + \bar{J}_\nu\,, \quad (5.5)$$

and the Bianchi identity remains the same one for the photon field, namely, $\partial_\mu \widetilde{f}^{\mu\nu} = 0$. The action principle in relation to $\widetilde{\phi}$ in (5.2) yields the axion field equation evaluated at the EM background :

$$\left(\Box + m^2\right) \widetilde{\phi} = -\frac{1}{2} g \, \widetilde{F}_{B\mu\nu} \, f^{\mu\nu} \,. \tag{5.6}$$

Since we consider a uniform magnetic background field, the $c_2$- and $d_3$-coefficients of the expansion vanish for most examples of non-linear EDs known in the literature, such as Euler-Heisenberg, Born-Infeld, ModMax, Logarithmic and some others, where the corresponding non-linear Lagrangian densities depend on the square of the $\mathcal{G}$-invariant. These considerations simplify the results that we shall work out in the next Sections ahead. The usual axionic ED coupled to the CFJ-term is recovered whenever $d_1 \to 0$, $d_2 \to 0$ and $c_1 \to 1$ in all the cases of non-linear ED mentioned previously.

## 5.2 The dispersion relations in presence of a uniform magnetic field

In this Section, we obtain the dispersion relations of the axion and photon fields in a uniform magnetic background. Thus, we can take $\mathbf{E} = \mathbf{0}$ in the equations of the section 5.1. Thereby, from now on, all the coefficients defined in (5.4) are not space-time dependent; they actually depend only on the magnetic vector $\mathbf{B}$. We start the description of the field propagating with the equations written in terms of $\mathbf{e}$ and $\mathbf{b}$, in the presence of constant and uniform magnetic background field. For the analysis of the free wave propagation, we just consider the linear terms in $\mathbf{e}$, $\mathbf{b}$ and $\widetilde{\phi}$, as well as, the equations with no source, $\mathbf{\bar{J}} = \mathbf{0}$ and $\bar{\rho} = 0$. Under these conditions, the electrodynamics equations in terms of the propagating vector field are:

$$\nabla \cdot \mathbf{D} = \mathbf{v} \cdot \mathbf{b}, \tag{5.7a}$$
$$\nabla \times \mathbf{e} + \frac{\partial \mathbf{b}}{\partial t} = \mathbf{0}, \tag{5.7b}$$
$$\nabla \cdot \mathbf{b} = 0, \tag{5.7c}$$
$$\nabla \times \mathbf{H} + \mathbf{v} \times \mathbf{e} = v^0 \mathbf{b} + \frac{\partial \mathbf{D}}{\partial t}, \tag{5.7d}$$

where the vectors $\mathbf{D}$ and $\mathbf{H}$ are, respectively, given by

$$\mathbf{D} = c_1 \mathbf{e} + d_2 \mathbf{B} (\mathbf{B} \cdot \mathbf{e}) + g \, \widetilde{\phi} \, \mathbf{B}, \tag{5.8a}$$
$$\mathbf{H} = c_1 \mathbf{b} - d_1 \mathbf{B} (\mathbf{B} \cdot \mathbf{b}). \tag{5.8b}$$

The scalar field equation (5.6) in terms of the magnetic background field leads to

$$\left(\Box + m^2\right) \widetilde{\phi} = g \left(\mathbf{e} \cdot \mathbf{B}\right) . \tag{5.9}$$

We substitute the plane wave solutions of $\mathbf{e}$, $\mathbf{b}$ and $\widetilde{\phi}$ in the field equations (5.7)(a-d) and (5.9). Eliminating conveniently the amplitudes of $\mathbf{b}$ and $\widetilde{\phi}$ in terms of the electric field amplitude, the wave equation in the momentum space is read below :

$$M^{ij}(\omega, \mathbf{k}) \, e_0^{\,j} = 0 , \tag{5.10}$$

where $e_0^{\,j}$ ($j = 1, 2, 3$) are the components of the electric amplitude, and the matrix elements $M^{ij}$ are given by

$$M^{ij}(\omega, \mathbf{k}) = a \, \delta^{ij} + b \, k^i k^j + c \, B^i B^j + d \left(\mathbf{B} \cdot \mathbf{k}\right) \left(B^i k^j + B^j k^i\right) - i \, \epsilon^{ijm} \left(v^0 \, k^m - \omega \, v^m\right) , \tag{5.11}$$

whose the coefficients $a$, $b$, $c$ are defined by

$$a = \omega^2 - \mathbf{k}^2 + d \left(\mathbf{k} \times \mathbf{B}\right)^2 , \tag{5.12a}$$

$$b = 1 - d \, \mathbf{B}^2 , \tag{5.12b}$$

$$c = \xi(\omega, \mathbf{k}) \, \omega^2 - d \, \mathbf{k}^2 , \tag{5.12c}$$

$$\xi(\omega, \mathbf{k}) = f + \frac{g_a^2}{\mathbf{k}^2 - \omega^2 + m^2} , \tag{5.12d}$$

in which $d := d_1/c_1$, $f := d_2/c_1$ and $g_a := \sqrt{g^2/c_1}$ for simplicity in the equations. Thus, the non-linearity evaluated on the magnetic background is manifested in the parameters $d$ and $f$, and the coupling constant $g_a$ corrects the axion coupling constant with the coefficient $c_1$. Notice that the $b$-coefficient depends only on the magnetic background, but the other ones depend on the $\omega$-frequency and on the $\mathbf{k}$-wave vector.

Back to the expressions of $\mathbf{D}$ and $\mathbf{H}$ in (5.8a)-(5.8b) with the plane wave solutions, the components of $\mathbf{D}$ and $\mathbf{H}$ in terms of the electric and magnetic amplitudes can be written as

$$D_i = \epsilon_{ij}(\mathbf{k}, \omega) \, e_j \quad \text{and} \quad H_i = (\mu_{ij})^{-1} \, b_j , \tag{5.13}$$

where $\epsilon_{ij}$ and $(\mu_{ij})^{-1}$ are the permittivity and permeability (inverse) tensors, respectively,

$$\epsilon_{ij}(\mathbf{k}, \omega) = c_1 \, \delta_{ij} + c_1 \, \xi(\mathbf{k}, \omega) \, B_i B_j , \tag{5.14a}$$

$$(\mu_{ij})^{-1} = c_1 \, \delta_{ij} - d_1 \, B_i B_j . \tag{5.14b}$$

The permeability tensor is obtained by computing the inverse of (5.14b)

$$\mu_{ij} = \frac{1}{c_1} \frac{(1 - d\,\mathbf{B}^2)\,\delta_{ij} + d\,B_i\,B_j}{1 - d\,\mathbf{B}^2} \ . \tag{5.15}$$

Notice that the electric permittivity depends on the $\omega$-frequency and the $\mathbf{k}$-wave vector due to the axion coupling $g \neq 0$. Also, the definition of these tensors do not include the components of the CFJ 4-vector $v^\mu$. Thereby, this LSV scenario does not contribute with the physical properties of the tensors.

According to the works of refs. [109, 110], the $v^0$-component may induce contributions that violate the causality and stability. For this reason, we shall adopt a space-like CFJ 4-vector, *i.e.*, $v^0 = 0$ in the matrix element $M^{ij}$ from eq. (5.11). The dispersion relations come from the non-trivial solutions to the wave equation (5.11). The condition for non-trivial solutions is $\det M^{ij} = 0$; for the space-like case of the CFJ background, it is reduced to an $\omega$-polynomial equation:

$$a^3 + a^2 \left[ b\,\mathbf{k}^2 + 2d\,(\mathbf{B} \cdot \mathbf{k})^2 \right] + ac \left[ a\,\mathbf{B}^2 + b(\mathbf{B} \times \mathbf{k})^2 \right] - a\,d^2\,(\mathbf{B} \cdot \mathbf{k})^2\,(\mathbf{B} \times \mathbf{k})^2$$
$$- \left[ a\mathbf{v}^2 + c(\mathbf{B} \cdot \mathbf{v})^2 \right] \omega^2 - (\mathbf{v} \cdot \mathbf{k}) \left[ b(\mathbf{v} \cdot \mathbf{k}) + 2d\,(\mathbf{B} \cdot \mathbf{k})(\mathbf{B} \cdot \mathbf{v}) \right] \omega^2 = 0 \ . \tag{5.16}$$

Back to eqs. (5.12a)-(5.12d), notice that the coefficient $a$ takes into account non-linearity by means of the piece $d\,(\mathbf{B} \times \mathbf{k})^2 = d\,[\mathbf{B}^2\,\mathbf{k}^2 - (\mathbf{B} \cdot \mathbf{k})^2]$, $b$ incorporates non-linearity in the piece $d\,\mathbf{B}^2$. The coefficient $c$, on the other hand, splits into non-linearity (the $d$- and $f$-terms) and axionic (the axion mass, $m$, and the coupling constant, $g_a$) effects: $c = (f\omega^2 - d\,\mathbf{k}^2) - g_a^2\,\omega^2/(\omega^2 - \mathbf{k}^2 - m^2)$. Finally, the LSV appears represented by the background vector, $\mathbf{v}$. It is worthy to remark that no term in the dispersion relation (5.16) couples the three effects together. We actually mean that non-linearity, axionic and LSV effects interfere with one another only in pairs. No term is present in which parameters of the three different effects appear grouped together in a product. However, below, in discussing the effective masses of axion-photon coupled system, it will become clear that three effects mix up to give these effective masses.

We would like to point out that, if only non-linearity is considered, the dispersion relations do not exhibit momentum dependence, so no dispersive effect shows up. By introducing the axion sector, the profile changes and dispersive effects emerges. Now, if LSV is also considered in addition to non-linearity and axionic physics, not only dispersion is enforced, but another fact should be highlighted: photonic dispersion relations show that the modulus of the LSV background vector, $|\mathbf{v}| = v$, along with the axion mass endows the photon with an effective mass, $m_{eff}$. For instance, let us consider the rest frame ($k=0$ with the identification $\omega^2 = m_{eff}^2$), as well as $\mathbf{B} = B\,\hat{\mathbf{z}}$ and $\mathbf{v} = v\,\hat{\mathbf{y}}$ in eq. (5.16). Therefore, it is possible to show that the effective masses for the photon and the axion correspond to the roots of the equation

$$(1 + f\,B^2)\,m_{eff}^4 - \left[(1 + f\,B^2)\,m^2 + g_a^2\,B^2 + \frac{v^2}{c_1^2}\right] m_{eff}^2 + \frac{m^2\,v^2}{c_1^2} = 0\,. \qquad (5.17)$$

The solutions of this equation are

$$\begin{aligned} m_{eff(1)}^2 &= \frac{(1 + f\,B^2)\,m^2 + g_a^2\,B^2 + v^2/c_1^2}{2\,(1 + f\,B^2)} \\ &\quad - \frac{\sqrt{((1 + f\,B^2)m^2 + g_a^2\,B^2 + v^2/c_1^2)^2 - 4m^2 v^2\,(1 + f\,B^2)/c_1^2}}{2\,(1 + f\,B^2)}\,, \qquad (5.18a) \\ m_{eff(2)}^2 &= \frac{(1 + f\,B^2)\,m^2 + g_a^2\,B^2 + v^2/c_1^2}{2\,(1 + f\,B^2)} \\ &\quad + \frac{\sqrt{((1 + f\,B^2)m^2 + g_a^2\,B^2 + v^2/c_1^2)^2 - 4m^2 v^2\,(1 + f\,B^2)/c_1^2}}{2\,(1 + f\,B^2)}\,. \qquad (5.18b) \end{aligned}$$

In the limit $g_a \to 0$, the first solution (5.18a) is reduced to

$$m_{eff(1)} = \frac{v}{\sqrt{c_1(c_1 + d_2\,B^2)}}\,, \qquad (5.19)$$

whereas, when $v \to 0$, the effective mass is null. Consequently, in the uncoupled limit ($g_a \to 0$), the CFJ parameter ($v$) gives an effective mass with the correction of the non-linearity evaluated at the magnetic background field. Considering the same limit of $v \to 0$, the second solution (5.18b) is reduced to the expression

$$m_{eff(2)} = \sqrt{m^2 + \frac{g^2\,B^2}{c_1(c_1 + d_2\,B^2)}}\,, \qquad (5.20)$$

where the axion mass, and the coupling constant have a fundamental role for the effective mass. Thus, the axion mass is corrected by the coupling constant and by the magnetic field. In the uncoupled limit, $g_a \to 0$, the second root (5.18b) yields the axion mass, i.e., $m_{eff(2)} = m$.

Another analysis that supports the aforementioned effective masses ($m_{eff}$) and the scenario of the non-linear ED unified to the LSV and axions consists in finding the group velocity solutions. For simplicity, we consider the wave vector as $\mathbf{k} = k\,\hat{\mathbf{z}}$, and the same situation in which $\mathbf{B} = B\,\hat{\mathbf{z}}$ and $\mathbf{v} = v\,\hat{\mathbf{y}}$. Bearing this in mind, the eq. (5.16) leads to

$$(\omega^2 - k^2)\left[(1 + f\,B^2)(\omega^2 - k^2 - m^2) - g_a^2\,B^2 - \frac{v^2}{c_1^2}\right] + \frac{m^2\,v^2}{c_1^2} = 0\,. \tag{5.21}$$

Differentiating eq. (5.21) with respect to the frequency and wave vector components, the group velocity is given by the relation

$$v_g = \frac{d\omega}{dk} = \frac{k}{\omega}\,, \tag{5.22}$$

where $\omega$ corresponds to a possible root of the eq. (5.21). From the two solutions for the group velocity, the uncoupled limit $g_a \to 0$ yields the following results

$$v_{g(1)}(k) = \frac{k}{\sqrt{k^2 + m_{eff(1)}^2}} \quad , \quad v_{g(2)}(k) = \frac{k}{\sqrt{k^2 + m^2}}\,, \tag{5.23}$$

with $m_{eff(1)}$ being described by the eq. (5.19) and $m \neq 0$. On the other hand, when the LSV parameter is absent ($v \to 0$), we arrive at the solutions:

$$v_{g(1)}(k) = 1 \quad , \quad v_{g(2)}(k) = \frac{k}{\sqrt{k^2 + m_{eff(2)}^2}}\,. \tag{5.24}$$

where $m_{eff(2)}$ corresponds to the eq. (5.20).

Therefore, from these group velocities, one can also recognize that $m_{eff(1)}$ and $m_{eff(2)}$ denote the photon and axion effective masses, respectively.

Now let us return to eq. (5.16). The general solution is quite involved in view of the coefficients (5.12a)-(5.12d). For simplicity, we consider the two cases below :

(a) The case of the vectors $\mathbf{B}$, $\mathbf{k}$ and $\mathbf{v}$ perpendiculars among themselves : $\mathbf{B}\cdot\mathbf{k} = \mathbf{B}\cdot\mathbf{v} = \mathbf{k}\cdot\mathbf{v} = 0$. Considering this condition, the equation (5.16) is reduced to :

$$\omega_\perp^2\left[\omega_\perp^2 - k^2 + d\,B^2\,k^2\right]\left[(1 + \xi\,B^2)\,\omega_\perp^2 - k^2 - v^2\right] = 0\,, \tag{5.25}$$

where we denote the perpendicular frequency $\omega_\perp$, and $B$, $k$ and $v$ are the magnitudes of the previous vectors. The first solution is $\omega_\perp = 0$, and the non-trivial solutions from (5.25) are given by

$$\omega_{1\perp}(k) = k\,\sqrt{1 - d\,B^2}\,, \tag{5.26a}$$

$$\omega_{2\perp}(k) = \sqrt{\frac{k^2 + v^2 + g_a^2\,B^2 + (1 + f\,B^2)(k^2 + m^2) - \Gamma}{2\,(1 + f\,B^2)}}\,, \tag{5.26b}$$

$$\omega_{3\perp}(k) = \sqrt{\frac{k^2 + v^2 + g_a^2\,B^2 + (1 + f\,B^2)(k^2 + m^2) + \Gamma}{2\,(1 + f\,B^2)}}\,, \tag{5.26c}$$

where

$$\Gamma = \sqrt{[k^2 + v^2 + g_a^2 B^2 + (1 + f B^2)(k^2 + m^2)]^2 - 4(1 + f B^2)(k^2 + m^2)(k^2 + v^2)}. \quad (5.27)$$

The analysis of the limits to establish comparisons with the results in the literature is immediate. The limits $f \to 0$ and $c_1 = 1$ yield the dispersion relations of the axionic ED coupled to CFJ term in the presence of an external magnetic field. Furthermore, if we also take $g_a \to 0$, the dispersion relations are reduce to $\omega_{2\perp}(k) = \sqrt{(k^2 + v^2)(1 + f B^2)^{-1}}$ and $\omega_{3\perp}(k) = \sqrt{k^2 + m^2}$ for $m > v$. Note that in $\omega_{2\perp}(k)$, occurs the characteristic effect of CFJ, where the Lorentz-breaking parameter gives a small mass for the photon. These results confirm the roots of the eq. (5.17) in the rest frame ($k = 0$). The usual Maxwell limit reduces all the frequencies to : $\omega_{1\perp}(k) = \omega_{2\perp}(k) = k$ and $\omega_{3\perp}(k) = \sqrt{k^2 + m^2}$. The refractive (perpendicular) index are defined by

$$n_{i\perp}(\mathbf{k}) = \frac{|\mathbf{k}|}{\omega_{i\perp}(\mathbf{k})}, \quad (i = 1, 2, 3). \quad (5.28)$$

(b) The second case consists in considering $\mathbf{v}$ orthogonal to both $\mathbf{B}$ and $\mathbf{k}$, but $\mathbf{B}$ parallel to $\mathbf{k}$ : $\mathbf{B} \cdot \mathbf{v} = \mathbf{k} \cdot \mathbf{v} = 0$ and $\mathbf{B} \cdot \mathbf{k} = B k$. In this case, the equation (5.16) is

$$\omega_\parallel^2 \left(\omega_\parallel^2 - k^2\right) \left[\left(1 + \xi B^2\right)\left(\omega_\parallel^2 - k^2\right) - v^2\right] = 0, \quad (5.29)$$

where $\omega_\parallel$ is now the frequency for $\mathbf{B}$ parallel to $\mathbf{k}$. The trivial solution is $\omega_\parallel = 0$, and the others solutions are read below :

$$\omega_{1\parallel}(k) = k, \quad (5.30a)$$

$$\omega_{2\parallel}(k) = \sqrt{\frac{(2k^2 + m^2)(1 + f B^2) + v^2 + g_a^2 B^2 - \Lambda}{2(1 + f B^2)}}, \quad (5.30b)$$

$$\omega_{3\parallel}(k) = \sqrt{\frac{(2k^2 + m^2)(1 + f B^2) + v^2 + g_a^2 B^2 + \Lambda}{2(1 + f B^2)}}, \quad (5.30c)$$

where,

$$\Lambda = \sqrt{[g_a^2 B^2 + m^2(1 + f B^2)]^2 + 2v^2 g_a^2 B^2 - 2v^2 m^2 (1 + f B^2) + v^4}. \quad (5.31)$$

The first solution (5.30a) is the usual photon DR due to $\mathbf{B} \times \mathbf{k} = \mathbf{0}$ in the $a$-parameter in (5.12a). The limits of $f \to 0$ and $c_1 = 1$ also recover the DRs of the axionic ED coupled to CFJ term in the presence of the external magnetic field $B$. The limit $g_a \to 0$, when the axion is decoupled from the CFJ ED, the DRs are reduced to the results : $\omega_{2\parallel}(k) = \sqrt{k^2 + v^2 \, (1 + f \, B^2)^{-1}}$ and $\omega_{3\parallel}(k) = \sqrt{k^2 + m^2}$ for $m > v$. This confirm the same results recovered in the case (a). The correspondent refractive (parallel) index are defined by

$$n_{i\parallel}(\mathbf{k}) = \frac{|\mathbf{k}|}{\omega_{i\parallel}(\mathbf{k})} \; , \; (i = 1, 2, 3) \; . \tag{5.32}$$

where we must substitute the DRs (5.30a)-(5.30c). Notice that, in both the cases (a) and (b), the refractive index of the medium depends on the modulus $\mathbf{k}$, so, consequently, it depends on the wavelength, as $\lambda = 2\pi/|\mathbf{k}|$.

To close this section, in possess of the set of dispersion relations (5.26) and (5.30), we call back one of the motivations to do this work, namely, to keep track of how the three different physical scenarios we bring together in the action of eq. (5.1) interfere with one another, which is manifested by means of the terms coupling the parameters of the different scenarios. Keeping in mind that the coefficients $c_1$, $f$ and $d$ express the non-linearity and that $g_a$ incorporates the axion-photon coupling and the coefficient $c_1$, the presence of the denominator $(1 + fB^2)$, common to all frequency solutions, in combination with the terms in $m^2$, $v^2$, $g_a B^2$, $fB^2 m^2$ and $fB^2 m^2 v^2$, as it appears in eqs. (5.26a)-(5.26c) and (5.30a)-(5.30c), shows in an explicit way how the three different physics mix among themselves to produce tiny effects in optical quantities like phase and group velocities and refraction indices. The explicit forms of the coefficients in terms of the non-linear electrodynamic models of Euler-Heisenberg, Born-Infeld and ModMax will be shown in the next section, namely, by equations (5.39),(5.45) and (5.51), respectively.

## 5.3 The birefringence phenomenon

Birefringence is an optical property of an anisotropic medium expressed by the dependence of the refractive index on the polarization and direction of propagation of an electromagnetic wave. Just to recall, the polarization conventionally refers to the configuration of the electric field of the wave. However, in the previous Section, we have worked out refraction indices associated to the propagation of the waves in two situations: perpendicular and parallel to the background magnetic field: $\mathbf{k} \cdot \mathbf{B} = 0$ and $\mathbf{k} \cdot \mathbf{B} = |\mathbf{k}||\mathbf{B}|$, respectively, with no reference to the polarization established by the electric field. Eqs. (5.28) and (5.32) explicitly show how the non-linearity – manifested by the external magnetic field – the axion parameters and the LSV vector interfere with one another in the expressions for the perpendicular and parallel refraction indices. And we would like to stress that we are here adopting the point of view that the phenomenon of birefringence manifests itself by the difference between the refractive indices of eqs. (5.28) and (5.32) , as defined below,

$$\Delta n_{ij}(\mathbf{k}) = n_{i\parallel}(\mathbf{k}) - n_{j\perp}(\mathbf{k}) \ , \ (i,j = 1,2,3) \ , \tag{5.33}$$

where we are contemplating the cases in which $i = j$ and $i \neq j$; in general, $\Delta n_{ij} \neq 0$, and it depends on the wavelength, which characterizes dispersive propagation. Notice also that $\Delta n_{ij} \neq \Delta n_{ji}$ according to the definition (5.33). The difference between the refraction indices in these situations is exclusively due to the choice of the wave propagation direction with respect to the external **B**-field.

Substituting the results from the previous section, the variation of refractive index in the case of $i = j$ are read

$$\Delta n_{11} = 1 - \frac{1}{\sqrt{1 - d\, B^2}} \ , \tag{5.34a}$$

$$\Delta n_{22}(k) \simeq 1 - \sqrt{1 + f\, B^2} - \frac{g_a^2\, B}{2} \frac{B\, \sqrt{1 + f\, B^2}}{m^2 + f\, B^2\, (k^2 + m^2)}$$
$$+ \frac{v^2}{2k^2} \left( \sqrt{1 + f\, B^2} - \frac{1}{1 + f\, B^2} \right) \ , \tag{5.34b}$$

$$\Delta n_{33}(k) \simeq \frac{g_a^2\, B}{2} \frac{B\, k}{(1 + f\, B^2)(k^2 + m^2)^{3/2}} \frac{k^2}{m^2 + f\, B^2 (k^2 + m^2)} \ , \tag{5.34c}$$

where we have considered that $g_a$ is very weak in comparison with the squared inverse of the magnetic background ($g_a^2\, B \ll 1$), and the wave number ($k = 2\pi/\lambda$) is much greater than the CFJ parameter, *i. e.*, $k \gg v$. The birefringence effects for $i \neq j$ are read below :

$$\Delta n_{12}(k) \simeq 1 - \sqrt{1 + f B^2} \left[ 1 + \frac{v^2}{2k^2} + \frac{g_a^2 B^2}{2(m^2 + fB^2(k^2 + m^2))} \right], \tag{5.35a}$$

$$\Delta n_{13}(k) \simeq 1 - \frac{k}{\sqrt{k^2 + m^2}} + \frac{k}{\sqrt{k^2 + m^2}} \frac{g_a^2 B^2/2}{f B^2 (k^2 + m^2) + m^2}, \tag{5.35b}$$

$$\Delta n_{23}(k) \simeq 1 - \frac{k}{\sqrt{k^2 + m^2}} - \frac{v^2}{2k^2(1 + f B^2)} + \frac{g_a^2 B}{2\sqrt{k^2 + m^2}} \frac{B k}{m^2 + f B^2(k^2 + m^2)}, \tag{5.35c}$$

$$\Delta n_{21}(k) \simeq 1 - \frac{1}{\sqrt{1 - d B^2}} - \frac{v^2}{2k^2(1 + f B^2)}, \tag{5.35d}$$

$$\Delta n_{31}(k) \simeq \frac{k}{\sqrt{k^2 + m^2}} - \frac{1}{\sqrt{1 - d B^2}} + \frac{g_a^2 B}{2} \frac{B k}{m^2 + f B^2 (k^2 + m^2)}, \tag{5.35e}$$

$$\Delta n_{32}(k) \simeq \frac{k}{\sqrt{k^2 + m^2}} - \sqrt{1 + f B^2} + \frac{v^2}{2k^2(1 + f B^2)}$$
$$- \frac{g_a^2 B}{2} \left[ \frac{B k}{(1 + f B^2)(k^2 + m^2)} + \frac{B \sqrt{1 + f B^2}}{m^2 + f B^2(k^2 + m^2)} \right]. \tag{5.35f}$$

Turning off the magnetic background ($B \to 0$), birefringence emerges in all the results with $\lim_{B \to 0} \Delta n_{ij} \neq 0$, for $i \neq j$, and $\lim_{B \to 0} \Delta n_{ij} = 0$ for $i = j$. Therefore, in this limit of $B \to 0$, the birefringence phenomenon appears only due to the axion mass and the CFJ parameter. Only the expression (5.34a) does not depend on the wavelength. For the usual Maxwell ED coupled to the axion and the CFJ term, the limits of $c_1 \to 1$, $d \to 0$ and $f \to 0$ yield the results below :

$$\Delta n_{11} = 0, \tag{5.36a}$$

$$\Delta n_{22} \simeq -\frac{g^2 B^2}{2m^2}, \tag{5.36b}$$

$$\Delta n_{33} \simeq \frac{g^2 B^2}{(k^2 + m^2)^{3/2}} \frac{k^2}{m^2 - v^2}, \tag{5.36c}$$

$$\Delta n_{12} = -\Delta n_{21} \simeq -\frac{v^2}{2k^2}, \tag{5.36d}$$

$$\Delta n_{13} = -\Delta n_{31} \simeq 1 - \frac{k}{\sqrt{k^2 + m^2}}, \tag{5.36e}$$

$$\Delta n_{23} = -\Delta n_{32} \simeq 1 + \frac{v^2}{2k^2} - \frac{k}{\sqrt{k^2 + m^2}}. \tag{5.36f}$$

The limit $g_a \to 0$, for which we have a non-linear ED coupled to the CFJ term, the results

(5.34a)-(5.35f) are reduced to

$$\Delta n_{11} = 1 - \frac{1}{\sqrt{1 - d\, B^2}}, \tag{5.37a}$$

$$\Delta n_{22}(k) \simeq 1 - \sqrt{1 + f\, B^2} + \frac{v^2}{2k^2}\left(\sqrt{1 + f\, B^2} - \frac{1}{1 + f\, B^2}\right), \tag{5.37b}$$

$$\Delta n_{33}(k) = 0, \tag{5.37c}$$

$$\Delta n_{12}(k) \simeq 1 - \sqrt{1 + f\, B^2}\left(1 + \frac{v^2}{2k^2}\right), \tag{5.37d}$$

$$\Delta n_{13}(k) \simeq 1 - \frac{k}{\sqrt{k^2 + m^2}}, \tag{5.37e}$$

$$\Delta n_{23}(k) \simeq 1 - \frac{k}{\sqrt{k^2 + m^2}} - \frac{v^2}{2k^2(1 + f\, B^2)}, \tag{5.37f}$$

$$\Delta n_{21}(k) \simeq 1 - \frac{1}{\sqrt{1 - d\, B^2}} - \frac{v^2}{2k^2(1 + f\, B^2)}, \tag{5.37g}$$

$$\Delta n_{31}(k) \simeq \frac{k}{\sqrt{k^2 + m^2}} - \frac{1}{\sqrt{1 - d\, B^2}}, \tag{5.37h}$$

$$\Delta n_{32}(k) \simeq \frac{k}{\sqrt{k^2 + m^2}} - \sqrt{1 + f\, B^2} + \frac{v^2}{2k^2(1 + f\, B^2)}. \tag{5.37i}$$

In this case, the non-linearity plays a key role in the birefringence phenomenon. We shall discuss ahead birefringence by contemplating three non-linear electrodynamic models: Euler-Heisenberg, Born-Infeld and ModMax ED.

(a) The Euler-Heisenberg ED is described by the Lagrangian:

$$\mathcal{L}_{EH}(\mathcal{F}, \mathcal{G}) = \mathcal{F} + \frac{2\alpha^2}{45 m_e^4}\left(4\mathcal{F}^2 + 7\mathcal{G}^2\right), \tag{5.38}$$

where $\alpha = e^2 = (137)^{-1} = 0.00729$ is the fine structure constant, and $m_e = 0.5\,\text{MeV}$ is the electron mass. Taking this Lagrangian and applying the expansion presented in Section (5.1), the coefficients read as below:

$$d^{EH} \simeq \frac{16\alpha^2}{45 m_e^4} \quad \text{and} \quad f^{EH} \simeq \frac{28\alpha^2}{45 m_e^4}, \tag{5.39}$$

for a weak magnetic field. Substituting these coefficients in (5.34a)-(5.34c), we obtain

$$\Delta n_{11}^{(EH)} \simeq -\frac{8\alpha^2 B^2}{45 m_e^4} \,, \tag{5.40a}$$

$$\Delta n_{22}^{(EH)} \simeq -\frac{14\alpha^2 B^2}{45 m_e^4} - \frac{g^2 B^2}{2m^2} \,, \tag{5.40b}$$

$$\Delta n_{33}^{(EH)} \simeq \frac{g^2 B^2}{(k^2 + m^2)^{3/2}} \frac{k^2}{m^2 - v^2} \,, \tag{5.40c}$$

$$\Delta n_{12}^{(EH)} \simeq -\frac{14\alpha^2 B^2}{45 m_e^4} + \frac{v^2}{2k^2} - \frac{g^2 B^2}{2m^2} \,, \tag{5.40d}$$

$$\Delta n_{21}^{(EH)} \simeq -\frac{8\alpha^2 B^2}{45 m_e^4} + \frac{v^2}{2k^2} \,, \tag{5.40e}$$

$$\Delta n_{13}^{(EH)} \simeq -\Delta n_{31}^{(EH)} = 1 - \frac{k}{\sqrt{k^2 + m^2}} \,, \tag{5.40f}$$

$$\Delta n_{23}^{(EH)} \simeq -\Delta n_{32}^{(EH)} = \frac{k}{\sqrt{k^2 + v^2}} - \frac{k}{\sqrt{k^2 + m^2}} \,. \tag{5.40g}$$

Using the parameters previously defined, the solution (5.40a) yields the numeric value

$$\frac{|\Delta n_{11}^{(EH)}|}{B^2} \simeq \frac{8\alpha^2}{45 m_e^4} = 5.3 \times 10^{-24}\,\mathrm{T}^{-2} \,, \tag{5.41}$$

that is of the same order of the result presented by the PVLAS-FE experiment for vacuum magnetic birefringence, *i.e.*, $\Delta n_{PVLAS-FE}/B^2 = (19 \pm 27) \times 10^{-24}\,\mathrm{T}^{-2}$ [40]. The solution $\Delta n_{13}^{(EH)}$ is finite in both the limits of $B \to 0$, and $B \to \infty$. Turning off the magnetic background, the axion mass and the $v$-parameter contribute to the birefringence as follows:

$$\Delta n_{13}^{(EH)B \to 0} = \begin{cases} 1 - \frac{k}{\sqrt{k^2+m^2}} \simeq \frac{m^2}{2k^2} \,, & \text{if } m > v \,, \\ \\ 1 - \frac{k}{\sqrt{k^2+v^2}} \simeq \frac{v^2}{2k^2} \,, & \text{if } v > m \,. \end{cases} \tag{5.42}$$

In the case of $\Delta n_{21}^{(EH)}$, the CFJ $v$-parameter contributes to the birefringence when $B \to 0$:

$$\Delta n_{21}^{(EH)B \to 0} = \begin{cases} -1 + \frac{k}{\sqrt{k^2+v^2}} \simeq -\frac{v^2}{2k^2} \,, & \text{if } m > v \,, \\ \\ -1 + \frac{k}{\sqrt{k^2+m^2}} \simeq -\frac{m^2}{2k^2} \,, & \text{if } v > m \,. \end{cases} \tag{5.43}$$

In this same limit, the variation $\Delta n_{31}^{(EH)}$ is $\Delta n_{31}^{(EH)} \simeq -m^2/(2k^2)$ if $m > v$, and $\Delta n_{31}^{(EH)} \simeq -v^2/(2k^2)$ if $v > m$. These conditions constrain the CFJ parameter to depend on the range of the axion mass. This is the case of a ultra-light axion (ULA), candidate to DM, with mass lower-bounded according to $m \gtrsim 10^{-22}\,\text{eV}\,(2\sigma)$ from non-linear clustering [103], to be compared with $v \lesssim 10^{-23} - 10^{-25}$ GeV [90].

(b) The Born-Infeld ED is governed by the Lagrangian :

$$\mathcal{L}_{BI}(\mathcal{F},\mathcal{G}) = \beta^2 \left[ 1 - \sqrt{1 - 2\frac{\mathcal{F}}{\beta^2} - \frac{\mathcal{G}^2}{\beta^4}} \right] , \qquad (5.44)$$

The coefficients of the expansion around the magnetic background are

$$d^{BI} = \frac{1}{\beta^2 + B^2} \quad \text{and} \quad f^{BI} = \frac{1}{\beta^2} , \qquad (5.45)$$

in which both the coefficients go to zero when $\beta \to \infty$. Substituting these results in (5.33), the solutions for the birefringence in the BI theory (when $g^2 B \ll 1$) are given by

$$\Delta n_{11}^{(BI)} = \Delta n_{12}^{(BI)} = 1 - \sqrt{1 + \frac{B^2}{\beta^2}} , \qquad (5.46a)$$

$$\Delta n_{22}^{(BI)} \simeq 1 - \sqrt{1 + \frac{B^2}{\beta^2}} - \frac{g^2 B^2}{2} \frac{1}{m^2 + (k^2 + m^2) B^2/\beta^2}$$

$$+ \frac{v^2}{2k^2} \left( \sqrt{1 + \frac{B^2}{\beta^2}} - \frac{\beta^2}{\beta^2 + B^2} \right) , \qquad (5.46b)$$

$$\Delta n_{33}^{(BI)} \simeq \frac{g^2 B^2}{(k^2 + m^2)^{3/2}} \sqrt{1 + \frac{B^2}{\beta^2}} \frac{k^2}{(1 + B^2/\beta^2) m^2 - v^2}$$

$$\times \frac{m^2 - v^2}{m^2 - v^2 + (k^2 + m^2) B^2/\beta^2} , \qquad (5.46c)$$

$$\Delta n_{13}^{(BI)} = \Delta n_{23}^{(BI)} \simeq 1 - \frac{k}{\sqrt{k^2 + m^2}} , \qquad (5.46d)$$

$$\Delta n_{31}^{(BI)} = \Delta n_{32}^{(BI)} = \frac{k}{\sqrt{k^2 + m^2}} - \sqrt{1 + \frac{B^2}{\beta^2}} . \qquad (5.46e)$$

The limit $\beta \to \infty$ recovers the results (5.36a)-(5.36f). For a weak magnetic background, *i. e.*, $\beta \gg B$, the birefringence effect is residual in $B^2/\beta^2$ :

$$\Delta n_{11}^{(BI)} \simeq -\frac{B^2}{2\beta^2} , \qquad (5.47a)$$

$$\Delta n_{22}^{(BI)} \simeq -\frac{B^2}{2\beta^2} + \frac{3v^2 B^2}{4k^2 \beta^2} - \frac{g^2 B^2}{2m^2} , \qquad (5.47b)$$

$$\Delta n_{33}^{(BI)} \simeq \frac{g^2 B^2}{(k^2 + m^2)^{3/2}} \frac{k^2}{m^2 - v^2} . \qquad (5.47c)$$

In the limit $B \to 0$, the birefringence vanishes in (5.46a)-(5.46c). For the propagation effects, we use $\sqrt{\beta} = 16$ MeV associated with the electron's self-energy in BI ED. In this case, the solution (5.47a) has the numeric value

$$\frac{|\Delta n_{11}^{(BI)}|}{B^2} \simeq 3.2 \times 10^{-24} \, \text{T}^{-2} \,, \tag{5.48}$$

that is the same order of the PVLAS-FE experiment. When $B \to 0$, the variation $\Delta n_{23}^{(BI)}$ depend only on the axion mass and $v$-parameter :

$$\Delta n_{23}^{(BI)} \stackrel{B \to 0}{\simeq} \frac{\sqrt{2}\, k}{\sqrt{2k^2 + m^2 + v^2 - |m^2 - v^2|}} - \frac{\sqrt{2}\, k}{\sqrt{2k^2 + m^2 + v^2 + |m^2 - v^2|}} \simeq \frac{|m^2 - v^2|}{2k^2} \,, \tag{5.49}$$

if $k^2 \gg (m^2, v^2)$.

(c) The modified Maxwell (ModMax) ED is set by the Lagrangian

$$\mathcal{L}_{MM}(\mathcal{F}, \mathcal{G}) = \cosh\gamma\, \mathcal{F} + \sinh\gamma \sqrt{\mathcal{F}^2 + \mathcal{G}^2} \,, \tag{5.50}$$

where $\gamma$ is a real and positive parameter of this theory. The coefficients of the expansion in the magnetic background, in this case, are

$$d^{MM} = 0 \quad \text{and} \quad f^{MM} = 2\, e^\gamma \, \frac{\sinh\gamma}{B^2} \,. \tag{5.51}$$

Thus, the variation of the refractive index for a weak axion-coupling constant are read below :

$$\Delta n_{11}^{(MM)} = 0 \,, \tag{5.52a}$$

$$\Delta n_{22}^{(MM)} \simeq \frac{k\, e^\gamma}{\sqrt{e^{2\gamma} k^2 + v^2}} - \frac{k\, e^\gamma}{\sqrt{k^2 + v^2}}$$
$$+ \frac{1}{2} \frac{k}{\sqrt{k^2 + v^2}} \frac{g^2\, B^2\, e^{2\gamma}}{k^2 + v^2 - e^{2\gamma}(k^2 + m^2)} \,, \tag{5.52b}$$

$$\Delta n_{33}^{(MM)} \simeq -\frac{g^2\, B^2}{2(k^2 + m^2)^{3/2}} \frac{m^2 - v^2}{e^{2\gamma}\, m^2 - v^2}$$
$$\times \frac{e^\gamma\, k^3}{k^2 + v^2 - e^{2\gamma}(k^2 + m^2)} \,, \tag{5.52c}$$

$$\Delta n_{12}^{(MM)} \simeq 1 - e^\gamma \,, \tag{5.52d}$$

$$\Delta n_{13}^{(MM)} \simeq \Delta n_{23}^{(MM)} \simeq 1 - \frac{k}{\sqrt{k^2 + m^2}} \,, \tag{5.52e}$$

$$\Delta n_{21}^{(MM)} \simeq -e^{-2\gamma} \frac{v^2}{2k^2} \,, \tag{5.52f}$$

$$\Delta n_{32}^{(MM)} \simeq \frac{k}{\sqrt{k^2 + m^2}} - e^\gamma \,. \tag{5.52g}$$

The results (5.36a)-(5.36f) also are recovered in the limit $\gamma \to 0$. Notice that, with $\gamma \neq 0$, the birefringence remains in the second solution (5.52b) when $B \to 0$ :

$$\Delta n_{22}^{(MM)} \simeq \frac{k\, e^{\gamma}}{\sqrt{e^{2\gamma}\, k^2 + v^2}} - \frac{k\, e^{\gamma}}{\sqrt{k^2 + v^2}} . \tag{5.53}$$

This particular result is the case in which the ModMax ED is added to the CFJ term without the presence of the axion. The result (5.52f) shows the birefringence solution that depends directly on the $v$-CFJ parameter, and it goes to zero when $v \to 0$. When the magnetic field is null, the solution $\Delta n_{21}^{(MM)}$ is

$$\Delta n_{21}^{(MM)} \stackrel{B \to 0}{=} -1 + \frac{\sqrt{2}\, k}{\sqrt{2k^2 + m^2 + e^{-2\gamma}\, v^2 - e^{-2\gamma}|m^2 e^{2\gamma} - v^2|}} , \tag{5.54}$$

that depend on the conditions $m\, e^{\gamma} > v$ and $m\, e^{\gamma} < v$ :

$$\Delta n_{21}^{(MM)} = -1 + \frac{k}{\sqrt{k^2 + m^2}} \simeq -\frac{m^2}{2k^2} , \tag{5.55a}$$

$$\Delta n_{21}^{(MM)} = -1 + \frac{k}{\sqrt{k^2 + e^{-2\gamma}\, v^2}} \simeq -e^{-2\gamma} \frac{v^2}{2k^2} , \tag{5.55b}$$

respectively. This result confirms (5.52f). When the CFJ $v$-parameter predominates in relation to the axion mass, and with this condition, the birefringence is null in the limit $v \to 0$. In the case of $\Delta n_{23}^{(MM)}$, the birefringence is null in an intense magnetic field. The variation $\Delta n_{31}^{(MM)}$ is finite on both the limits $B \to 0$ and $B \to \infty$. When the magnetic background is intense, $\Delta n_{31}^{(MM)} = -1$, that does not depend on the any parameter of the theory.

Using the PVLAS-FE result, $\Delta n_{PVLAS-FE}/B^2 = (19 \pm 27) \times 10^{-24}\, \text{T}^{-2}$, in (5.52c), for a ULA of mass $m_a \simeq 10^{-13}$ eV, with the gamma-ModMax parameter $\gamma = 0.1$, the axion-photon coupling taken as $g_{a\gamma} = 8.8 \times 10^{-13}\, \text{GeV}^{-1}$, with a wavelength of $k = 0.25$ eV, we obtain the result :

$$v = 9.97 \times 10^{-23}\, \text{GeV} , \tag{5.56}$$

which is compatible with the CFJ parameter known in the literature [90].

Chapter 6

# Conclusions and Future Perspectives

In this thesis, we investigate the properties of the electromagnetic (EM) wave propagation in a general non-linear electrodynamics coupled to the axion field. Our approach involves expanding the nonlinear sector up to second order in the propagating fields in a uniform and constant EM background field. We obtain the correspondent field equations for the propagating EM and axion fields. Next, we investigate the dispersion relations and group velocities in the presence of magnetic and electric background fields, separately. The refractive indices associated with these dispersion relations depend on the wavelength due to contributions involving the axion mass ($m$) and coupling constant ($g$) in the magnetic background field. For the case of the electric background, the dependence on the wavelength occurs when $m \neq 0$. The permittivity and permeability tensors are calculated as function of the EM background and wave propagation frequencies. We study the conditions for these tensors to be positive through the eigenvalues of the permittivity and permeability matrices.

We apply all the results for the Born-Infeld (BI) electrodynamics coupled to the axion field. Consequently, we obtain the wave propagation properties for the BI-axion model in magnetic and electric background fields. The results of the usual Maxwell electrodynamics are recovered when the BI parameter is very large, and the coupling with the axion goes to zero. The magnetic permeability is manifestly positive in the BI-axion model with a magnetic background, while one of the solutions for the electric permittivity to be positive imposes the condition (4.55) on the $\omega$-frequency. Otherwise, if the $\omega$-frequency satisfies the constraints

$$\omega < -\sqrt{\mathbf{k}^2 + m^2 + \frac{g^2\,\mathbf{B}^2}{\sqrt{1+\mathbf{B}^2/\beta^2}}} \quad \text{and} \quad \omega > \sqrt{\mathbf{k}^2 + m^2 + \frac{g^2\,\mathbf{B}^2}{\sqrt{1+\mathbf{B}^2/\beta^2}}}\,, \qquad (6.1)$$

the medium can behave as a metamaterial. Similarly, in the axion-BI model with an electric background, the electric permittivity is positive if $\beta > |\mathbf{E}|$, and one of the solutions for the magnetic permeability to be positive constrains the $\omega$-frequency in eq. (4.64).

In addiction, we have investigated the birefringence phenomena in the BI-axion model for the situations with magnetic and electric background fields, separately. We obtain the variation of the refractive indices (parallel and perpendicular to the electric propagating amplitude) as function of the $k$-wave number, $\omega$-frequency, $\beta$-parameter and background fields in eqs. (4.67) and (4.74). Therefore, since we have three solutions for the $\omega$-frequencies, the variation of the refractive index has also three possible solutions for the birefringence. In both the cases with magnetic or electric background fields, in the weak field regime, the variation of the refractive index is residual in $g^2$, see eqs. (4.68a) and (4.75a). The third solution points out for the birefringence as a function of the $\beta$-parameter, axion mass and background fields, as described in eqs. (4.68b) and (4.75b). As we expect from usual electrodynamics, the absence of birefringence in these results is attained in the $\beta \to \infty$ limit, and for the axion mass approaching zero. In the case of the birefringence with magnetic background field, we use the results of the PVLAS-FE experiment, with $\sqrt{\beta} = 100$ GeV, $\lambda = 1064$ nm and $m = 1$ meV, to obtain the axion coupling constant $g = 9.065 \times 10^{-9}\,\text{GeV}^{-1}$. This result agrees with the upper bound $g < 6.4 \times 10^{-8}\,\text{GeV}^{-1}\,(95\%\,\text{C.L.})$ in the PVLAS-FE experiment. The birefringence in the electric background case gives the result $g^2/(2m^2)$ in the regime of weak electric field. This result must be interesting in connection with the optical Kerr effect to constrain the parameter space $(m, g)$ for the axion-like particle.

In the chapter 5, we propose a general non-linear electrodynamics coupled to a scalar axion to which we adjoin the Carrol-Field-Jackiw (CFJ) term. We expand the Lagrangian of the model around a uniform electromagnetic background field up to second order in the photon field. The CFJ term introduces a background 4-vector $v^\mu = (v^0, \mathbf{v})$, that consequently, breaks the Lorentz symmetry in the theory. The case with only an uniform magnetic background field ($\mathbf{B}$) is analyzed where the properties of the wave propagation are discussed. Thereby, we calculate the dispersion relations of the model for a space-like ($v^0 = 0$) CFJ term. The wave propagation is affected by three vectors $\mathbf{B}$, $\mathbf{k}$ (wave vector) and $\mathbf{v}$. The dispersion relations are obtained for two cases : (a) when $\mathbf{B}$, $\mathbf{k}$ and $\mathbf{v}$ are perpendicular among themselves, (b) when $\mathbf{v}$ is perpendicular to $\mathbf{B}$ and $\mathbf{k}$, but $\mathbf{B}$ and **k** are parallel vectors. These results allow us to defined the refractive index of this medium, and the birefringence phenomenon is discussed under these conditions. Since there are three different solutions for the dispersion relations, we discuss the possible cases of birefringence $\Delta n_{ij}$, with $i,j = 1,2,3$. We apply the birefringence results for three cases of non-linear ED well known in the literature : Euler-Heisenberg, Born-Infeld, and the ModMax ED. When the non-linearity is null, the birefringence effect emerges due to the axion coupling with the magnetic background. In some situations, when the magnetic field is turned off, the birefringence is due to the CFJ parameter, the axion mass and the parameter of the non-linear ED.

One of the solutions of Euler-Heisenberg ED exhibits the birefringence result $\Delta n_{11}^{(EH)}/B^2 \simeq 69.4 \times 10^{-24}\,\text{T}^{-2}$, that is compatible with the PVLAS-FE experiment for vacuum magnetic birefringence, i. e., $\Delta n_{PVLAS-FE}/B^2 = (19 \pm 27) \times 10^{-24}\,\text{T}^{-2}$. The third solution (5.40c) shows the birefringence positive as function of the magnetic background field. In the case of the Born-Infeld ED, one of solutions for the birefringence yields $|\Delta n_{11}^{(BI)}|/B^2 \simeq 3.2 \times 10^{-24}\,\text{T}^{-2}$, when the Born-Infeld parameter is bounded by the finite electron self-energy. This numeric value is of the same order of the PVLAS-FE experiment result. In the case of the ModMax ED, the birefringence of $\Delta n_{33}^{(MM)}$ assumes negative values depending on the magnetic background field. When the solutions of $\Delta n_{ij}$, for $i \neq j$, are analyzed, the CFJ spatial-parameter $(v)$ plays a fundamental rule in the case of $\Delta n_{21}^{(MM)}$, in the ModMax ED. In the range of $v > e^\gamma m$, where $m$ is the axion mass, and $\gamma$ is the ModMax parameter, the birefringence emerges thanks to the $v$-parameter.

In our purpose of investigating how different new physics interfere with one another through the photon sector, we point out that, in an interesting recent article, Li and Ma [111] pursue an inspection on the effects stemming from Loop Quantum Gravity (LQG) corrections to both the photon and fermioninc matter sectors of Electrodynamics. Among these corrections, there appears a non-linear (actually, cubic) term in the extended Ampère-Maxwell equation. Though modulated by LQG parameters, very strong external magnetic fields at the astrophysical or those generated in relativistic heavy ion colliders may be sufficient to enhance the associated LQG effects and, therefore, one can compute how these latter effects contribute to the axion physics through the photon-axion coupling, as we have considered here.

Therefore, considering still our motivation to relate non-linear photon effects with axion physics, we recall that we have here considered as electromagnetic backgrounds only constant and uniform fields. It remains to be contemplated, for example, situations with non-uniform external electric/magnetic fields that will be exchanging energy and momentum with the photon-axion system, and to compute the modified dispersion relations, the corresponding group velocities, refractive indices and birefringence which will become space-dependent as a consequence of the non-uniformity of the background.

The content contained in this work generated the following works [115, 116], The first one deals with the investigations of axions in non-linear electrodynamic scenarios discussed in chapter 4, while the second one deals with the inclusion of the effects of Lorentz violation in the non-linear axionic scenario treated in chapter 5. As a perspective, we are motivated by the recent work of CMS [117], that searches axion-like particles mediating nonresonant ZZ or ZH production, we seek to introduce the axion to the weak hypercharge and Higgs sectors of the standard model. In this sense, we hope to investigate the anomalous vertices that will arise from the interaction of the axion with particles of the standard model and look for possible limits in the coupling parameters.

Overall, we anticipate that research on axions will gain even more momentum in the upcoming years, particularly with the commencement of LHC Run 3. This phase will witness enhancements in beam luminosity and detector sensitivity, facilitating the exploration of uncharted scientific territories. For instance, the recent work [118] provides some perspectives on non-resonant process research involving ALPs at the LHC.

Concerning astrophysical quests for ALPs, it is worth noting significant advances in this area as well. The James Webb Space Telescope (JWST) may effectively improve existing bounds in the literature. Furthermore, there are considerable efforts to understand how dark matter works based on JWST observations, and ALPs may play a central role in this discussion,as highlighted in references such as: [119, 120, 121, 122].

An equally crucial dimension is the growing correlation between high-energy models and condensed matter systems, such as topological materials. In such systems, axions can emerge as quasiparticles, leading to very rich physical effects, such as fractional Hall effect, chiral anomaly, and Casimir-Lifshitz effect [123]. Finally, with regard to a unified model where phenomena such as axion physics, nonlinear electrodynamics, and Lorentz symmetry violation may occur simultaneously, we believe that topological materials are considered the optimal setting for exploring these interconnected phenomena. Although a system incorporating axions and nonlinear electrodynamics is quite natural due to the intense magnetic fields required for the production of these particles. We hope that this work can be useful for future researchers who may study the topics addressed here.

# Appendix

# Appendix A

# The energy-momentum tensor discursion

In this Appendix, we address the recent discussion about the definition of Poynting vector in axionic electrodynamics. According to ref. [124], one could define two possible Poynting vectors in terms of the constitutive relations and electromagnetic fields, namely, $\mathbf{S}_{DB} \sim \mathbf{D} \times \mathbf{B}$ or $\mathbf{S}_{EH} \sim \mathbf{E} \times \mathbf{H}$, and the authors claim that the choice leads to different phenomenological results. This issue has also been discussed in refs. [126, 125], where the authors point out the relevance of considering terms of the order $O(g^2)$, to analyze the corresponding conservation law. Here, we follow our own path to tackle the question: we avoid to start off from any definition; instead, we work exclusively with the field equations to naturally identify the expressions for the Poynting vector and the momentum density transported by the waves.

In what follows, we work out the energy-momentum tensor for the non-linear ED model coupled to the axion field in an electromagnetic background, as described by the Lagrangian (4.6). For this purpose, we contract the eq. (5.5) with $f^{\nu\alpha}$ and using the Bianchi identity, we arrive at

$$\partial^\mu \left[ c_1 f_{\mu\nu} f^{\nu\alpha} - \frac{1}{2} Q_{B\mu\nu\kappa\lambda} f^{\kappa\lambda} f^{\nu\alpha} + g \widetilde{\phi} \widetilde{F}_{B\mu\nu} f^{\nu\alpha} - \delta_\mu{}^\alpha \left( -\frac{1}{4} c_1 f_{\rho\sigma}^2 + \frac{1}{8} Q_{B\rho\sigma\kappa\lambda} f^{\rho\sigma} f^{\kappa\lambda} \right. \right.$$
$$\left. \left. - \frac{1}{2} g \widetilde{\phi} \widetilde{F}_{B\rho\sigma} f^{\rho\sigma} \right) \right] = J_\nu f^{\nu\alpha} + \frac{1}{4} (\partial^\alpha c_1) f_{\mu\nu}^2 + \frac{1}{4} (\partial^\alpha c_2) \widetilde{f}_{\mu\nu} f^{\mu\nu} - \frac{1}{8} (\partial^\alpha Q_{B\mu\nu\kappa\lambda}) f^{\mu\nu} f^{\kappa\lambda}$$
$$+ \frac{1}{2} g (\partial^\alpha \widetilde{\phi}) \widetilde{F}_{B\,\mu\nu} f^{\mu\nu} + \frac{1}{2} g \widetilde{\phi} \left( \partial^\alpha \widetilde{F}_{B\mu\nu} \right) f^{\mu\nu} - (\partial^\mu H_{B\mu\nu}) f^{\nu\alpha} . \quad (A.1)$$

Now, we multiply the axion field equation (5.6) by $\partial^\alpha \widetilde{\phi}$ and, after some algebraic manipulations, we end up with

$$\partial^\mu \left\{ \partial_\mu \widetilde{\phi} \, \partial^\alpha \widetilde{\phi} - \delta_\mu{}^\alpha \left[ \frac{1}{2} (\partial_\nu \widetilde{\phi})^2 - \frac{1}{2} m^2 \widetilde{\phi}^2 \right] \right\} = -\frac{1}{2} g \widetilde{F}_{B\mu\nu} f^{\mu\nu} \partial^\alpha \widetilde{\phi} . \quad (A.2)$$

The sum of the equations (A.1) with (A.2) yields the result

$$\partial^\mu \Theta_\mu{}^\alpha = \Omega^\alpha \,, \tag{A.3}$$

where the energy-momentum tensor of the system is given by

$$\begin{aligned}\Theta_\mu{}^\alpha &= (\partial_\mu \widetilde{\phi})(\partial^\alpha \widetilde{\phi}) + c_1 f_{\mu\nu} f^{\nu\alpha} - \frac{1}{2} Q_{B\mu\nu\kappa\lambda} f^{\kappa\lambda} f^{\nu\alpha} + g\,\widetilde{\phi}\,\widetilde{F}_{B\mu\nu} f^{\nu\alpha} \\ &\quad - \delta_\mu{}^\alpha \left[ -\frac{1}{4} c_1 f_{\rho\sigma}^2 + \frac{1}{8} Q_{B\rho\sigma\kappa\lambda} f^{\rho\sigma} f^{\kappa\lambda} + \frac{1}{2}(\partial_\rho \widetilde{\phi})^2 - \frac{1}{2} m^2 \widetilde{\phi}^2 - \frac{1}{2} g\,\widetilde{\phi}\,\widetilde{F}_{B\rho\sigma} f^{\rho\sigma} \right] \end{aligned} \tag{A.4}$$

and $\Omega_\alpha$ corresponds to the dissipative terms, namely,

$$\begin{aligned}\Omega^\alpha &= J_\nu f^{\nu\alpha} + \frac{1}{2} g\,\widetilde{\phi} \left( \partial^\alpha \widetilde{F}_{B\mu\nu} \right) f^{\mu\nu} - (\partial^\mu H_{B\mu\nu}) f^{\nu\alpha} \\ &\quad + \frac{1}{4} (\partial^\alpha c_1) f_{\mu\nu}^2 + \frac{1}{4} (\partial^\alpha c_2) \widetilde{f}_{\mu\nu} f^{\mu\nu} - \frac{1}{8} (\partial^\alpha Q_{B\mu\nu\kappa\lambda}) f^{\mu\nu} f^{\kappa\lambda} \,. \end{aligned} \tag{A.5}$$

The energy-momentum tensor (A.4) exhibits the well-known contributions of non-linear electrodynamics and axion field. Furthermore, due the presence of an electromagnetic background, we also obtain axion-photon mixing terms related to $\widetilde{\phi}\,\widetilde{F}_{B\mu\nu} f^{\nu\alpha}$ and $\widetilde{\phi}\,\delta_\mu{}^\alpha\,\widetilde{F}_{B\rho\sigma} f^{\rho\sigma}$. We highlight that our result does not include the influence of magnetic monopoles because we used the Bianchi identity. Notice that $\Omega^\alpha$ contains the usual contribution $J_\nu f^{\nu\alpha}$ and other terms involving derivatives of the background fields. For the particular case of vanishing axion field, we recover the results discussed in ref. [64]. We point out that the energy-momentum tensor is not symmetric, $\Theta^{\mu\alpha} \neq \Theta^{\alpha\mu}$, due to the background terms $Q_{B\mu\nu\kappa\lambda} f^{\kappa\lambda} f^{\nu\alpha}$ and $g\,\widetilde{\phi}\,\widetilde{F}_{B\mu\nu} f^{\nu\alpha}$. This fact also happens in scenarios with Lorentz symmetry violation (see, for instance, refs. [127, 128]). If we consider a uniform and constant electromagnetic background with $J^\mu = 0$, the eq. (A.3) leads to the conservation law $\partial^\mu \Theta_\mu{}^\alpha = 0$. When $\alpha = 0$, we obtain

$$\partial_t u + \nabla \cdot \mathbf{S} = 0 \,, \tag{A.6}$$

where $u := \Theta^{00}$ denotes the energy density,

$$\begin{aligned}u &= \frac{1}{2}(\partial_t \widetilde{\phi})^2 + \frac{1}{2}(\nabla \widetilde{\phi})^2 + \frac{1}{2} m^2 \widetilde{\phi}^2 + \frac{1}{2} c_1 (\mathbf{e}^2 + \mathbf{b}^2) \\ &\quad + \frac{1}{2} d_1 (\mathbf{e} \cdot \mathbf{E})^2 + \frac{1}{2} d_2 (\mathbf{e} \cdot \mathbf{B})^2 - \frac{1}{2} d_1 (\mathbf{b} \cdot \mathbf{B})^2 \\ &\quad - \frac{1}{2} d_2 (\mathbf{b} \cdot \mathbf{E})^2 + d_3 (\mathbf{e} \cdot \mathbf{E})(\mathbf{e} \cdot \mathbf{B}) + d_3 (\mathbf{b} \cdot \mathbf{E})(\mathbf{b} \cdot \mathbf{B}) \\ &\quad + g\,\widetilde{\phi}\,(\mathbf{b} \cdot \mathbf{E}) + g\,\widetilde{\phi}\,(\mathbf{e} \cdot \mathbf{B}) \,, \end{aligned} \tag{A.7}$$

and **S** corresponds to the Poynting vector, whose components are defined by $S^i := \Theta^{i0}$, such that

$$\begin{aligned} S^i &= (\partial^i \widetilde{\phi})(\partial_t \widetilde{\phi}) + c_1(\mathbf{e} \times \mathbf{b})^i + d_1 (\mathbf{e} \cdot \mathbf{E})(\mathbf{e} \times \mathbf{B})^i \\ &\quad - d_1 (\mathbf{b} \cdot \mathbf{B})(\mathbf{e} \times \mathbf{B})^i - d_2 (\mathbf{e} \cdot \mathbf{B})(\mathbf{e} \times \mathbf{E})^i - d_2 (\mathbf{b} \cdot \mathbf{E})(\mathbf{e} \times \mathbf{E})^i \\ &\quad + d_3 (\mathbf{e} \cdot \mathbf{B})(\mathbf{e} \times \mathbf{B})^i - d_3 (\mathbf{e} \cdot \mathbf{E})(\mathbf{e} \times \mathbf{E})^i + d_3 (\mathbf{b} \cdot \mathbf{E})(\mathbf{e} \times \mathbf{B})^i \\ &\quad - d_3 (\mathbf{b} \cdot \mathbf{B})(\mathbf{e} \times \mathbf{B})^i - g\widetilde{\phi}\,(\mathbf{e} \times \mathbf{E})^i \,. \end{aligned} \quad (A.8)$$

For $\alpha = j$, the conservation law is written as

$$\partial_t \mathbf{P} + \nabla \cdot \overleftrightarrow{\mathbf{T}} = \mathbf{0}\,, \quad (A.9)$$

where **P** stands for the momentum, with the components $P^i := \Theta^{0i}$ given by

$$\begin{aligned} P^i &= (\partial_t \widetilde{\phi})(\partial^i \widetilde{\phi}) + c_1 (\mathbf{e} \times \mathbf{b})^i - d_1 (\mathbf{e} \cdot \mathbf{E})(\mathbf{E} \times \mathbf{b})^i \\ &\quad - d_1 (\mathbf{b} \cdot \mathbf{B})(\mathbf{E} \times \mathbf{b})^i - d_2 (\mathbf{e} \cdot \mathbf{B})(\mathbf{B} \times \mathbf{b})^i \\ &\quad + d_2 (\mathbf{b} \cdot \mathbf{E})(\mathbf{B} \times \mathbf{b})^i - d_3 (\mathbf{e} \cdot \mathbf{B})(\mathbf{E} \times \mathbf{b})^i \\ &\quad - d_3 (\mathbf{e} \cdot \mathbf{E})(\mathbf{B} \times \mathbf{b})^i - d_3 (\mathbf{b} \cdot \mathbf{B})(\mathbf{B} \times \mathbf{b})^i \\ &\quad + d_3 (\mathbf{b} \cdot \mathbf{E})(\mathbf{E} \times \mathbf{b})^i + g\widetilde{\phi}\,(\mathbf{B} \times \mathbf{b})^i \,, \end{aligned} \quad (A.10)$$

and $\overleftrightarrow{\mathbf{T}}$ denotes the stress tensor, whose components $(\overleftrightarrow{\mathbf{T}})^{ij} := \Theta^{ij}$ yield the expression

$$\begin{aligned} (\overleftrightarrow{\mathbf{T}})^{ij} &= (\partial^i \widetilde{\phi})(\partial^j \widetilde{\phi}) - c_1(e^i e^j + b^i b^j) - d_1(\mathbf{e}\cdot\mathbf{E})\,E^i e^j - d_2(\mathbf{e}\cdot\mathbf{B})\,B^i e^j \\ &\quad + d_1(\mathbf{b}\cdot\mathbf{B})\,E^i e^j - d_2(\mathbf{b}\cdot\mathbf{E})\,B^i e^j - d_1(\mathbf{e}\cdot\mathbf{E})\,b^i B^j + d_2(\mathbf{e}\cdot\mathbf{B})\,b^i E^j \\ &\quad + d_1(\mathbf{b}\cdot\mathbf{B})\,b^i B^j + d_2(\mathbf{b}\cdot\mathbf{E})\,b^i E^j - d_3(\mathbf{e}\cdot\mathbf{B})\,E^i e^j - d_3(\mathbf{e}\cdot\mathbf{E})\,B^i e^j \\ &\quad - d_3(\mathbf{b}\cdot\mathbf{E})\,E^i e^j + d_3(\mathbf{b}\cdot\mathbf{B})\,B^i e^j - d_3(\mathbf{e}\cdot\mathbf{B})\,b^i B^j + d_3(\mathbf{e}\cdot\mathbf{E})\,b^i E^j \\ &\quad - d_3(\mathbf{b}\cdot\mathbf{E})\,b^i B^j - d_3(\mathbf{b}\cdot\mathbf{B})\,b^i E^j + g\widetilde{\phi}\,(b^i E^j - B^i e^j)\,. \end{aligned} \quad (A.11)$$

The presence of an electromagnetic background introduces some interesting features. Firstly, the energy density of the model acquires new contributions involving the axion field, given by $g\widetilde{\phi}\,(\mathbf{b}\cdot\mathbf{E})$ and $g\widetilde{\phi}\,(\mathbf{e}\cdot\mathbf{B})$. In addition, from the eqs. (A.8) and (A.10), we conclude that the Poynting vector does not coincide with the linear momentum. However, as already expected, both equations lead to the same expression if the electromagnetic background is switched of.

Finally, it should be mentioned that using the above definitions for $u, \mathbf{S}, \mathbf{P}$ and $\overleftrightarrow{\mathbf{T}}$, based on the conservation laws (A.6) and (A.9), the results will be consistent.

# Referências Bibliográficas